\documentclass[11pt]{article}
\usepackage{axodraw}
\usepackage{epsfig}
\usepackage{amsfonts}
\usepackage{amsmath}
\usepackage{bm,bbm}
\usepackage{cite}
\allowdisplaybreaks[1]

\input pix.sty

\newcommand{\ch}[1]{\chi_\rmii{$#1$}}
\newcommand{\chF}{\chi_\rmii{F}}
\newcommand{\chB}{\chi_\rmii{B}}
\newcommand{\muA}{\mu_\rmii{\!$A$}}
\newcommand{\muY}{\mu_\rmii{$Y$}}
\newcommand{\nS}{n_\rmii{$S$}}
\newcommand{\nG}{n_\rmii{$G$}}
\newcommand{\mW}{m_\rmii{$W$}}
\newcommand{\mZ}{m_\rmii{$Z$}}
\newcommand{\mWt}{m_\rmii{$\widetilde W$}}
\newcommand{\mZt}{m_\rmii{$\widetilde Z$}}
\newcommand{\mQt}{m_\rmii{$\widetilde Q$}}
\newcommand{\mH}{m_\rmii{$H$}}

\newcommand{\ko}{k_0}
\newcommand{\km}{k_-}
\newcommand{\kp}{k_+}
\newcommand{\qm}{q_-}
\newcommand{\qp}{q_+}
\newcommand{\aL}{a^{ }_\rmii{L}}
\newcommand{\aR}{a^{ }_\rmii{R}}
\renewcommand{\eq}{eq.~}
\renewcommand{\eqs}{eqs.~}
\renewcommand{\se}{sec.~}
\renewcommand{\ses}{secs.~}
\renewcommand{\fig}{fig.~}
\renewcommand{\figs}{figs.~}

\newcommand{\Nc}{N_{\rm c}}

\newcommand{\gammaE}{\gamma_\rmii{E}}

\newcommand{\rmO}{{\mathcal{O}}}
\newcommand{\bmu}{\bar\mu}
\newcommand{\CF}{C^{ }_\rmii{F}}
\newcommand{\CA}{C^{ }_\rmii{A}} 

\def\lsi{\raise0.3ex\hbox{$<$\kern-0.75em\raise-1.1ex\hbox{$\sim$}}}
\def\gsi{\raise0.3ex\hbox{$>$\kern-0.75em\raise-1.1ex\hbox{$\sim$}}}
\newcommand{\lsim}{\mathop{\lsi}}
\newcommand{\gsim}{\mathop{\gsi}}

\newcommand{\nF}{n_\rmii{F}}
\newcommand{\nB}{n_\rmii{B}}
 \renewcommand{\nF}[1]{n_\rmii{F{#1}}} 
 \renewcommand{\nB}[1]{n_\rmii{B{#1}}} 

\newcommand{\rmii}[1]{{\mbox{\tiny\rm{#1}}}}

\newcommand{\re}{\mathop{\mbox{Re}}}
\newcommand{\im}{\mathop{\mbox{Im}}}

\newcommand{\Tint}[1]{{\hbox{$\sum$}\!\!\!\!\!\!\!\int\,}_{\!\!\!\!\raise-0.9ex\hbox{$\scriptstyle{#1}$}}}
\newcommand{\Tinti}[1]{{{\Sigma}\!\!\!\!\raise0.3ex\hbox{$\int$}_\rmii{${#1}$}}}

\newcommand{\bi}{\begin{itemize}}
\newcommand{\ei}{\end{itemize}}
\newcommand{\hide}[1]{ }
\newcommand{\bsl}[1]{\,\slash\!\!\!\!{#1}\,}

\def\TAsc(#1,#2)(#3,#4,#5)%
{\SetWidth{2.0}\CArc(#1,#2)(#3,#4,#5)\SetWidth{1.0}}
\def\Lwidth{3}

\def\TAgl(#1,#2)(#3,#4,#5){\SetWidth{2.0}\PhotonArc(#1,#2)(#3,#4,#5){\Lwidth}%
{6.283 #3 mul 360 div #4 #5 sub #4 #5 sub mul sqrt mul Tdensity mul}%
\SetWidth{1.0}}
\def\TLgl(#1,#2)(#3,#4){\SetWidth{2.0}\Photon(#1,#2)(#3,#4){\Lwidth}
{#1 #3 sub #1 #3 sub mul #2 #4 sub #2 #4 sub mul add sqrt Tdensity mul}%
\SetWidth{1.0}}

\renewcommand{\picc}[1]{\;\parbox[c]{60pt}{\begin{picture}(60,30)(0,0)
\SetWidth{1.0}\SetScale{1.3} #1 \end{picture}}\; }
\def\Lwidth{1.3}

\renewcommand{\pic}[1]{\;\parbox[c]{30pt}{\begin{picture}(30,30)(0,-3)
\SetWidth{1.0}\SetScale{0.8} #1 \end{picture}}\;}
\renewcommand{\picb}[1]{\;\parbox[c]{45pt}{\begin{picture}(45,30)(0,-3)
\SetWidth{1.0}\SetScale{0.8} #1 \end{picture}}\;}
\def\procG{\picb{%
 \Lqu(0,30)(15,15)%
 \Lgl(0,0)(15,15)%
 \Lqu(15,15)(29,15)%
 \Lsc(31,14)(45,0)%
 \Line(29,15)(44,30)%
 \Line(31,14)(45,28)%
}}
\def\procH{\pic{%
 \Lqu(0,30)(15,26)%
 \Lgl(0,0)(15,4)%
 \Lsc(15,4)(15,24)%
 \Lsc(15,4)(30,0)%
 \Line(15,24)(30,28)%
 \Line(14,26)(30,30)%
}}
\def\procI{\pic{%
 \Lqu(0,0)(15,4)%
 \Lsc(0,30)(15,26)%
 \Lqu(15,4)(15,24)%
 \Lgl(15,4)(30,0)%
 \Line(15,24)(30,28)%
 \Line(14,26)(30,30)%
}}
\def\procJ{\pic{%
 \Lqu(0,30)(15,26)%
 \Lsc(0,0)(15,4)%
 \Lsc(15,4)(15,24)%
 \Lgl(15,4)(30,0)%
 \Line(15,24)(30,28)%
 \Line(14,26)(30,30)%
}}
\def\nprocG{\picb{%
 \Lqu(0,30)(15,15)%
 \Lgl(0,0)(15,15)%
 \Lqu(15,15)(30,15)%
 \Lgl(30,15)(45,0)%
 \Line(30,15)(37.25,21.75)%
 \GCirc(37.25,21.75){2.5}{0}
 \Line(36.5,22.5)(44,30)%
 \Line(38,21)(45,28)%
}}
\def\nprocH{\pic{%
 \Lqu(0,30)(15,25)%
 \Lgl(0,0)(15,4)%
 \Lgl(15,4)(15,25)%
 \Lgl(15,4)(30,0)%
 \Line(15,25)(22.25,27)%
 \GCirc(22.25,27){2.5}{0}
 \Line(22.5,26)(30,28)%
 \Line(22,28)(30,30)%
}}
\def\nprocI{\pic{%
 \Lqu(0,0)(15,4)%
 \Lgl(0,30)(15,25)%
 \Lqu(15,4)(15,25)%
 \Lgl(15,4)(30,0)%
 \Line(15,25)(22.25,27)%
 \GCirc(22.25,27){2.5}{0}
 \Line(22.5,26)(30,28)%
 \Line(22,28)(30,30)%
}}
\def\nprocIp{\picb{%
 \Lqu(0,30)(30,4)%
 \Lgl(0,0)(15,12)%
 \Lgl(20,16)(30,25)%
 \Lqu(30,4)(30,25)%
 \Lgl(30,4)(45,0)%
 \Line(30,25)(37.25,27)%
 \GCirc(37.25,27){2.5}{0}
 \Line(37.5,26)(45,28)%
 \Line(37,28)(45,30)%
}}
\def\nprocJ{\pic{%
 \Lqu(0,30)(15,25)%
 \Lsc(0,0)(15,4)%
 \Lgl(15,4)(15,25)%
 \Lsc(15,4)(30,0)%
 \Line(15,25)(22.25,27)%
 \GCirc(22.25,27){2.5}{0}
 \Line(22.5,26)(30,28)%
 \Line(22,28)(30,30)%
}}
\def\nprocK{\picb{%
 \Lsc(0,30)(15,15)%
 \Lsc(0,0)(15,15)%
 \Lgl(15,15)(30,15)%
 \Laqu(30,15)(45,0)%
 \Line(30,15)(37.25,21.75)%
 \GCirc(37.25,21.75){2.5}{0}
 \Line(36.5,22.5)(44,30)%
 \Line(38,21)(45,28)%
}}
\def\nprocL{\picb{%
 \Lgl(0,30)(15,15)%
 \Lgl(0,0)(15,15)%
 \Lgl(15,15)(30,15)%
 \Laqu(30,15)(45,0)%
 \Line(30,15)(37.25,21.75)%
 \GCirc(37.25,21.75){2.5}{0}
 \Line(36.5,22.5)(44,30)%
 \Line(38,21)(45,28)%
}}
\def\nprocM{\pic{%
 \Lgl(0,30)(15,25)%
 \Lgl(0,0)(15,4)%
 \Lqu(15,4)(15,25)%
 \Laqu(15,4)(30,0)%
 \Line(15,25)(22.25,27)%
 \GCirc(22.25,27){2.5}{0}
 \Line(22.5,26)(30,28)%
 \Line(22,28)(30,30)%
}}
\def\nprocMp{\picb{%
 \Lgl(0,30)(30,4)%
 \Lgl(0,0)(15,12)%
 \Lgl(20,16)(30,25)%
 \Lqu(30,4)(30,25)%
 \Laqu(30,4)(45,0)%
 \Line(30,25)(37.25,27)%
 \GCirc(37.25,27){2.5}{0}
 \Line(37.5,26)(45,28)%
 \Line(37,28)(45,30)%
}}
\def\nprocN{\picb{%
 \Lqu(0,30)(15,15)%
 \Laqu(0,0)(15,15)%
 \Lgl(15,15)(30,15)%
 \Laqu(30,15)(45,0)%
 \Line(30,15)(37.25,21.75)%
 \GCirc(37.25,21.75){2.5}{0}
 \Line(36.5,22.5)(44,30)%
 \Line(38,21)(45,28)%
}}
\def\nprocO{\pic{%
 \Lqu(0,30)(15,25)%
 \Laqu(0,0)(15,4)%
 \Lgl(15,4)(15,25)%
 \Laqu(15,4)(30,0)%
 \Line(15,25)(22.25,27)%
 \GCirc(22.25,27){2.5}{0}
 \Line(22.5,26)(30,28)%
 \Line(22,28)(30,30)%
}}
\def\nprocP{\pic{%
 \Lqu(0,30)(15,25)%
 \Lqu(0,0)(15,4)%
 \Lgl(15,4)(15,25)%
 \Lqu(15,4)(30,0)%
 \Line(15,25)(22.25,27)%
 \GCirc(22.25,27){2.5}{0}
 \Line(22.5,26)(30,28)%
 \Line(22,28)(30,30)%
}}
\def\nprocPp{\picb{%
 \Lqu(0,30)(30,4)%
 \Lqu(0,0)(15,12)%
 \Line(20,16)(30,25)%
 \Lgl(30,4)(30,25)%
 \Lqu(30,4)(45,0)%
 \Line(30,25)(37.25,27)%
 \GCirc(37.25,27){2.5}{0}
 \Line(37.5,26)(45,28)%
 \Line(37,28)(45,30)%
}}
\def\xproc{\picb{%
 \Laqu(26,14)(50,0)%
 \Lsc(0,15)(24,15)%
 \Line(24,15)(49,30)%
 \Line(26,14)(50,28)%
}}
\def\xprocG{\picb{%
 \Lgl(13,5)(13,15)%
 \Lgl(0,1)(13,5)%
 \Lgl(13,5)(26,1)%
 \Laqu(26,14)(50,0)%
 \Lsc(0,15)(24,15)%
 \Line(24,15)(49,30)%
 \Line(26,14)(50,28)%
}}
\def\xprocH{\pic{%
 \Lgl(33,-1.6)(38,7)%
 \Lgl(20,2)(33,-1.6)%
 \Lgl(33,-1.6)(42,-10.6)%
 \Laqu(26,14)(38,7)%
 \Laqu(38,7)(50,0)%
 \Lsc(0,15)(24,15)%
 \Line(24,15)(49,30)%
 \Line(26,14)(50,28)%
}}
\def\xnproc{\picb{%
 \Laqu(25,15)(50,0)%
 \Lgl(0,15)(25,15)%
 \Line(25,15)(37.25,21.75)%
 \Line(36.5,22.5)(49,30)%
 \Line(38,21)(50,28)%
 \GCirc(37.25,21.75){2.5}{0}
}}
\def\xnprocG{\picb{%
 \Laqu(25,15)(50,0)%
 \Lgl(0,15)(25,15)%
 \Line(25,15)(37.25,21.75)%
 \Line(36.5,22.5)(49,30)%
 \Line(38,21)(50,28)%
 \GCirc(37.25,21.75){2.5}{0}
 \Lgl(13,5)(13,15)%
 \Lgl(0,1)(13,5)%
 \Lgl(13,5)(26,1)%
}}
\def\xnprocH{\pic{%
 \Lgl(33,-1.6)(38,7)%
 \Lgl(20,2)(33,-1.6)%
 \Lgl(33,-1.6)(42,-10.6)%
 \Laqu(25,15)(38,7)%
 \Laqu(38,7)(50,0)%
 \Lgl(0,15)(25,15)%
 \Line(25,15)(37.25,21.75)%
 \Line(36.5,22.5)(49,30)%
 \Line(38,21)(50,28)%
 \GCirc(37.25,21.75){2.5}{0}
}}

\makeatletter \@addtoreset{equation}{section} \makeatother
\renewcommand{\theequation}{\arabic{section}.\arabic{equation}}
\makeatletter
\renewcommand\section{\@startsection {section}{1}{\z@}%
                                   {-5.5ex \@plus -1ex \@minus -.2ex}
                                   {2.3ex \@plus.2ex}%
                                   {\normalfont\large\bfseries}}
\renewcommand\subsection{\@startsection{subsection}{2}{\z@}%
                                     {-3.25ex\@plus -1ex \@minus -.2ex}%
                                     {1.5ex \@plus .2ex}%
                                     {\normalfont\normalsize\bfseries}}
\renewcommand\thesection {\@arabic\c@section}
\renewcommand\thesubsection   {\thesection.\@arabic\c@subsection}
\renewcommand{\@seccntformat}[1]{%
\csname the#1\endcsname.\hspace{1.0em}}
\makeatother

\begin{document}

\flushbottom

\begin{titlepage}

\begin{flushright}
\vspace*{1cm}
\end{flushright}
\begin{centering}
\vfill

{\Large{\bf
 Neutrino dynamics below the electroweak crossover
}} 

\vspace{0.8cm}

J.~Ghiglieri and M.~Laine
 
\vspace{0.8cm}

{\em
AEC, 
Institute for Theoretical Physics, 
University of Bern, \\ 
Sidlerstrasse 5, CH-3012 Bern, Switzerland \\}

\vspace*{0.8cm}

\mbox{\bf Abstract}
 
\end{centering}

\vspace*{0.3cm}
 
\noindent
We estimate the thermal masses and damping rates of active
($m < $~eV) and sterile ($M \sim$~GeV) neutrinos with  
thermal momenta $k\sim 3T$ at temperatures below 
the electroweak crossover (5~GeV $ < T < 160$~GeV). 
These quantities fix the equilibration or ``washout'' rates
of Standard Model lepton number densities. Sterile neutrinos interact via 
direct scatterings mediated by Yukawa couplings, and via their overlap 
with active neutrinos. Including all leading-order reactions
we find that the washout rate generally
exceeds the Hubble rate for 5~GeV $ < T < 30$~GeV. Therefore
it is challenging to generate a large lepton asymmetry facilitating
dark matter computations operating at $T < 5$~GeV, whereas the generation 
of a baryon asymmetry at $T > 130$~GeV remains an option. 
Our differential rates are tabulated in a form suitable for studies 
of specific scenarios with given neutrino Yukawa matrices. 

\vfill

 
\vspace*{1cm}
  
\noindent
June 2016

\vfill

\end{titlepage}

%
\section{Introduction}

There has been recent interest in the cosmological role that 
right-handed (sterile) neutrinos with masses in the GeV range could play.
The dynamics of such particles may lead
to a generation of a lepton asymmetry, which could
explain the observed baryon asymmetry 
of the universe~\cite{ars,as}.
The lepton asymmetry generation could continue after
it cannot be converted into a baryon asymmetry any more ($T \lsim 130$~GeV); 
if the resulting lepton asymmetry is several orders of magnitude larger 
than the baryon asymmetry, it might be resonantly converted into keV scale
sterile neutrino dark matter at $T \lsim 1$~GeV~\cite{sf}.
Computations of lepton asymmetry generation 
have been carried out for two almost degenerate 
generations of sterile neutrinos~\cite{singlet,shintaro,shuve,abada,n1,n2,n3},
as well as for 
three possibly less degenerate generations~\cite{dg,val}. If the 
masses fall below the 0.1~GeV range (but above the 0.1 MeV range), 
the production is so efficient that such particles should already
have been observed through the total energy density that they 
carry~\cite{neff}. If the masses are higher but still relatively close 
to this lower bound, roughly 0.5...5~GeV ($m^{ }_K ... m^{ }_B$), 
these particles
can be searched for with the planned SHiP experiment at CERN~\cite{ship}. 
Therefore, there is a need to refine the theoretical understanding
concerning the behaviour of such particles within the ultrarelativistic
plasma that filled the early universe. 

Computing reliably the lepton asymmetry generated in 
a specific scenario is extremely challenging, because both CP violation 
and complicated plasma physics play a role. At the same time
it is relatively straightforward to establish a constraint on whether
any lepton asymmetry would be washed out after its generation
(cf.\ e.g.\ refs.~\cite{rb,sd} for reviews). This is 
in analogy with attempts to explain baryon asymmetry through 
an electroweak phase transition: the washout constraint
$\Delta \langle \phi^\dagger \phi \rangle \gsim T^2/2$, where $\phi$
denotes a Higgs doublet and $\Delta$ a discontinuity across the 
transition temperature, already rules out a large class of theories, 
including the Standard Model which has no genuine transition. The goal 
of the present paper is to establish similar washout constraints for 
generating a large lepton asymmetry.

More precisely, we discuss a number of quantities that characterize the 
behaviour of GeV scale right-handed neutrinos within a plasma. 
One is called their
``production rate'': it tells how fast the particles are being produced
if their initial density is much below the equilibrium value. Another
is their ``equilibration rate'': it tells how fast the particles can 
adjust their number density if they are initially close to 
equilibrium but the temperature of the equilibrium
ensemble evolves, as is the case in the early universe. Despite being
conceptually different, it turns out that these quantities
can be related to each other~\cite{equil}. The main quantity that
we consider is the lepton number ``washout rate'', which tells how
fast any initial lepton asymmetry is being depleted in the presence
of right-handed neutrinos. This rate can also be related
to the rates mentioned above~\cite{washout}. In the following, 
we frequently refer to the ``production rate'', with the relations to 
the other quantities specified in \se\ref{se:setup}.

A number of previous computations of the right-handed neutrino
production rate are worth mentioning. For small masses ($M \ll \pi T$) 
the production rate has been computed at low temperatures
$T < 5$ GeV both for vanishing~\cite{numsm} and 
non-vanishing~\cite{dmpheno,hirata} lepton asymmetries. 
At high temperatures $T > 160$~GeV it has been computed
in the non-relativistic $M \gg \pi T$~\cite{salvio,nonrel,nb}, 
relativistic $M \sim \pi T$~\cite{mh,rel}, and 
ultrarelativistic $M \ll \pi T$ regimes~\cite{bb1,bb2}. Computations
in the ultrarelativistic regime are challenging, because
they require a nested resummation of the loop expansion in order to
generate a consistent weak-coupling series.  
An interpolation applicable for any $M/(\pi T)$ has also been suggested
for $T > 160$~GeV~\cite{interpolation}. 

The status as outlined above means that there
is a gap in our understanding  in the range 5~GeV$\,\lsim T \lsim 160$~GeV. 
A rough estimate was presented in appendix~A of ref.~\cite{singlet}, 
however this was not based on a controlled computation but just
included Born level $1\leftrightarrow 2$ processes with vacuum 
masses evolved through a changing Higgs expectation value. 
The main goal of the present paper is to 
fill the gap 5~GeV$\,\lsim T \lsim 160$~GeV.

The paper is organized as follows. 
The basic observables considered are defined and the structures of 
the corresponding results are outlined in \se\ref{se:setup}. 
In \se\ref{se:dir12} we discuss right-handed neutrino 
production through direct $1\leftrightarrow 2$ processes, 
as well as the so-called LPM resummation of the 
$1+n\leftrightarrow 2+n$ reactions that contribute at the same
order in the soft regime $M, \mW \ll \pi T$. 
In \se\ref{se:dir22} the production through direct
$2\leftrightarrow 2$ scatterings is considered. 
Sec.~\ref{se:indirect} is devoted to ``indirect'' production, via
an overlap with left-handed neutrinos, pointing out 
the importance of $2\leftrightarrow 2$ scatterings mediated by soft
gauge boson exchange at high temperatures and of 
$1\to 2$ gauge boson decays at low temperatures. 
Numerical results are collected in \se\ref{se:num}, 
and we conclude in \se\ref{se:concl}. A number of technical
details and remarks concerning NLO effects are relegated to five appendices. 

%
\section{Summary of the setup and main results}
\la{se:setup}

One of the physical quantities that we are interested in is the production
rate of right-handed neutrinos. 
The produced right-handed neutrinos have a momentum
$k \equiv |\vec{k}|$ and a mass $M$. The corresponding
on-shell four-momentum is denoted by  $\mathcal{K} = (\ko,\vec{k})$, 
where $\ko = \sqrt{k^2 + M^2}$. In light of the scenario relevant
for SHiP~\cite{ship}, we consider here small masses, $M \le 16$~GeV, 
and high temperatures, $T \ge 5$~GeV. Then the right-handed neutrinos
can be considered to be ``ultrarelativistic'', 
with momenta $k\sim \pi T$ and masses $M \ll \pi T$.

Let $h^{ }_{I a}$ be  
a neutrino Yukawa coupling, $a\in\{1,2,3\}$ 
a left-handed lepton generation index, 
and $I \in \{ 1,2,3\}$ a right-handed neutrino generation index, 
defined in a basis in which the Majorana mass matrix is real and diagonal:
\be
 \mathcal{L} = \mathcal{L}^{ }_\rmii{SM} 
 + \fr12 \sum_{I} \bar{N}^{ }_I 
 \bigl( i \gamma^\mu\partial_\mu - M^{ }_I \bigr) {N}^{ }_I
 - \sum_{I,a}
 \Bigl(
  \bar{N}^{ }_I \, h^{ }_{I a} \tilde{\phi}^\dagger  \aL\, \ell^{ }_a
  +
  \bar{\ell}^{ }_{a} \, \aR\, \tilde{\phi}\, h^{*}_{I a} N^{ }_I
 \Bigr)
 \;. \la{L_Majorana}
\ee
Here
$\tilde{\phi} = i \sigma_2 \phi^*$ is a Higgs doublet;   
$\aL, \aR$ are chiral projectors; 
and $\ell^{ }_a = (\nu\, e)^T_a$ is a left-handed lepton doublet.
For notational simplicity we normally suppress the generation index in 
a Majorana mass, i.e.\ $M^{ }_I \to M$.

We consider time scales large enough that all Standard Model (SM)
degrees of freedom are in thermal equilibrium. However quantities
whose interactions involve the $h^{ }_{I a}$, notably
right-handed neutrino phase space distributions
($\equiv f^{ }_{I\vec{k}}$) and 
Standard Model lepton densities ($\equiv n^{ }_a$), 
can be out of equilibrium. The task is to determine
the equilibration rates of these observables. We note that $n^{ }_a$
can be carried both by neutral and by charged leptons. 

For $T \ge 5$~GeV, 
all Standard Model leptons can be considered degenerate and massless
($\pi T \gg m_\tau \approx 1.8$~GeV). 
Then the dynamical information concerning the 
rates of interest is contained in the 2-point function of the
operator to which the right-handed neutrinos couple~\cite{equil}. 
For computational
convenience we first define the corresponding
imaginary-time correlator, 
\be
 \Pi^{ }_\rmii{E}(K) \equiv
 \tr \Bigl\{ 
 i \bsl{K} \! \int_0^{1/T} \!\! {\rm d}\tau \int_\vec{x} e^{i K\cdot X}
 \, \Bigl\langle
  (\tilde{\phi}^\dagger  \aL\, \ell)(X) \, 
  (\bar{\ell}\, \aR\, \tilde{\phi} ) (0)
 \Bigr\rangle^{ }_T \Bigr\}
 \;, \la{Pi}
\ee
where
$K \equiv (k_n,\vec{k})$ and $k_n$ is a fermionic
Matsubara frequency. Moreover, 
$X \equiv (\tau,\vec{x})$ denotes 
a Euclidean space-time coordinate  
and $\langle ... \rangle^{ }_T$
an equilibrium expectation value. 
The retarded correlator $\Pi^{ }_\rmii{R}$
can be expressed as an analytic continuation of $\Pi^{ }_\rmii{E}$ as
\be
 \Pi^{ }_\rmii{R}(\mathcal{K}) = 
 \left. \Pi^{ }_\rmii{E}(K)
 \right|^{ }_{k_n \to -i [\ko + i 0^+]}
 \;, \la{relation} 
\ee
and the rate observables are proportional to 
the spectral function $\rho \equiv \im \Pi^{ }_\rmii{R}$.

We choose a normalization for the phase 
distribution function $f^{ }_{I\vec{k}}$
such that the total number density of right-handed neutrinos, 
summed over the two spin states, reads
\be
 n^{ }_I = \int_\vec{k} 
 \, 2 f^{ }_{I\vec{k}}
 \;, \quad
 \int_\vec{k} \equiv  \int \! \frac{{\rm d}^3\vec{k}}{(2\pi)^3}
 \;. 
 \la{nI}
\ee
Denoting $\ko \equiv E^{ }_I$ for the mass $M^{ }_I$, 
it can be shown that~\cite{equil}
\be
 \dot{f}^{ }_{I\vec{k}} = \gamma^{ }_{I \vec{k}}\, 
 \bigl( \nF{}(E^{ }_I) - f^{ }_{I\vec{k}} \bigr) + 
 \rmO\bigl[ \bigl(\nF{} - f^{ }_{I\vec{k}}\bigr)^2, n_a^2 \bigr]
 \;, \la{fI_prod}
\ee
where the right-hand side was expanded to leading order
in small lepton densities, $\dot{f}$ refers to a covariant time derivative
in an expanding phase space background, 
and $\nF{}$ denotes the Fermi distribution
(similarly, $\nB{}$ denotes
a Bose distribution).\footnote{%
 The generalization of \eq\nr{fI_prod}
 to finite lepton densities can be found 
 in \eq(2.21) of  ref.~\cite{dmpheno}. Similarly, 
 the generalization of \eq\nr{na_prod} 
 to off-equilibrium right-handed neutrinos
 can be found in \eq(2.24) of ref.~\cite{dmpheno}. 
 }
The coefficient $\gamma^{ }_{I \vec{k}}$ can be called the 
(spin-averaged) ``{\em equilibration rate}'', and is given by
\be 
 \gamma^{ }_{I\vec{k}} = \sum_{a} 
 \frac{|h^{ }_{Ia}|^2 \im \Pi^{ }_\rmii{R}(\mathcal{K})}{E^{ }_I}
 + \rmO(h^4)
 \;. \la{gammaI}
\ee
This relation applies to all orders in Standard Model couplings. 

Normally, when referring to the right-handed neutrino 
{\em ``production rate''}, 
it is assumed that their number density is small, 
$f^{ }_{I\vec{k}} \ll \nF{}$. 
For this case \eqs\nr{nI} and \nr{fI_prod} imply that
\be
 \dot{n}^{ }_I
 =  
 \sum_{a} 
 \int_\vec{k} 
 \frac{2 \nF{}(E^{ }_I)|h^{ }_{Ia}|^2 \im \Pi^{ }_\rmii{R}(\mathcal{K})}
 {E^{ }_I}
 + \rmO(h^4, n^{ }_I)
 \;. \la{nI_prod}
\ee

The same processes by which right-handed neutrinos equilibrate or are
produced also violate lepton densities carried
by Standard Model particles. 
Because lepton numbers are violated, their equilibrium values vanish. 
Close to equilibrium,
the lepton densities evolve as 
\be
  \dot{n}^{ }_a = - \gamma^{ }_{ab}\, n^{ }_b +  
  \rmO\bigl[n_a (\nF{} - f^{ }_{I\vec{k}}) , n_a^3 \bigr]
  \;, \la{na_prod}
\ee
where the matrix of decay coefficients, 
or {\em ``washout rates''}, can be written as~\cite{washout} 
\be
 \gamma^{ }_{ab} = - \sum_I  \int_{\vec{k}}
 \frac{2 \nF{}'(E^{ }_I) |h^{ }_{Ia}|^2 
 \im \Pi^{ }_\rmii{R}(\mathcal{K})}{E^{ }_I}
 \; \Xi^{-1}_{ab} + \rmO(h^4)
 \;. \la{gamma_ab}
\ee
Here 
$\Xi^{ }_{ab} = \partial n^{ }_a / \partial {\mu^{ }_b} |_{\mu^{ }_b = 0}
 \sim T^2$
is a susceptibility matrix related to lepton densities.
It was determined up to 
next-to-next-to-leading order
(NNLO) in Standard Model couplings 
at $T \gsim 160$~GeV in ref.~\cite{susc},  
and leading-order results valid for $T \lsim 160$~GeV are
given in appendix~A. 
We note that $\Xi$ is non-diagonal, 
because the plasma
as a whole is charge neutral, so that changes in the number
densities of different lepton flavours are correlated. 

As is clear from \eqs\nr{gammaI}, \nr{nI_prod} and \nr{gamma_ab}, 
the dynamical information that we need is contained in the function
$\im\Pi^{ }_\rmii{R}$, obtained from \eq\nr{relation}. 
We now turn to its determination. 

In order to carry out a theoretically consistent computation, 
power-counting rules need to be established 
for the various scales appearing in the problem. 
We denote by $h_t$ the renormalized top Yukawa coupling; 
by $\Nc \equiv 3$ the number of colours;  
by $g^{ }_1$, $g^{ }_2$ the hypercharge and weak gauge couplings; 
and by $\lambda$ the Higgs self-coupling. 
The notation $g^2$ refers generically
to the couplings $g_1^2, g_2^2, h_t^2, \lambda$ which are taken
to be parametrically of the same order of magnitude, 
and ``small'' in the sense that $g^2 \ll \pi^2$.

Suppose that we are at a temperature $T < 160$~GeV so that, in gauge-fixed
perturbation theory, the neutral component of the Higgs field has an 
expectation value. The expectation value is denoted by $v$; 
at $T=0$, $v\simeq 246$~GeV. 
We mainly consider a regime in which $v\lsim T$, even though 
the case $\mW \gsim \pi T$, 
i.e.\ $v \gsim \pi T / g$, is considered as well. 
For $v \lsim T$ vacuum masses 
$\sim gv$ are of the same order as thermal masses $\sim gT$ but much
smaller than typical momenta $k \sim \pi T$. In other words, all particles
can be considered to be ultrarelativistic. Based on 
various numerical tests, this regime is numerically applicable 
in a rather broad temperature range, 
\be
 30\,\mbox{GeV}\,\lsim\, T \,\lsim\, 160\,\mbox{GeV}
 \;. \la{regime}
\ee
At lower temperatures, Higgs and gauge bosons become non-relativistic
and need to be decoupled from the computation (the top quark becomes
non-relativistic already at a somewhat higher temperature). 

%
\begin{figure}[t]
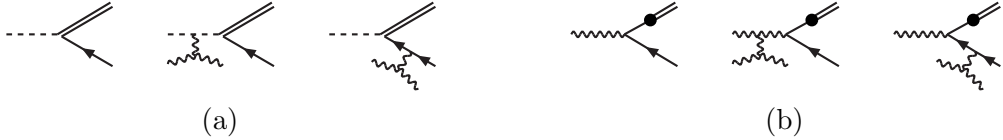


\hspace*{0.0cm}%
\begin{minipage}[c]{14cm}
\begin{eqnarray*}
&& 
 \xproc
 \hspace*{0.35cm}
 \xprocG
 \hspace*{0.35cm}
 \xprocH
 \hspace*{1.95cm}
 \xnproc
 \hspace*{0.35cm}
 \xnprocG
 \hspace*{0.35cm}
 \xnprocH
 \\[3mm] 
&& 
 \hspace*{2.7cm}
 \mbox{(a)}
 \hspace*{7.0cm}
 \mbox{(b)}
\end{eqnarray*}
\end{minipage}

\vspace*{0mm}

\caption[a]{\small 
 (a) Examples of $1+n\leftrightarrow 2+n$ processes for 
 the direct generation of right-handed neutrinos 
 from a Yukawa interaction. 
 (b) Examples of $1+n\leftrightarrow 2+n$ processes for 
 the generation of left-handed neutrinos which 
 subsequently oscillate into right-handed ones. 
 Arrowed, dashed, and wiggly lines correspond to 
 Standard Model fermions, scalars, and gauge bosons, respectively, whereas
 right-handed neutrinos are denoted by a double line. The closed blob includes
 a Higgs expectation value. 
} 
\la{fig:lpm}
\end{figure}
%

%
\begin{figure}[t]
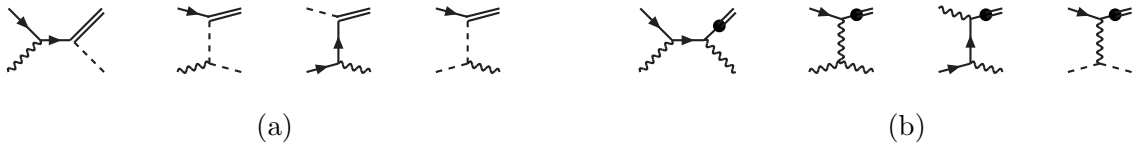


\begin{eqnarray*}
&& 
 \hspace*{-0.6cm}
 \procG
 \hspace*{0.45cm}
 \procH
 \hspace*{0.45cm}
 \procI
 \hspace*{0.45cm}
 \procJ
 \hspace*{1.45cm}
 \nprocG
 \hspace*{0.45cm}
 \nprocH
 \hspace*{0.45cm}
 \nprocI
 \hspace*{0.45cm}
 \nprocJ
 \\[3mm] 
&& 
 \hspace*{2.8cm}
 \mbox{(a)}
 \hspace*{7.9cm}
 \mbox{(b)}
\end{eqnarray*}

\vspace*{-4mm}

\caption[a]{\small 
 (a) Examples of $2\to 2$ processes for 
 the direct generation of right-handed neutrinos 
 from a Yukawa interaction. 
 (b) Examples of $2\to 2$ processes for 
 the generation of left-handed neutrinos which 
 subsequently oscillate into right-handed ones.
 The notation is as in \fig\ref{fig:lpm}.
 The complete set for case~(a) is shown in 
 \fig{1} of ref.~\cite{interpolation} and for case~(b) in 
 \fig\ref{fig:processes} below.
} 
\la{fig:types}
\end{figure}
%

In the regime of \eq\nr{regime}, 
there are two types of contributions to $\im\Pi^{ }_\rmii{R}$. 
First, the Higgs
field $\tilde\phi$ in \eq\nr{Pi} 
can represent a propagating mode 
(Goldstone or Higgs). This leads to the same processes
as have previously been considered in the symmetric phase~\cite{bb1,bb2}; 
examples of $1+n\leftrightarrow 2+n$ processes are shown 
in \fig\ref{fig:lpm}(a) and of $2\leftrightarrow2$ processes 
in \fig\ref{fig:types}(a).   
Second, the Higgs field could be replaced by its expectation value, 
$\tilde\phi \simeq (v\, 0)^T/\sqrt{2}$. Then we are left to 
consider processes experienced by an active (left-handed) neutrino. 
Examples of amplitudes are illustrated in 
\figs\ref{fig:lpm}(b) and \ref{fig:types}(b). 
We refer to first type as
a ``direct'' contribution and to the second as an 
``indirect'' one. 

When amplitudes such as those in \figs\ref{fig:lpm} 
and \ref{fig:types} are squared, 
there are no interference terms between the 
direct and indirect sets, provided that we
adopt a class of gauges (such as the $R_\xi$ gauge)
in which scalar and gauge fields do not transform
to each other. Then the rate can be written as 
\be 
 \im\Pi^{ }_\rmii{R} \; = \; 
 \left. \im\Pi^{ }_\rmii{R} \right|^{ }_\rmi{direct} \; + \; 
 \left. \im\Pi^{ }_\rmii{R} \right|^{ }_\rmi{indirect}
 \;, \la{splitup}
\ee
where the ``direct'' processes are like in sets~(a) 
of \figs\ref{fig:lpm} and \ref{fig:types}.
Like in the symmetric phase~\cite{bb1,bb2}, 
the direct term has the parametric magnitude
$
 \left. \im\Pi^{ }_\rmii{R} \right|^{ }_\rmi{direct} \sim g^2 T^2
$
(recalling that $|h^{ }_{I a}|^2$ has been factored out). 
In contrast the indirect term has a more complicated structure
(cf.\ \se\ref{ss:general}), 
\be
 \left. \im \Pi^{ }_\rmii{R} \right|^{ }_\rmi{indirect}
 \; = \; 
 \frac{v^2}{2}\, 
 \frac{M^2 \ko \Gamma }{(M^2 - m_\ell^2)^2 + \ko^2 \Gamma^2}
 \;, \la{indirect}
\ee
where $m_\ell$ is the active neutrino thermal mass in the
ultrarelativistic regime (cf.\ \eq\nr{mell}),\footnote{%
 When we exit the ultrarelativistic regime, $-m_\ell^2$
 is replaced by a more complicated function,  
 cf.\ \se\ref{ss:potential}.
 } 
and $\Gamma$ is its
thermal width (cf.\ \eq\nr{def_Gamma}).
The term in \eq\nr{indirect} is proportional
to $v^2$ because it originates from processes induced by 
electroweak symmetry breaking, and to~$M^2$ because in the massless
limit helicity and fermion number conservation would forbid  
transitions between left and right-handed states.

 We note in passing that, in an alternative language, 
 the combination
 $
  {|h^{ }_{Ia}|^2 v^2 M^2 }/ \{2[(M^2 - m_\ell^2)^2 + \ko^2 \Gamma^2]\}
 $
 originating from \eq\nr{indirect}
 can be interpreted as a medium-modified mixing angle squared. This  
 weights the part of the interaction rate $\ko\Gamma$ 
 of the weak eigenstates that is transmitted to 
 the sterile mass eigenstates. The interaction rate $\ko\Gamma$ also appears
 in the denominator of the effective mixing angle, thereby 
 contributing towards the ``unitarity'' of the conversion process. 
 Nevertheless, because $M^2$ and $m_\ell^2$ can cancel  
 against each other, the medium-modified mixing angle can 
 be much larger than the vacuum one. 

Now, an essential ingredient in our analysis is the determination 
of the active neutrino interaction rate $\Gamma$. We find that, because of 
strong infrared enhancement, $\Gamma \sim g^2 T/\pi$
in the regime of \eq\nr{regime}, 
cf.\ \eq\nr{indirect_HTL_gauge_soft}.
For $\ko \sim k \sim \pi T$ we thus get 
\be
 \left. \im \Pi^{ }_\rmii{R} \right|^{ }_\rmi{indirect}
 \; \sim \; 
 \frac{v^2}{2}\, 
 \frac{M^2 g^2 T^2 }{(M^2 - g^2 T^2)^2 + g^4 T^4}
 \;. \la{indirect_est}
\ee
This implies that for $M \sim gT$ 
we get 
$
 \left. \im \Pi^{ }_\rmii{R} \right|^{ }_\rmi{indirect}
 \sim v^2 
 \sim T^2 
$, 
i.e.\ the indirect production dominates  
over the direct one. The direct production dominates 
only if we go to the symmetric phase 
($v\lsim  gT/\pi$ if $M\sim gT$, or $v\lsim T$ if $M \sim g^2 T/\pi$).

In order to consolidate these findings, we proceed to discuss 
$
 \left. \im \Pi^{ }_\rmii{R} \right|^{ }_\rmi{direct}
$
and
$
 \left. \im \Pi^{ }_\rmii{R} \right|^{ }_\rmi{indirect}
$.
We start from the former, considering 
the $1+n\leftrightarrow 2+n$ and 
$2\to 2$ contributions in turn, 
and return to the indirect processes in \se\ref{se:indirect}.  

%
\section{Direct $1+n\leftrightarrow 2+n$ scatterings}
\la{se:dir12}

%
\subsection{LPM resummation in the symmetric phase}
\la{ss:symmetric}

An essential ingredient in the physics of the processes illustrated
in \fig\ref{fig:lpm}(a) is the proper inclusion of the so-called
Landau-Pomeranchuk-Migdal (LPM) resummation. Because of phase space
suppression, the rate of the $1\leftrightarrow 2$ process (after factoring
out $|h^{ }_{I a}|^2$) is $\sim m^2 \sim g^2 T^2$, where $m$ denotes
vacuum or thermal masses. On the other hand, adding gauge scatterings to the
$1\leftrightarrow 2$ result leads to no further suppression, because the 
exchanged gauge boson is soft, with a virtuality $\sim g^2 T^2$. 
Therefore all the gauge scatterings need to be resummed in order 
to obtain the correct leading-order result. 

Starting with the symmetric phase, 
the basic equations for the LPM resummation can be summarized
as follows~\cite{bb1}. We define  
\be
 \hat{H} \; \equiv \; - \frac{M^2}{2\ko} + 
 \frac{m_\ell^2 - \nabla_\perp^2}{2\omega_1} + 
 \frac{m_\phi^2 - \nabla_\perp^2}{2\omega_2} - i \, \Gamma(y) 
 \quad y \equiv |\vec{y}^{ }_\perp| 
 \;, \la{H}
\ee
where $\nabla^{ }_\perp$ is a two-dimensional gradient operating in 
directions orthogonal to $\vec{k}$, 
and the thermal masses of hard particles (with $k \gg m$) read 
\be
 m_\ell^2 = \frac{(g_1^2 + 3 g_2^2)T^2}{16}
 \;, \quad
 m_\phi^2 =  -\frac{\mH^2}{2} + 
  \Bigl( g_1^2 + 3 g_2^2 + 4 h_t^2  + 8 \lambda  
  \Bigr) \frac{T^2}{16}
 \;, \la{mell}
\ee
where $m^{ }_\rmii{$H$}\approx 125$~GeV is the physical Higgs mass. 
Soft gauge scatterings are represented by a thermal width which reads 
\be
 \Gamma({y}) = \frac{T}{8\pi} \sum_{i=1}^{2} d_i\, g_i^2
 \biggl[ \ln\biggl( \frac{m^{ }_\rmii{E$i$} y}{2} \biggr)
 + \gammaE + K_0 \bigl( m^{ }_\rmii{E$i$} y \bigr) \biggr]
 \;, \la{V}
\ee
where $d_1 \equiv 1$, $d_2 \equiv 3$, 
and $K_0$ is a modified Bessel function. The Debye masses associated
with the hypercharge and SU(2) gauge fields are defined as 
\be
 m^{2}_\rmii{E1} \; \equiv \; 
 \Bigl( \fr{\nS}6 + \frac{5\nG}{9} \Bigr) g_1^2 T^2 
 \;, \quad
 m^{2}_\rmii{E2} \; \equiv \; 
 \Bigl( \fr23 + \fr{\nS}6 + \frac{\nG}{3} \Bigr) g_2^2 T^2
 \;. \la{Debye}
\ee
Here $\nS \equiv 1$ is the number of Higgs doublets and 
$\nG \equiv 3$ the number of fermion generations.
The Debye masses appear frequently 
in the remainder of this paper. 
The Hamiltonian plays a role in the inhomogeneous equations
\be
 (\hat{H} + i 0^+)\, g(\vec{y}) \, = \, 
  \delta^{(2)}(\vec{y}) \;, \quad 
 (\hat{H} + i 0^+)\, \vec{f}(\vec{y}) \, = \, 
  -\nabla^{ }_\perp \delta^{(2)}(\vec{y}) 
 \;. \la{Seq}
\ee
{}From the solutions of these equations, the LPM-resummed contribution 
to $\im\Pi^{ }_\rmii{R}$ reads
\ba
 {\im\Pi_\rmii{R}^{\rmii{LPM},\rmi{symmetric}}} 
 & \equiv & 
  \frac{1}{4\pi} 
 \int_{-\infty}^{\infty} \! {\rm d}\omega_1 \, 
 \int_{-\infty}^{\infty} \! {\rm d}\omega_2 \;
 \delta(\ko - \omega_1 - \omega_2) \, 
 \bigl[ 1 - \nF{}(\omega_1) + \nB{}(\omega_2) \bigr] 
 \nn  
 & \times & 
  \, \frac{\ko}{\omega_2} 
 \lim_{\vec{y} \to \vec{0}}
 \biggl\{
   \frac{M^2}{\ko^2} \im\, \bigl[g(\vec{y} )\bigr]  + 
   \frac{1}{\omega_1^2} \im\, \bigl[\nabla_\perp\cdot \vec{f}(\vec{y} )\bigr] 
 \biggr\}
 \;. \la{lpm_symm}
\ea

%
\subsection{LPM resummation in the broken phase}
\la{ss:lpm_broken}

In the broken phase, the scalar sector splits up into Higgs 
and Goldstone modes. The contribution of the Goldstone modes depends 
strongly on the gauge choice; at tree-level, it is straightforward to 
verify that both the ``direct'' and ``indirect'' contributions are
gauge-dependent, but their sum is gauge-independent. Once LPM resummation
is incorporated, it is complicated to carry out 
computations in a general gauge, because this implies the presence 
of many different masses and correspondingly a large matrix of gauge
and scalar states mixed by gauge interactions. In the following
we restrict ourselves to the Feynman $R_\xi$ gauge, which minimizes 
the number of different states and masses.
In this gauge, the Goldstone modes correspond to  
the physical $W^\pm$ and $Z^0$ bosons, and we denote 
\be
 m_{\phi_0}^2 \; \equiv \; 2 \lambda v^2 \;, \quad
 m_{\phi_3}^2 \; \equiv \; \mZ^2 = \frac{(g_1^2 + g_2^2)v^2}{4} \;, \quad
 m_{\phi_1}^2 \; \equiv \; m_{\phi_2}^2 \; \equiv \; 
 \mW^2 = \frac{g_2^2 v^2}{4} 
 \;. \la{scalar_masses}
\ee

With non-degenerate scalar masses, 
the Green's functions in \eq\nr{Seq} split up into
several components, $g_0, ... , g_3$, 
and similarly for $\vec{f}$.
The LPM-resummed
$1\leftrightarrow 2$ contribution can be 
expressed as a generalization of \eq\nr{lpm_symm}, 
\ba
 {\im\Pi_\rmii{R}^{\rmii{LPM},\rmi{broken}}} 
 & \equiv & 
 \frac{1}{16\pi} 
 \int_{-\infty}^{\infty} \! {\rm d}\omega_1 \, 
 \int_{-\infty}^{\infty} \! {\rm d}\omega_2 \;
 \delta(\ko - \omega_1 - \omega_2) \, 
 \bigl[ 1 - \nF{}(\omega_1) + \nB{}(\omega_2) \bigr] 
 \nn  
 & \times & 
  \, \frac{\ko}{\omega_2} 
 \lim_{\vec{y} \to \vec{0}} \sum_{\mu = 0}^{3}
 \biggl\{
   \frac{M^2}{\ko^2} \im\, \bigl[g^{ }_\mu(\vec{y} )\bigr]  + 
   \frac{1}{\omega_1^2}
   \im\, \bigl[\nabla_\perp\cdot \vec{f}^{ }_\mu(\vec{y} )\bigr] 
 \biggr\}
 \;. \la{lpm_brok}
\ea
Here the contributions from $\mu = 1$ and $\mu = 2$ are equal, 
given that the charged scalar fields $\phi_1$ and $\phi_2$ are degenerate.  
The task now is to determine the Hamiltonian $\hat{H}$ for this situation. 

As a first step, we introduce notation for defining the gauge field
propagators in the broken phase. Because the temporal gauge field components  
get thermal masses, given by \eq\nr{Debye}, the temporal
and spatial gauge fields mix differently. In fact,  
the self-energies contain 
more structure than just thermal masses; in general the mixing is
momentum-dependent. However, 
because of a sum rule derived in ref.~\cite{sum1} and a more general 
argument presented in ref.~\cite{sum2}, the quantities of our interest
(see below) can be reduced to (static) 
Matsubara zero mode propagators.
Within the regime of validity of the 
Hard Thermal Loop (HTL) description~\cite{htl1,htl2}, 
the static self-energies are momentum independent. Therefore 
mixing can be described by constant angles, separate
for spatial and temporal gauge fields. 

The standard weak mixing angle can be defined as 
\be
 \sin(2\theta^{ }_\rmii{}) 
 \; \equiv \; \frac{2g^{ }_1 g^{ }_2}{g_1^2 + g_2^2}
 \;. \la{theta_w}
\ee
Denoting
$
 s^{ }_\rmii{} \equiv \sin \theta^{ }_\rmii{}
$, 
$
 c^{ }_\rmii{} \equiv \cos \theta^{ }_\rmii{}
$ and adopting a convention according to which a covariant derivative
acting on the Higgs doublet reads
$
  D_\mu\phi  \; \equiv \;  
 \bigl(
   \partial_\mu + \frac{i g^{ }_1}{2}  B^{ }_\mu
  - i g^{ }_2 %
   T^{a} A^{a}_\mu 
 \bigr)\, \phi 
$, where 
$
 \tr(T^a T^b) = \delta^{ }_{ab}/2
$, 
the spatial gauge field components can be diagonalized as
\be
 A_i^3 = c_\rmii{}^{ } Z_i^{ } - s_\rmii{}^{ } Q_i^{ } 
 \;, \quad
 B_i^{ }  = s_\rmii{}^{ } Z_i^{ } + c_\rmii{}^{ } Q_i^{ }
 \;, \quad
 i \in \{ 1,2,3 \} 
 \;, 
\ee
where $Q_\mu^{ }$ is the massless photon field and $B^{ }_\mu$
is the hypercharge field. The mixing angle of the static 
temporal components is denoted by $\tilde\theta$, 
and is given by
\be
 \sin(2\tilde\theta) \; \equiv \;
 \frac{\sin(2\theta) \mZ^2 }{\sqrt{\sin^2(2\theta) \mZ^4 +
 [\cos(2\theta) \mZ^2 + m_\rmii{E2}^2 - m_\rmii{E1}^2]^2}}
 \;. \la{thetat}
\ee
Denoting
$
 \tilde s^{ }_\rmii{} \equiv \sin \tilde\theta^{ }_\rmii{}
$, 
$
 \tilde c^{ }_\rmii{} \equiv \cos \tilde\theta^{ }_\rmii{}
$, 
the zero components are diagonalized by 
\be
 A_0^3 = \tilde c_\rmii{}^{ } \tilde Z_0^{ }
     - \tilde s_\rmii{}^{ } \tilde Q_0^{ } 
 \;, \quad
 B_0^{ }  = \tilde s_\rmii{}^{ } \tilde Z_0^{ }
     + \tilde c_\rmii{}^{ } \tilde Q_0^{ }
 \;. \la{temporal}
\ee
All diagonal fields have non-zero masses because of the 
thermal corrections in \eq\nr{Debye}: 
\ba
  \mWt^2 & \equiv & \mW^2 + m_\rmii{E2}^2
  \;, \quad
  \mZt^2 \; \equiv \;  m_{+}^2 
  \;, \quad
  \mQt^2 \; \equiv \;  m_{-}^2  
  \;, \\ 
  m_{\pm}^2 & \equiv & 
 \frac{1}{2} 
 \Bigl\{
   \mZ^2 + m_\rmii{E1}^2 + m_\rmii{E2}^2 \pm 
   \sqrt{\sin^2(2\theta) \mZ^4 +
   [\cos(2\theta) \mZ^2 + m_\rmii{E2}^2 - m_\rmii{E1}^2]^2}
 \Bigr\}
 \;. 
\ea

The gauge field combinations to which
neutral and charged left-handed leptons couple, respectively, are
$
 Z^{ }_0 \equiv c A^3_0 + s B^{ }_0
$, 
$
 Z'_0 \equiv - c A^3_0 + s B^{ }_0 
$. 
In the diagonal basis the corresponding propagators become
\ba
 \langle Z^{ }_0 Z^{ }_0 \rangle & = & 
 \cos^2(\theta - \tilde\theta) 
 \,  \langle \tilde Z^{ }_0 \tilde Z^{ }_0 \rangle
 + 
 \sin^2(\theta - \tilde\theta) 
 \,  \langle \tilde Q^{ }_0 \tilde Q^{ }_0 \rangle
 \;,  \la{z0z0} \\ 
 \langle Z'_0 Z'_0 \rangle & = & 
 \cos^2(\theta + \tilde\theta) 
 \,  \langle \tilde Z^{ }_0 \tilde Z^{ }_0 \rangle
 + 
 \sin^2(\theta + \tilde\theta) 
 \,  \langle \tilde Q^{ }_0 \tilde Q^{ }_0 \rangle
 \;. \la{zp0zp0}
\ea
These structures appear frequently below. 

Now, we return to the thermal width, 
denoted by $\Gamma$ in \eq\nr{H}. Choosing 
$\vec{k}$ to point in the $x_3$-direction, nearly light-like particles
couple to the gauge field components $A_0$ and $A_3$. In a thermal plasma, 
the soft scatterings mediated by these components are not identical, and 
the final result originates from the difference of the two contributions. 
Because of a sum rule~\cite{sum1,sum2}, 
the result can most simply
be expressed in terms of the static Matsubara zero-mode
sector ($q_n = 0$) related to these gauge potentials.
In the static limit the propagators of temporal components
can be expressed as in \eqs\nr{z0z0} and \nr{zp0zp0}. 
We define the widths related to $W^\pm$, $Z$, and $Z'$ exchanges as
\ba
 \Gamma^{ }_\rmii{$W$}(y) & \equiv & 
 \frac{g_2^2 T}{4}
 \! \int_{\vec{q}_\perp} 
 \!\! e^{i \vec{q}_\perp \cdot\vec{y}}
 \, \biggl[ \frac{1}{q_\perp^2 + \mW^2} 
 - \frac{1}{q_\perp^2 + \mWt^2 } \biggr]
 \;, \la{VWy} \\ 
 \Gamma^{ }_\rmii{$Z$}(y) & \equiv & 
 \frac{(g_1^2 + g_2^2)T}{4}
 \! \int_{\vec{q}_\perp} 
 \!\! e^{i \vec{q}_\perp \cdot\vec{y}}
 \, \biggl[ \frac{1}{q_\perp^2 + \mZ^2}
    - \frac{\cos^2(\theta-\tilde\theta)}
   {q_\perp^2 + \mZt^2 } 
    - \frac{\sin^2(\theta-\tilde\theta)}
   {q_\perp^2 + \mQt^2 } \biggr]
 \;, \la{VZy} \\ 
 \Gamma^{ }_\rmii{$Z'$}(y) & \equiv & 
 \frac{(g_1^2 + g_2^2)T}{4}
 \! \int_{\vec{q}_\perp} 
 \!\! e^{i \vec{q}_\perp \cdot\vec{y}}
 \, \biggl[ \frac{\cos^2(2\theta)}{q_\perp^2 + \mZ^2}
   + \frac{\sin^2(2\theta)}{q_\perp^2 }
    - \frac{\cos^2(\theta+\tilde\theta)}
  {q_\perp^2 + \mZt^2 } 
    - \frac{\sin^2(\theta+\tilde\theta)}
  {q_\perp^2 + \mQt^2 } \biggr]
 \;, \hspace*{6mm} \nn \la{Gamma_Zpy}
\ea
where 
$
 \int_{\vec{q}_\perp} 
 \equiv \int \frac{{\rm d}^{2-2\epsilon}\vec{q}_\perp}
 {(2\pi)^{2-2\epsilon}}
$
and 
$q_\perp \equiv |\vec{q}_\perp|$.
Dimensional regularization has been used for defining the value
of an infrared divergent integral in \eq\nr{Gamma_Zpy}, related to 
soft photon exchange, even 
though this divergence soon drops out
(cf.\ the discussion below \eq\nr{V3x3}). 
The full width matrix, 
in the space of neutral and charged scalars and leptons that
participate in the production of right-handed neutrinos, ordered as 
$\nu\phi^{ }_0$,
$\nu\phi^{ }_3$,
$e\phi^{ }_1$, 
$e\phi^{ }_2$, 
reads
\be
 \Gamma^{ }_{4\times4} \; = \; 
 \left( 
 \begin{array}{cccc}
 2 \Gamma^{ }_\rmii{$W$}(0) + \Gamma^{ }_\rmii{$Z$}(0) &
 - \Gamma^{ }_\rmii{$Z$}(y) & 
 - \Gamma^{ }_\rmii{$W$}(y) & - \Gamma^{ }_\rmii{$W$}(y) \\ 
 - \Gamma^{ }_\rmii{$Z$}(y) &   
 2 \Gamma^{ }_\rmii{$W$}(0) + \Gamma^{ }_\rmii{$Z$}(0) &
 - \Gamma^{ }_\rmii{$W$}(y) & - \Gamma^{ }_\rmii{$W$}(y) \\ 
 - \Gamma^{ }_\rmii{$W$}(y) & - \Gamma^{ }_\rmii{$W$}(y) & 
 2 \Gamma^{ }_\rmii{$W$}(0) + \Gamma^{ }_\rmii{$Z'$}(0) &
 - \Gamma^{ }_\rmii{$Z'$}(y) \\ 
 - \Gamma^{ }_\rmii{$W$}(y) & - \Gamma^{ }_\rmii{$W$}(y)  &
 - \Gamma^{ }_\rmii{$Z'$}(y) & 2
  \Gamma^{ }_\rmii{$W$}(0) + \Gamma^{ }_\rmii{$Z'$}(0) \\ 
 \end{array}
 \right)
 \;. \la{V4x4}
\ee
The arguments $0$ and $y$ correspond to 
self-energy and exchange contributions, respectively. The
combination $ 2 \Gamma^{ }_\rmii{$W$}(0) + \Gamma^{ }_\rmii{$Z$}(0) $ 
corresponds to the active neutrino width or interaction rate, 
re-derived in some more detail in \se\ref{ss:soft_t} 
(cf.\ \eq\nr{indirect_HTL_gauge_soft}).

Given that the pairs $e \phi^{ }_1$ and $e \phi^{ }_2$ are 
degenerate, we can choose one of them as a representative. Then, 
the matrix in \eq\nr{V4x4} can be reduced into a $3\times3$ form, 
\be
 \Gamma^{ }_{3\times3} \; = \; 
 \left( 
 \begin{array}{ccc}
 2 \Gamma^{ }_\rmii{$W$}(0) + \Gamma^{ }_\rmii{$Z$}(0) &
 - \Gamma^{ }_\rmii{$Z$}(y) & 
 - 2 \Gamma^{ }_\rmii{$W$}(y) \\ 
 - \Gamma^{ }_\rmii{$Z$}(y) & 
  2 \Gamma^{ }_\rmii{$W$}(0) + \Gamma^{ }_\rmii{$Z$}(0) &
 - 2 \Gamma^{ }_\rmii{$W$}(y) \\ 
 - \Gamma^{ }_\rmii{$W$}(y) & - \Gamma^{ }_\rmii{$W$}(y) & 
 2 \Gamma^{ }_\rmii{$W$}(0) + \Gamma^{ }_\rmii{$Z'$}(0)
   - \Gamma^{ }_\rmii{$Z'$}(y) \\ 
 \end{array}
 \right)
 \;. \la{V3x3}
\ee
A nice consequence of this reduction is that infrared
divergences related to photon exchange cancel 
in the combination $\Gamma^{ }_\rmii{$Z'$}(0)  - \Gamma^{ }_\rmii{$Z'$}(y)$.
To be explicit, 
\ba
 \Gamma^{ }_\rmii{$W$}(y) & = & 
 \frac{g_2^2T}{8\pi}
 \Bigl[
   K^{ }_0(\mW y) - 
   K^{ }_0(\mWt y)
 \Bigr]
 \;, \la{Vfirst} \\ 
 \Gamma^{ }_\rmii{$W$}(0) & = & 
 \frac{g_2^2T}{8\pi}
 \ln\frac{ \mWt }{ \mW }
 \;, \\ 
 \Gamma^{ }_\rmii{$Z$}(y) & = & 
 \frac{(g_1^2 + g_2^2)T}{8\pi}
 \Bigl[
   K^{ }_0(\mZ y) - 
   \cos^2(\theta-\tilde\theta) K^{ }_0(\mZt y) - 
   \sin^2(\theta-\tilde\theta) K^{ }_0(\mQt y)
 \Bigr]
 \;, \hspace*{6mm} \\ 
 \Gamma^{ }_\rmii{$Z$}(0) & = & 
 \frac{(g_1^2 + g_2^2)T}{8\pi}
 \Bigl[ \cos^2(\theta-\tilde\theta) 
      \ln\frac{ \mZt }{ \mZ } + 
        \sin^2(\theta-\tilde\theta) 
      \ln\frac{ \mQt }{ \mZ }
 \Bigr]
 \;, 
\ea
whereas the difference in the bottom-right component takes the form
\ba
 \Gamma^{ }_\rmii{$Z'$}(0) - \Gamma^{ }_\rmii{$Z'$}(y)  & = &  
 \frac{(g_1^2 + g_2^2)T}{8\pi} \biggl\{ 
    \cos^2(\theta+\tilde\theta)
    \Bigl[  K^{ }_0(\mZt y)
  + \ln\frac{\mZt y}{2}
           + \gammaE
    \Bigr]
 \nn & & \hspace*{1.8cm} 
  + \, 
    \sin^2(\theta+\tilde\theta)
    \Bigl[  K^{ }_0(\mQt y)
  + \ln\frac{\mQt y}{2}
           + \gammaE
    \Bigr]
 \nn & & \hspace*{1.8cm} 
  - \, 
    \cos^2(2\theta)
    \Bigl[  K^{ }_0(\mZ y) + \ln\frac{\mZ y}{2}
           + \gammaE
    \Bigr]
 \biggr\}
 \;. \la{Vlast}
\ea

With the width determined, let us generalize 
the Hamiltonian of \eq\nr{H} 
to contain a diagonal mass matrix,  
\be
 m_\phi^2 \; \to \; 
 \mathop{\mbox{diag}}(m_{\phi_0}^2,m_{\phi_3}^2,m_{\phi_1}^2) 
 \;. 
\ee
The Green's functions $g$ and $\vec{f}$ are generalized
to 3-component vectors. With the $3\times 3$ width $\Gamma^{ }_{3\times 3}$, 
we can then solve \eq\nr{Seq}, and 
insert the result into \eq\nr{lpm_brok}. 

As a crosscheck, we note that in the symmetric phase, 
the parameters appearing in \eqs\nr{Vfirst}--\nr{Vlast} behave as
$\mW \to 0, \mZ \to 0$, $\tilde\theta\to 0$, 
$\mWt \to m^{ }_\rmii{E2}$, 
$\mZt \to m^{ }_\rmii{E2}$, 
$\mQt \to m^{ }_\rmii{E1}$,
and $ \Gamma^{ }_\rmii{$Z'$}  \to \Gamma^{ }_\rmii{$Z$}$.
Moreover all the pairs 
$\nu\phi^{ }_0$,
$\nu\phi^{ }_3$,
$e\phi^{ }_1$ and 
$e\phi^{ }_2$ become degenerate, 
so we can reduce the $3\times 3$ matrix into a single function, 
\be
 \Gamma^{ }_{1\times 1} \; = \;  
 \lim_{ \mW, \mZ \to 0}
 \Bigl\{
 2 \bigl[
  \Gamma^{ }_\rmii{$W$}(0) - \Gamma^{ }_\rmii{$W$}(y)
 \bigr]
 + 
  \Gamma^{ }_\rmii{$Z$}(0) - \Gamma^{ }_\rmii{$Z$}(y)
 \Bigr\} 
 \;. 
\ee 
Noting that 
$
 \lim_{m\to 0} [
  K^{ }_0(m y) + \ln\frac{m y}{2}
           + \gammaE
 ] = 0 
$,
this agrees with \eq\nr{V}.

For a numerical solution, we make use of 
the general approach of ref.~\cite{original}, 
adapted to the problem at hand 
in ref.~\cite{interpolation}.
The idea is to express the solutions of the
inhomogeneous equations, \eq\nr{Seq}, in terms of the solutions 
of the homogeneous equation which are regular at origin. 
Choosing the normalizations of the regular solutions as 
\be
 u_{\ell,\mu}^{r}(\rho)
 \, = \, \rho^{1/2 + |\ell|} \, \bigl[ 1 + \rmO(\rho^2) \bigr]
 \;, \quad
 \mu \in \{0,1,2,3\}
 \;, \la{url_asymptotics}
\ee
where $\rho \equiv m^{ }_\rmii{E2}\, y$ and 
$\ell$ is an angular quantum number, we find
\ba
 \im\Pi_\rmii{R}^{\rmii{LPM},\rmi{broken}} 
 & = & 
  \frac{1}{4\pi^2} 
 \int_{-\infty}^{\infty} \! {\rm d}\omega \;
 \bigl[ 1 - \nF{}(\omega) + \nB{}(\ko - \omega) \bigr] 
 \nn  
 & \times & 
 \sum_{\mu = 0}^{3}  \int_0^\infty \! {\rm d}\rho \, 
 \biggl[
   \frac{\omega M^2}{4\ko^2}
   \im \biggl\{ \frac{ 1 }{[u_{0,\mu}^r(\rho)]^2} \biggr\} 
 +
   \frac{m_\rmii{E2}^2}{\omega}
   \im \biggl\{ \frac{1}{[u_{1,\mu}^r(\rho)]^2} \biggr\} 
 \biggr]
 \;. \la{lpm} \hspace*{6mm}
\ea
Here, as before, $\ko \equiv \sqrt{k^2 + M^2}$
and the kinematic range $M \ll k$ is assumed.
The numerical solution is straightforward, with a result
as illustrated in \fig\ref{fig:direct12} 
(the solid lines at high temperatures).\footnote{%
 Numerics can be sped up by realizing
 that the off-diagonal elements in \eq\nr{V3x3} fall off exponentially 
 for $y\gg \mW^{-1},\mWt^{-1},\mZ^{-1},\mZt^{-1}, \mQt^{-1}$. For
 large enough $\rho$ one can then switch to three separate solvers
 for the three independent $u_{\ell,\mu}^r(\rho)$. This is particularly
 advantageous for $T\lsim 60$~GeV, where a large 
 tree-level term is present, which requires integration to 
 large values of $\rho$ to reach the required accuracy.
}

\begin{figure}[t]

\hspace*{-0.1cm}
\centerline{%
  \epsfxsize=7.5cm\epsfbox{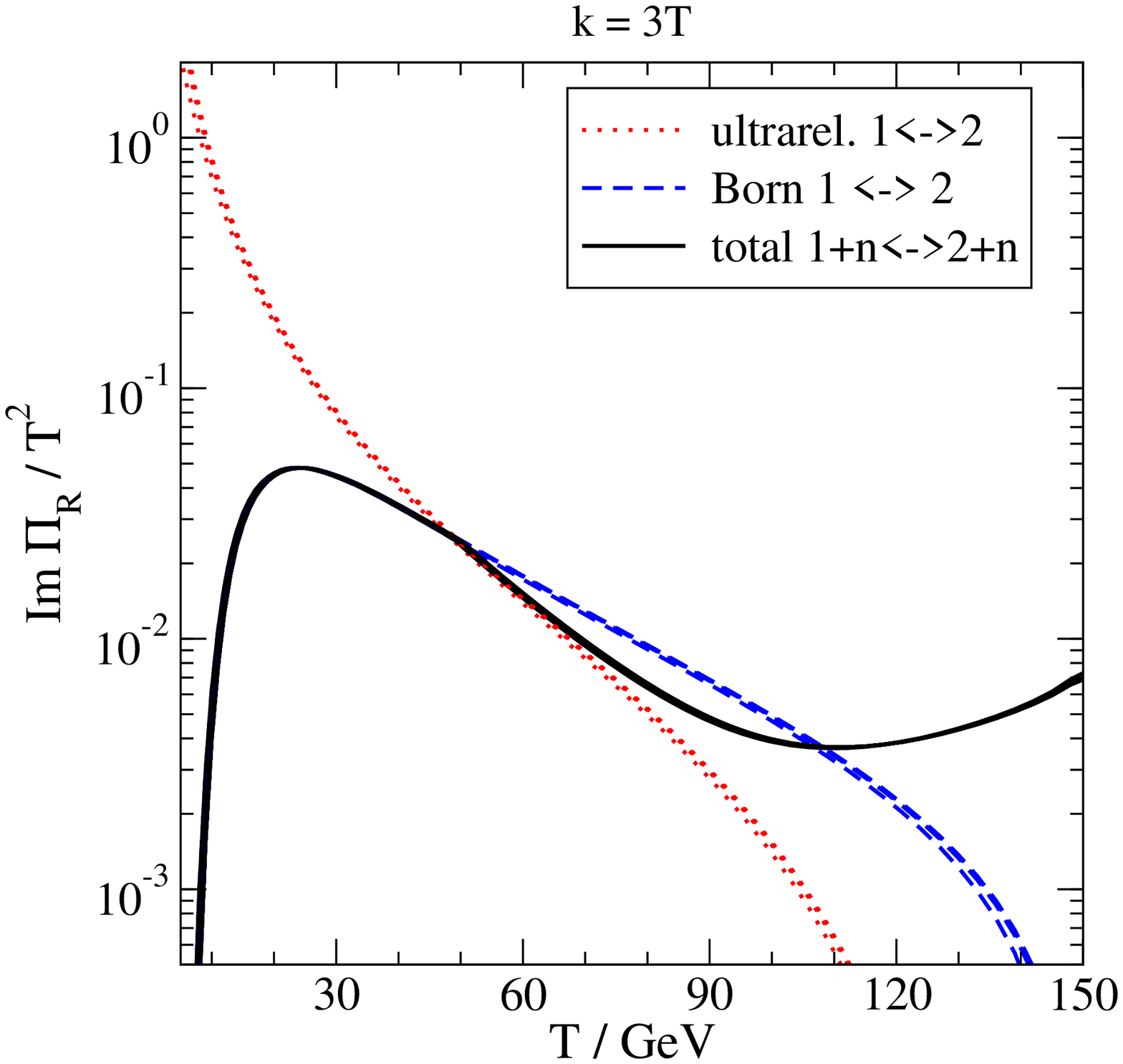}%
  \hspace{0.1cm}
  \epsfxsize=7.5cm\epsfbox{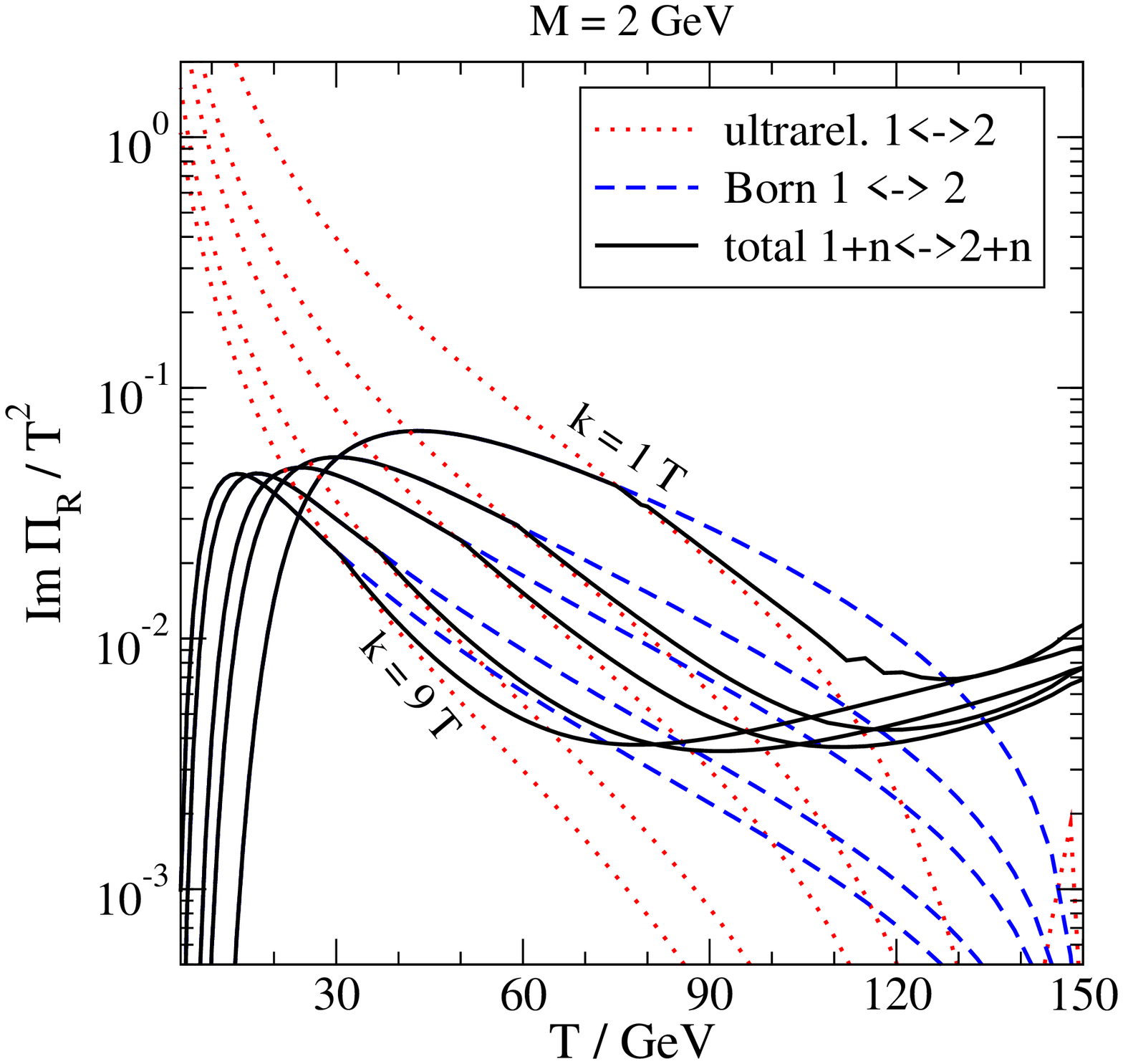}
}

\caption[a]{\small
 The direct $1+n \leftrightarrow 2+n$ 
 contribution to $\im \Pi^{ }_\rmii{R}/T^2$.
 Shown are the Born result with collinear kinematics and 
 thermal masses (``ultrarel.\ $1\leftrightarrow 2$''), 
 the naive Born result (``Born $1\leftrightarrow 2$'', cf.\ \eq\nr{tree}), 
 as well as the LPM result going over to the Born result
 at low $T$ (``total $1+n\leftrightarrow 2+n$'').
 The naive Born rate includes no (chirally invariant)
 thermal lepton mass~\cite{weldon} and therefore becomes too large
 in the regime where $m^{ }_\ell$ is substantial. 
 Left: fixed $k = 3 T$ and 
 masses $M/\mbox{GeV} \in\{ 0.5,1,2,4,8,16\}$.
 Right: fixed $M = 2$~GeV and 
 momenta $k/T \in \{1,2,3,6,9\}$. 
}

\la{fig:direct12}
\end{figure}

%
\subsection{Limit of low temperatures}
\la{ss:lowT12}

Once we go deep in the broken phase, 
the masses $m^{ }_{\phi_\mu}$ defined in \eq\nr{scalar_masses}
eventually become large, 
$m_{\phi_\mu}^2 / \ko \gg g^2 T / (8\pi)$.
Then $\Gamma^{ }_{3\times 3}$ represents a small correction compared
with the mass terms in \eq\nr{Seq}, and can be omitted. However, the collinear
approximation $ m_{\phi_\mu}^2 / \ko \ll \ko$ that is 
employed in the formalism of the LPM resummation also breaks down 
in the same regime. In this situation the rate is given
just by the $1\leftrightarrow 2$ processes, without any resummation
nor kinematic approximation. The hard thermal lepton mass $m_\ell$
can also be omitted at low temperatures. 
Then the result can be given in a closed form, 
\ba
 { \im\Pi_\rmii{R}^\rmi{Born} } 
 & = & 
 \sum_{\mu = 0}^{3} \, \mathcal{F}(m^{ }_{\phi_\mu})
 \;,  \la{tree} \\ 
 \mathcal{F}(m^{ }_{ }) & \equiv & 
 \frac{(M^2 - m_{ }^2)T}{32 \pi k}
 \ln\left\{
   \frac{\sinh\Bigl[ \frac{\kp + {m_{ }^2} / {(4\kp)} }{2T} \Bigr]
   \cosh\Bigl[ \frac{\kp -  {m_{ }^2} / {(4\km)}  }{2T} \Bigr] }
  {\sinh \Bigl[ \frac{\km + {m_{ }^2} / {(4\km)} }{2T} \Bigr]
  \cosh\Bigl[ \frac{\km  -  {m_{ }^2} / {(4\kp)}  }{2T} \Bigr]  }
 \right\} 
 \;,
 \la{calF}
\ea
where we have defined 
\be
 k^{ }_\pm \equiv \frac{\ko \pm k}{2}
 \;. \la{kpm}
\ee
Actually, $M \ll m^{ }_{\phi_\mu}$ so that \eq\nr{calF} could
be simplified by setting $M\to 0$
(cf.\ \eq\nr{tree3}).
In our numerical solution we switch from the LPM resummed result of 
\eq\nr{lpm} to the Born term of \eq\nr{tree} when the two results 
cross at low $T$, cf.\ \fig\ref{fig:direct12}. 

%
\section{Direct $2 \leftrightarrow 2$ scatterings}
\la{se:dir22}

%
\subsection{Ultrarelativistic regime}
\la{ss:direct22ur}

We now move on to discuss direct $2\to 2$ scatterings, illustrated
in \fig\ref{fig:types}(a). As long as we are in the ultrarelativistic
regime, $\mW\ll \pi T$, or $v \lsim T$, the masses of the ``real''
particles participating in these processes play no practical role, 
because all the scatterers have hard momenta $\sim \pi T$. 
Therefore, to leading order, the
computation can be directly taken over from the symmetric 
phase~\cite{bb2}. 
The techniques we employ here are similar to those in ref.~\cite{bb2}, 
except
for the treatment of soft momentum transfer, where a mild modification 
is adopted. 
In this section we describe the computation of the direct $2\to 2$
scatterings in some detail, in order to prepare the ground for the 
generalization to the indirect case in 
\ses\ref{ss:hard_t} and \ref{ss:soft_t}. 

In the $2\to 2$ scatterings of \fig\ref{fig:types}(a),  
the particles mediating $t$-channel exchange can have soft momenta. 
However, an explicit computation shows that 
only the lepton exchange is so infrared (IR) sensitive 
that the thermal lepton mass plays a role. The computation is organized
by first determining the contribution from hard momentum transfer
by using massless propagators, and subsequently treating the case
of soft momentum transfer more carefully. 

In naive massless
perturbation theory, the hard part can be written as 
\ba
 && \hspace*{-1.5cm} 2 \nF{}(\ko)
 \left. \im\Pi^{ }_\rmii{R} \right|^\rmi{hard}_\rmi{direct, $2\to 2$} 
 \nn 
 & = & 
 \int \! {\rm d}\Omega^{ }_{2\to 2} \; 
 \Bigl\{\; 
  \nB{}(p_1) \, \nB{}(p_2) \, \bigl[1 - \nF{}(k_1)\bigr]
  \,\fr12\, |\mathcal{M}^{ }_\rmi{a}|^2 
 \nn & & \hspace*{1.6cm} + \, 
   \nB{}(p_1) \, \nF{}(p_2) \, \bigl[1 + \nB{}(k_1)\bigr]
  \, \sum |\mathcal{M}^{ }_\rmi{b}|^2 
 \nn & & \hspace*{1.6cm} + \, 
   \nF{}(p_1) \, \nF{}(p_2) \, \bigl[1 - \nF{}(k_1)\bigr]
  \, \sum |\mathcal{M}^{ }_\rmi{c}|^2 
 \; \Bigr\} 
 \;. \la{boltzmann}
\ea
Here ${\rm d}\Omega^{ }_{n\to m}$ denotes the usual phase space 
integration measure with 4-momentum conservation, 
${\rm d}\Omega^{ }_{n\to m} \equiv 
 \Pi_{i=1}^{n} \frac{{\rm d}^3\vec{p}_i}{2 p_i (2\pi)^3 }
 \Pi_{j=1}^{m-1} \frac{{\rm d}^3\vec{k}_j}{2 k_j (2\pi)^3 }
 \, (2\pi)^4 \, \delta^{(4)}  
 ( \sum_{i=1}^n \mathcal{P}_i - \sum_{j=1}^m \mathcal{K}_j
 )
$. 
The three-momenta of incoming particles are denoted by $\vec{p}_i$, 
with $p_i \equiv |\vec{p}_i|$; those of outgoing particles are 
$\vec{k}_i$, with $\vec{k}_m \equiv \vec{k}$ the right-handed
neutrino momentum. The matrix elements squared read
\ba
 |\mathcal{M}^{ }_\rmi{a}|^2 & \equiv & 
 \bigl(g_1^2 + 3 g_2^2\bigr) 
 \, \biggl(  \frac{u}{t} + \frac{t}{u}  \biggr)
 \;, \la{2z2} \\ 
 \sum |\mathcal{M}^{ }_\rmi{b}|^2 & \equiv & 
 - \bigl(g_1^2 + 3 g_2^2\bigr) 
 \, \biggl(
  \frac{u}{s} 
 +  \frac{s}{u} 
 \biggr)
 \;, \\ 
 \sum |\mathcal{M}^{ }_\rmi{c}|^2 & \equiv & 
 6 h_t^2 \Nc
 \;. \la{M*M}
\ea 
Here 
$s \equiv (\mathcal{P}^{ }_1 + \mathcal{P}^{ }_2)^2$,
$t \equiv (\mathcal{K}^{ }_2 - \mathcal{P}^{ }_2)^2$, and
$u \equiv (\mathcal{K}^{ }_2 - \mathcal{P}^{ }_1)^2$.

\newcommand{\lnfplus}{\ln_\rmi{f}^+}
\newcommand{\lnfminus}{\ln_\rmi{f}^-}
\newcommand{\lnbplus}{\ln_\rmi{b}^+}
\newcommand{\lnbminus}{\ln_\rmi{b}^-}
\newcommand{\lifplus}[1]{{\mbox{li}}_\rmi{${#1}$f}^+}
\newcommand{\lifminus}[1]{{\mbox{li}}_\rmi{${#1}$f}^-}
\newcommand{\libplus}[1]{{\mbox{li}}_\rmi{${#1}$b}^+}
\newcommand{\libminus}[1]{{\mbox{li}}_\rmi{${#1}$b}^-}

The phase space integrals can be reduced into 2-dimensional integrals
as explained in ref.~\cite{bb2}. Different parametrizations are introduced
for $s$ and $t$-channel exchange
(the $u$-channel can be transformed into the $t$-channel 
by the exchange $p_1\leftrightarrow p_2$).
Defining the notation 
\ba
 && 
 \lnfminus \equiv \ln \Bigl( 1 + e^{-|\qm|/T} \Bigr)
 \;, \quad
 \lnfplus \equiv \ln \Bigl( 1 + e^{-\qp/T} \Bigr)
 \;, \la{not1} \\
 && 
 \lnbminus \equiv \ln \Bigl( 1 - e^{-|\qm|/T} \Bigr)
 \;, \quad
 \lnbplus \equiv \ln \Bigl( 1 - e^{-\qp/T} \Bigr)
 \;, \\
 &&
 \lifminus{i} \equiv {\mbox{Li}}^{ }_i \Bigl( - e^{-|\qm|/T} \Bigr)
 \;, \quad  \hspace*{3.2mm}
 \lifplus{i} \equiv {\mbox{Li}}^{ }_i \Bigl( - e^{-\qp/T} \Bigr)
 \;, \\
 && 
 \libminus{i} \equiv {\mbox{Li}}^{ }_i \Bigl( e^{-|\qm|/T} \Bigr)
 \;, \quad \hspace*{6mm}
 \libplus{i} \equiv {\mbox{Li}}^{ }_i \Bigl( e^{-\qp/T} \Bigr)
 \;, \la{notation}
\ea
where 
\be
 q^{ }_{\pm} \; \equiv \; \frac{q_0 \pm q}{2} 
 \;, \la{qpm}
\ee
the result reads
\ba
 && \hspace*{-1cm} 
 \left. (4\pi)^3\ko 
 \im\Pi^{ }_\rmii{R} \right|^\rmi{hard}_\rmi{direct, $2\to 2$} 
 \nn 
 & = & 
 \int_{\ko}^{\infty} \! {\rm d} \qp \int_0^{\ko} \! {\rm d} \qm 
 \, \Bigl\{ 
  \bigl[\nB{}(q_0) + \nF{}(q_0 - \ko) \bigr]  \,  \Phi_\rmi{$s$1} 
 + \, 
  \bigl[ \nF{}(q_0) + \nB{}(q_0 - \ko)  \bigr]  \,  \Phi_\rmi{$s$2} 
 \Bigr\} 
 \nn  
 & + & 
 \int_{0}^{\ko} \! {\rm d} \qp \int_{-\infty}^{0} \! {\rm d} \qm 
 \,  \Bigl\{  
  \bigl[1 - \nF{}(q_0) + \nB{}(\ko - q_0) \bigr]  \,  
  \Phi_\rmi{$t$2} 
 \Bigr\} 
 \;. \hspace*{5mm} \la{direct_hard}
\ea
Here $\Phi_\rmi{$s$1}$ refers to bosonic and 
$\Phi_\rmi{$s$2}$ to fermionic $s$-channel exchange, 
and 
$\Phi_\rmi{$t$2}$ to fermionic $t$-channel exchange;
the notation $\Phi_\rmi{$t$1}$ is reserved for bosonic
$t$-channel exchange
that does not appear here (or rather, appears as a diagram but
does not lead to non-trivial kinematic dependence). 
The functions appearing in \eq\nr{direct_hard} are
\ba
 \Phi_\rmi{$s$1} 
 & = & 
  6 h_t^2 \Nc  \, 
  \bigl[ q + 2 T (\lnfplus - \lnfminus) \bigr]
 \;, \\ 
 \Phi_\rmi{$s$2} 
 & = & 
 (g_1^2 + 3g_2^2) \, 
 \biggl\{
  \frac{q}{2} + \frac{T}{q} 
  \bigl[ 
   (\ko - \qm)(\lnfplus - \lnbminus) + (\ko - \qp)(\lnfminus - \lnbplus)
  \bigr] 
 \nn 
 & & \; 
   + \, \frac{T^2}{q^2} (2 \ko - q_0) 
    \bigl(  \libplus{2} + \lifminus{2} - \lifplus{2} - \libminus{2}  \bigr)
 \biggr\}
 \;, \\ 
 \Phi_\rmi{$t$2} 
 & = & 
 (g_1^2 + 3g_2^2) \, 
 \biggl\{
  \frac{T}{q} 
  \bigl[ 
   (\ko - \qm)(\lnfplus - \lnbminus) + (\ko - \qp)(\lnfminus - \lnbplus)
  \bigr] 
 \nn 
 & & \; 
   + \, \frac{T^2}{q^2} (2 \ko - q_0) 
    \bigl(  \libplus{2} + \libminus{2} - \lifplus{2} - \lifminus{2}  \bigr)
 \biggr\}
 \;.
\ea
The $s$-channel functions $  \Phi_\rmi{$s$1} $ and   $\Phi_\rmi{$s$2} $
remain finite in the whole integration range, 
whereas the $t$-channel function
$  \Phi_\rmi{$t$2} $ has a divergence in the vicinity of the origin
($q,q_0 \ll \ko$): 
\be
 \Phi_\rmi{$t$2} 
 =  (g_1^2 + 3g_2^2) \, 
 \frac{\ko \pi^2 T^2}{q^2}
  + \rmO\Bigl( \frac{1}{q} \Bigr) 
 \;.  \la{direct_div}
\ee
This divergence is not integrable and its proper treatment requires 
HTL resummation~\cite{htl1,htl2}, as we now explain. 

Suppose that we compute $\im \Pi^{ }_\rmii{R}$ within HTL resummed 
perturbation theory. The HTL scalar propagator has no cut, 
so the HTL result has no part which would correspond to $2\to 2$ scatterings
with soft Higgs exchange. Therefore the Higgs can be taken to be 
a ``hard'' external particle, and its thermal mass can be omitted. 
This leads to 
\be
 \left. \im \Pi^{ }_\rmii{R} \right|^\rmii{HTL}_\rmi{direct,\,$2\to 2$} = 
 \int_{-\infty}^{\infty} \! {\rm d}q_0 
 \int_\vec{q} \frac{-2 \mathcal{K}\cdot\rho_{\ell}(q_0,\vec{q})}
 {|\vec{k-q}|}
 \, \bigl[ 1 + \nB{}(\ko - q_0) - \nF{}(q_0) \bigr]
 \, \delta(\ko - q_0 - |\vec{k-q}|)
 \;, \la{direct_HTL_pre}
\ee 
where the lepton spectral function is given in \eqs\nr{rhoell}--\nr{rhos}.
The lepton spectral function is parametrized by the mass given 
in \eq\nr{mell}, which is purely of thermal origin, so that all 
left-handed leptons are degenerate. 
Setting $k = \ko$ and restricting to $q, q_0 \sim m_\ell \ll \ko$ where 
the HTL structures play a role, the constraint
$\delta(\ko - q_0 - |\vec{k-q}|)$ in \eq\nr{direct_HTL_pre}
leads to 
\be
 \vec{k}\cdot\vec{q} \; = \; 
 \ko q_0 + \frac{q^2 - q_0^2}{2} 
 \; = \; \ko q_0 + \rmO(m_\ell^2)
 \;. \la{appro_q}
\ee
Simplifying the integrand with this approximation but keeping the full
integration range (the reason should become clear in a moment), we get
\be
 \left. \im \Pi^{ }_\rmii{R} \right|^\rmii{HTL}_\rmi{direct,\,$2\to 2$} 
 \approx 
 \frac{1}{2\pi^2}
 \int_{-\infty}^{\ko} \! {\rm d}q_0 
 \int_{|q_0|}^{2\ko - q_0} \! {\rm d}q\, q\, q_0 \, 
 (\hat{\rho}_s - \hat{\rho}_0) (q_0,q) 
 \, \Bigl[ \nB{}(\ko) + \fr12\, \Bigr]
 \;, \la{direct_HTL}
\ee
where 
$\hat{\rho}_s$ and $\hat{\rho}_0$ 
are from \eqs\nr{rho0} and \nr{rhos} and
terms of $\rmO(m_\ell / \ko)$ have been omitted. 

Now, the contribution from hard momentum transfer, \eq\nr{direct_hard}, 
is IR divergent because of the term in \eq\nr{direct_div}. The reason
for this divergence is that the computation leading to \eq\nr{direct_hard}
did {\em not} incorporate HTL resummation. Fortunately we can correct for 
this mistake a posteriori. In order to do so, we need to subtract from 
\eq\nr{direct_hard} the ``would-be'' HTL contribution, which appears there
in a naive perturbative form. This is obtained from \eq\nr{direct_HTL}
by formally expanding in a weak coupling, i.e.\ by assuming 
$q,q_0 \gg m_\ell$. According to \eqs\nr{rho0} and \nr{rhos} this yields
\be
 \hat{\rho}_s - \hat{\rho}_0 \; \approx \; 
 \frac{m_\ell^2 [ - \frac{q_0}{2 q^2} + \frac{1}{2 q_0} ] \im L}
 {q_0^2 - q^2 - m_\ell^2}
 \; \approx \; 
 \frac{\pi m_\ell^2}{4 q^3 q_0}
 \;, \la{lepton_expand}
\ee
where terms of $\rmO(m_\ell^4)$ have been omitted. 
Within this approximation \eq\nr{direct_HTL} becomes
\be
 \left. \im \Pi^{ }_\rmii{R} \right|^{\rmii{HTL},\,\rmi{expanded}}
  _\rmi{direct,\,$2\to 2$} = 
 \frac{1}{8\pi}
 \int_{-\infty}^{\ko} \! {\rm d}q_0 
 \int_{|q_0|}^{2\ko - q_0} \! {\rm d}q\,  \frac{m_\ell^2}{q^2}
 \, \Bigl[ \nB{}(\ko) + \fr12\, \Bigr]
 \;. \la{direct_HTL_exp}
\ee
Taking note of \eq\nr{mell} and of the changes of integration variables 
\be
  \int_{\ko}^{\infty} \! {\rm d} \qp \int_0^{\ko} \! {\rm d} \qm 
 = \fr12
  \int_{\ko}^\infty \! {\rm d}q_0 \int_{|2\ko - q_0|}^{q_0} \! {\rm d}q 
 \;, \quad
 \int_{0}^{\ko} \! {\rm d} \qp \int_{-\infty}^{0} \! {\rm d} \qm  
 = \fr12
  \int_{-\infty}^{\ko} \! {\rm d}q_0 \int_{|q_0|}^{2\ko - q_0} \! {\rm d}q 
 \;, \la{range}
\ee 
this agrees exactly with \eq\nr{direct_div}. 

The philosophy thus is to subtract \eq\nr{direct_HTL_exp}
from the ``naive'' computation of \eq\nr{direct_hard}. Subsequently
the ``soft'' contribution from \eq\nr{direct_HTL} is added in its 
proper form. 

Let us now compute \eq\nr{direct_HTL} properly. The integral contains
two scales, $\ko$ and $m_\ell$, and we evaluate it in the approximation
$m_\ell \ll \ko$. The leading contribution originates from 
$q,q_0 \sim m_\ell$. In order to evaluate this contribution, it is 
advantageous to change integration variables from $q,q_0$ to 
$q_\perp, q_0$, where
\be
 q_\perp^2 \; \equiv \; q^2 - q_\parallel^2 \; \equiv \; 
 q^2 - \frac{(\vec{k}\cdot\vec{q})^2}{k_0^2} 
 \; \approx \; q^2 - q_0^2
 \;, \la{qperp}
\ee
where we made use of \eq\nr{appro_q}. 
Then 
\ba
 \left. \im \Pi^{ }_\rmii{R} \right|^{\rmii{HTL},\,\rmi{soft}}
 _\rmi{direct,\,$2\to 2$}
 & \approx & 
 \frac{1}{2\pi^2}
 \int_{-\infty}^{\ko} \! {\rm d}q_0 
 \int_{0}^{2\ko} \! {\rm d}q_\perp\, q_\perp\, q_0 \, 
 (\hat{\rho}_s - \hat{\rho}_0) \Bigl(q_0,\sqrt{q_0^2 + q_\perp^2}\Bigr) 
 \, \Bigl[ \nB{}(\ko) + \fr12\, \Bigr]
 \nn 
 & \approx &
 \int_0^{2\ko} \frac{{\rm d}q_\perp \, q_\perp}{2\pi}
 \underbrace{
 \int_{-\infty}^{\infty} \! \frac{{\rm d}q_0}{\pi} \, q_0 \, 
 (\hat{\rho}_s - \hat{\rho}_0) \Bigl(q_0,\sqrt{q_0^2 + q_\perp^2}\Bigr)
 }_{
  \fr12 \frac{m_\ell^2}{q_\perp^2 + m_\ell^2}
 }
 \, \Bigl[ \nB{}(\ko) + \fr12\, \Bigr]
 \nn 
 & = & 
 \frac{m_\ell^2}{8\pi} \ln\Bigl[ 1 + \Bigl(\frac{2\ko}{m_\ell} \Bigr)^2 \Bigr]
 \, \Bigl[ \nB{}(\ko) + \fr12\, \Bigr]
 \;, \la{direct_HTL_soft}
\ea
where we employed a sum rule derived in ref.~\cite{bb2}.\footnote{%
 Within $ \rmO\bigl( {m_\ell^4}/ {\ko^2} \bigr) $ accuracy
 the argument of the logarithm can be simplified,  
 cf.\ \eq\nr{direct_full}. If however the result is evaluated numerically
 for small $k \lsim g T$ where it is not leading-order correct but 
 represents an extrapolation,  
 it is advantageous to employ \eq\nr{direct_HTL_soft} 
 in order to avoid spurious negative expressions. We have adopted
 this recipe for our numerics.  
 } 
Putting everything together, we obtain 
\ba
 \left. \im\Pi^{ }_\rmii{R} \right|^{ }_\rmi{direct, $2\to 2$} 
 & = & 
 \left. \im\Pi^{ }_\rmii{R} \right|^\rmi{hard}_\rmi{direct, $2\to 2$} - 
 \left. \im \Pi^{ }_\rmii{R} \right|^{\rmii{HTL},\,\rmi{expanded}}
  _\rmi{direct,\,$2\to 2$} + 
 \left. \im \Pi^{ }_\rmii{R} \right|^{\rmii{HTL},\,\rmi{soft}}
 _\rmi{direct,\,$2\to 2$}
 \nn[3mm] 
 & = & 
 \frac{1}{(4\pi)^3\ko}
 \int_{\ko}^{\infty} \! {\rm d} \qp \int_0^{\ko} \! {\rm d} \qm 
 \Bigl\{ 
  \bigl[\nB{}(q_0) + \nF{}(q_0 - \ko) \bigr]  \,  \Phi_\rmi{$s$1} 
 \nn & & \quad
 + \, 
  \bigl[ \nF{}(q_0) + \nB{}(q_0 - \ko)  \bigr]  \,  \Phi_\rmi{$s$2} 
 \Bigr\} 
 \nn  
 & + & 
 \frac{1}{(4\pi)^3\ko}
 \int_{0}^{\ko} \! {\rm d} \qp \int_{-\infty}^{0} \! {\rm d} \qm  
 \Bigl\{ 
  \bigl[1 - \nF{}(q_0) + \nB{}(\ko - q_0) \bigr]  \,  
  \Phi_\rmi{$t$2} 
 \nn 
 & & 
  -  \, \Bigl[\nB{}(\ko) + \fr12\, \Bigr]
 \, (g_1^2 + 3g_2^2) \,  \frac{\ko \pi^2 T^2}{q^2}
 \, \Bigr\} 
 \nn 
 & + & \frac{m_\ell^2}{4\pi} \, \Bigl[\nB{}(\ko) + \fr12\, \Bigr] 
 \, \ln\Bigl( \frac{2\ko}{m_\ell} \Bigr)
 \; + \; 
 \rmO\Bigl( \frac{m_\ell^4}{\ko^2} \Bigr)
 \;. \la{direct_full}
\ea
This expression is IR finite and agrees with ref.~\cite{bb2}. 
Parametrically, 
$
  \left. \im\Pi^{ }_\rmii{R} \right|^{ }_\rmi{direct} 
 \sim g^2 T^2
$.
A numerical evaluation is shown in \fig\ref{fig:direct22}
with a dashed line (``ultrarel.\ $2\leftrightarrow 2$'').

\begin{figure}[t]

\hspace*{-0.1cm}
\centerline{%
  \epsfxsize=7.5cm\epsfbox{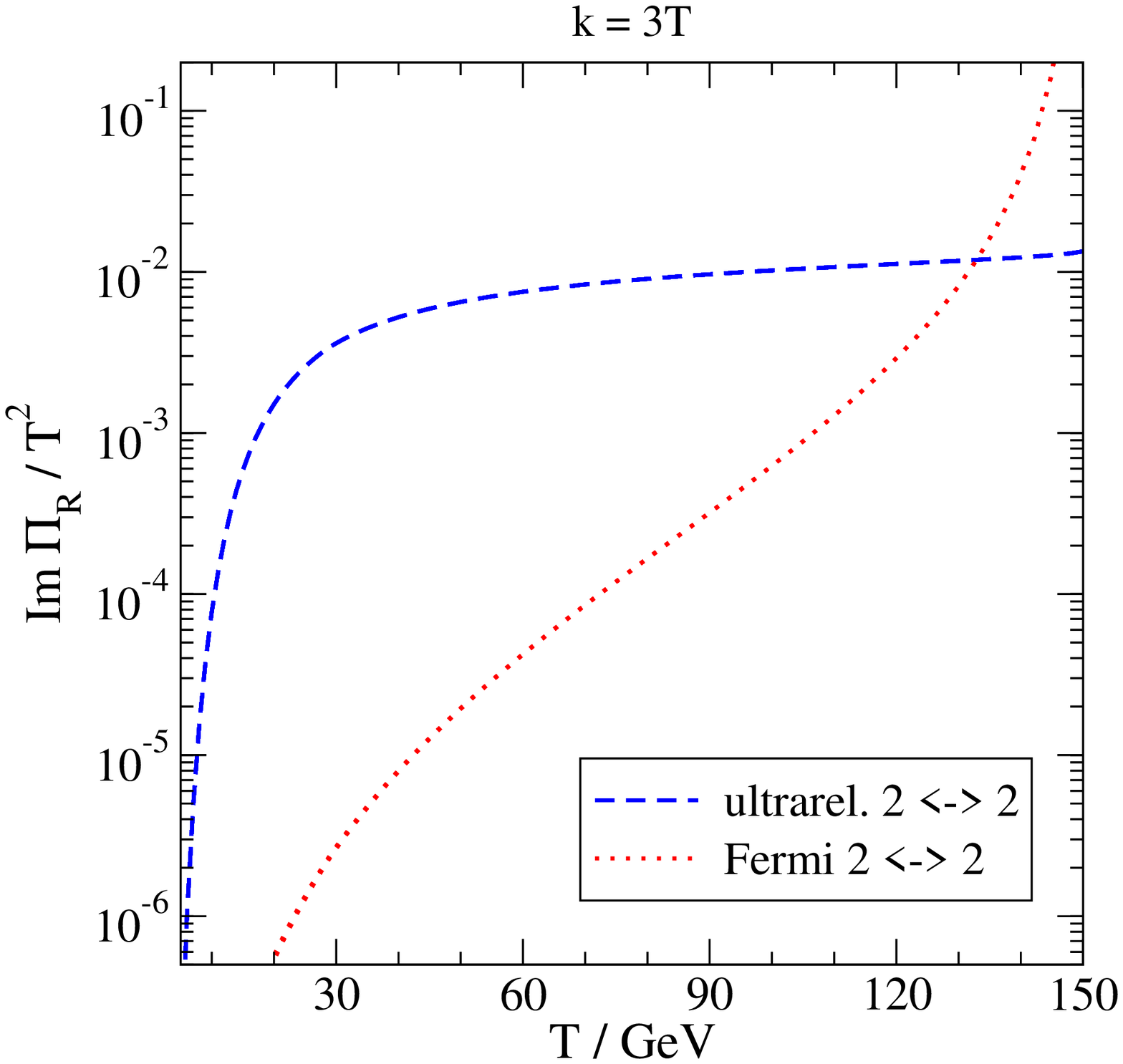}%
  \hspace{0.1cm}
  \epsfxsize=7.5cm\epsfbox{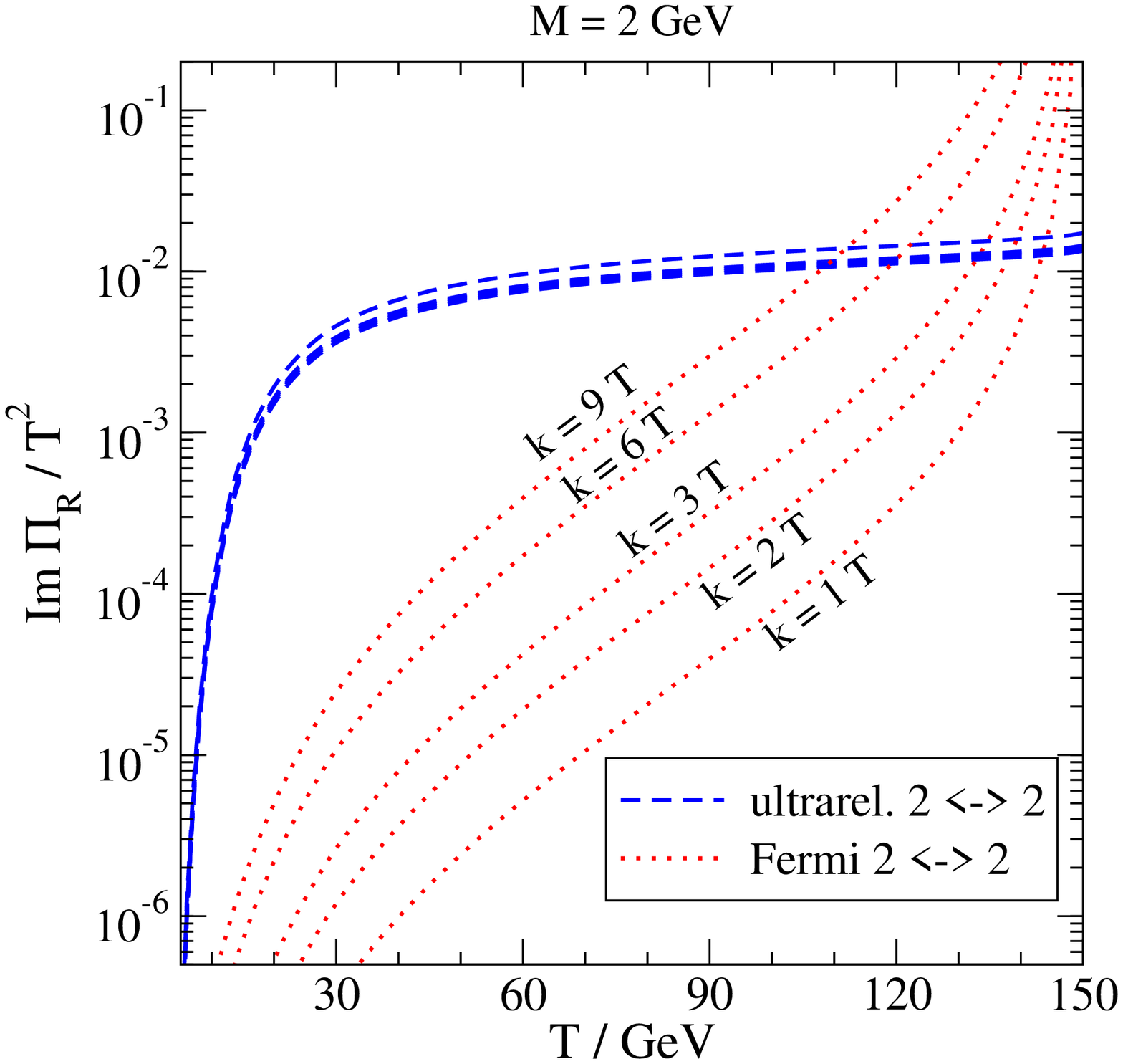}
}

\caption[a]{\small
 The direct $2 \to 2$ 
 contribution to $\im \Pi^{ }_\rmii{R}/T^2$.
 Shown are the result from $2\leftrightarrow 2$ scatterings
 treated in the ultrarelativistic approximation
 (``ultrarel.\ $1\leftrightarrow 2$'', cf.\ \eq\nr{direct_full}),  
 switched off at low $T$ as indicated at
 the end of \se\ref{ss:lowT22}, and the leading 
 low-$T$ contribution 
 (``Fermi $2\leftrightarrow 2$'', cf.\ \eq\nr{direct_Fermi}).
 The latter is only of academic interest, 
 because it is vanishingly small in its range
 of applicability, $T \lsim 30$~GeV. 
 Left: fixed $k = 3 T$ and 
 masses $M/\mbox{GeV} \in\{ 0.5,1,2,4,8,16\}$.
 Right: fixed $M = 2$~GeV.
}

\la{fig:direct22}
\end{figure}

%
\subsection{Limit of low temperatures}
\la{ss:lowT22}

All the $2\leftrightarrow 2$ scattering 
reactions depicted in \fig\ref{fig:types}(a), 
leading to \eqs\nr{2z2}--\nr{M*M}, involve a particle
in the initial state whose contribution becomes 
exponentially suppressed when $\mW\gsim \pi T$.
Therefore, the contribution of \eq\nr{direct_full}
rapidly switches off once we exit the regime of \eq\nr{regime}. Because of the 
resummations that were needed for obtaining \eq\nr{direct_full}
it is non-trivial to obtain a general expression which has
the correct high and low-temperature limits and is a smooth
function in between. However, we can easily determine
the low-temperature limit. The formally dominant contribution 
originates from Higgs mediated bottom quark scatterings. 
Accounting for these through a Fermi type computation 
and making use of the same notation as in \eqs\nr{not1}--\nr{direct_hard}, 
we obtain
\ba
 \left. \im\Pi^{ }_\rmii{R} \right|^\rmi{Fermi}_\rmi{direct, $2\to 2$} 
 & = & 
 \frac{1}{(4\pi)^3\ko}
 \int_{\ko}^{\infty} \! {\rm d} \qp \int_0^{\ko} \! {\rm d} \qm 
 \, 
  \bigl[\nB{}(q_0) + \nF{}(q_0 - \ko) \bigr]  \,  \Phi_\rmi{$s$1} 
 \nn  
 & + & 
 \frac{1}{(4\pi)^3\ko}
 \int_{0}^{\ko} \! {\rm d} \qp \int_{-\infty}^{0} \! {\rm d} \qm 
 \, 
  \bigl[1 + \nB{}(q_0) - \nF{}(\ko - q_0) \bigr]  \,  
  \Phi_\rmi{$t$1} 
 \;, \hspace*{5mm} \la{direct_Fermi}
\ea
where\footnote{%
 These results apply in the Feynman $R_\xi$ gauge. In a general gauge 
 the Goldstone mode part changes; the gauge dependence cancels 
 against similar $2\leftrightarrow 2$ indirect contributions, 
 of the type discussed in \se\ref{ss:indirect_Fermi}.  
 } 
\ba
 \Phi_\rmi{$s$1} 
 & = & 
  \sum_{\mu =0,3} \frac{h_b^2 \Nc}{m_{\phi_\mu}^4} \, (q_0^2 - q^2)^2  \, 
  \bigl[ q + 2 T (\lnfplus - \lnfminus) \bigr]
 \;, \\ 
 \Phi_\rmi{$t$1} 
 & = & 
    \sum_{\mu =0,3} \frac{h_b^2 \Nc}{m_{\phi_\mu}^4} \, (q_0^2 - q^2)^2  \, 
  \bigl[ 2 T (\lnfminus - \lnfplus) \bigr]
 \;. 
\ea
Here $h_b = g_2 m_b / (\sqrt{2} \mW) \simeq 0.03$ is the bottom
quark Yukawa coupling. 
The result has been illustrated with a dotted line 
in \fig\ref{fig:direct22} (``Fermi $2\leftrightarrow 2$'').
However, in practice this contribution is so small within its range
of validity ($T \lsim 30$~GeV) that it can be omitted. 

At the same time, it is 
important to switch off the massless $2\leftrightarrow 2$ scatterings
in this regime. We have done this by multiplying 
$
 \left. \im\Pi^{ }_\rmii{R} \right|^{ }_\rmi{direct, $2\to 2$}  
$
of \eq\nr{direct_full}
by a phenomenological factor $\kappa(\mW)$. 
For this we choose a ``susceptibility'' related 
to $W^\pm$ bosons (cf.\ \eq\nr{chib}), 
normalized to the massless limit: 
\be
 \kappa(\mW) 
 \; \equiv \; \frac{3}{\pi^2 T^3}
 \int_0^\infty \! {\rm d}p\, p^2 \nB{}(E^{ }_\rmii{$W$})
 [1 + \nB{}(E^{ }_\rmii{$W$})]
 \;.  
\ee
This has been included in the numerical 
``ultrarel.\ $2\leftrightarrow 2$'' results shown in \fig\ref{fig:direct22}.
In principle it would be interesting to carry out a consistent
computation for this regime, however in practice this is not necessary
because, as we will see, the indirect contribution dominates
the full result by many orders of magnitude at $T \lsim 30$~GeV. 

%
\section{Indirect contribution}
\la{se:indirect}

%
\subsection{General structure}
\la{ss:general}

We now proceed to the indirect contributions, 
illustrated in \figs\ref{fig:lpm}(b) and 
\ref{fig:types}(b). As a first step, 
let us justify the form of \eq\nr{indirect}. 

According to \eq\nr{Pi}, 
the indirect contribution reads
\be
 \left. \Pi^{ }_\rmii{E}(K) \right|^{ }_\rmi{indirect} =
 \frac{v^2}{2} \tr \bigl\{ i \bsl{K} \aL \nu(-K) \bar{\nu}(0) \aR \bigr\} 
 \;. 
\ee
It is advantageous to resum the neutrino 
propagator to all orders, so that the production rate remains
finite even for small virtualities. 
The inverse neutrino propagator is of the form 
\be
 \langle \nu(K) \bar{\nu}(0) \rangle ^{-1} = i\bsl{K} + i \bsl{\Sigma}(K)
 \;. 
\ee
If we make use of the property 
$
 \bsl{\Sigma}(-K) = - \bsl{\Sigma}(K)
$, valid in a CP-symmetric plasma,\footnote{%
 We reiterature that, like in \eq\nr{fI_prod}, 
 we work close to equilibrium, with vanishing lepton asymmetries. 
 } 
then 
\ba
 \left. \Pi^{ }_\rmii{E}(K) \right|^{ }_\rmi{indirect}
 & = &
 \frac{v^2}{2} \tr \biggl\{\aR i \bsl{K}
 \frac{1}{-i \bsl{K} - i \bsl{\Sigma}(K)} \biggr\} 
 \; = \; 
 - v^2\, \frac{K^2 + K\cdot\Sigma}{(K + \Sigma)^2}
 \;.  
\ea
After the analytic continuation
in \eq\nr{relation}, we write 
$
 \Sigma \to \re \Sigma + i \im \Sigma
$.
Then a few steps lead to\footnote{%
 Formally we assume here that 
 $ 
  M^2 \sim \mathcal{K}\cdot\re\Sigma \sim 
  \mathcal{K}\cdot\im\Sigma \sim g^2 T^2
 $.
 }  
\be
 \left. \im \Pi^{ }_\rmii{R}(\mathcal{K}) \right|^{ }_\rmi{indirect}
 = 
 \frac{v^2}{2} \frac{M^2\, 2 \mathcal{K}\cdot \im \Sigma}
 {(M^2 + 2 \mathcal{K}\cdot\re\Sigma)^2 + 4 (\mathcal{K}\cdot\im\Sigma)^2}
 \;, \la{master0}
\ee
where $M^2 = \mathcal{K}^2$. 

It is clear from the \eq\nr{master0} that an essential role
in the indirect production is played by the real and imaginary parts of the 
(retarded) active neutrino self-energy. 
In the regime 
in which weak gauge bosons are ultrarelativistic 
($\mZ \ll \pi T$), the self-energy has a 
Hard Thermal Loop form 
corresponding to \eqs\nr{rho0} and \nr{rhos}, 
\be
 \Sigma^{ }_\rmii{HTL}(\mathcal{K}) = 
 \Bigl( - \frac{m_\ell^2}{2} L, \frac{m_\ell^2\, \vec{k}}{2 k^2} (1 - \ko L)
 \Bigr)
 \;, \la{Sigma}
\ee
where $m_\ell^2$ is given by \eq\nr{mell} 
and $L \equiv \frac{1}{2k}\ln\frac{\ko + k}{\ko - k}$. 
Then we get for the real part 
\be
 2 \mathcal{K}\cdot \re\Sigma^{ }_\rmii{HTL} = - m_\ell^2
 \;.  \la{Vell}
\ee
When $\mZ \gsim \pi T$, the result changes;
up to a normalization, $ 2 \mathcal{K}\cdot \re\Sigma $
is then referred to as a finite-temperature matter potential, 
whose structure is reviewed in \se\ref{ss:potential}.  

As far as the imaginary part of the active neutrino 
self-energy goes, we can  write 
\be
 \im \Sigma \;\equiv\; \bigl[ a_i + \rmO(\mathcal{K}^2) \bigr]\, \mathcal{K} + 
 \biggl[ \frac{\Gamma}{2} + \rmO(\mathcal{K}^2) \biggr]  u
 \;, \la{def_Gamma}
\ee
where a Lorentz-violating term proportional to the four-velocity of 
the heat bath $u \equiv (1,\vec{0})$
has been singled out. Then 
\be
 \lim_{\mathcal{K}^2\to 0} 2 \mathcal{K}\cdot\im\Sigma = \ko \Gamma
 \;. \la{Gamma_def}
\ee 
Subsequently the final expression for $\im \Pi^{ }_\rmii{R}$, 
valid for $M \ll k$, takes the form in \eq\nr{indirect}.

%
\subsection{Finite-temperature matter potential}
\la{ss:potential}

An important role in the indirect contribution 
discussed above is played by the real part 
of the active neutrino self-energy, $2 \mathcal{K}\cdot \re\Sigma$, 
cf.\ \eq\nr{master0}, which up to a normalization is also called
the finite-temperature matter potential.  
We review here its general form in the 
phase in which the Higgs mechanism is operative.

We work in a regime in which the right-handed neutrinos and 
all active leptons are ultrarelativistic, $M, m_\tau \ll \pi T$.
Then $2 \mathcal{K}\cdot \re\Sigma$ is a function of two dimensionless
ratios, $k/(\pi T)$ and $m^{ }_\rmii{$G$}/(\pi T)$, where 
$m^{ }_\rmii{$G$}$ refers to weak gauge boson masses, i.e.\ $G \in \{ W,Z\}$.
Defining 
\be
 \lim_{\mathcal{K}^2 \to 0} 2 \mathcal{K}\cdot \re\Sigma 
 = 2 g_2^2\, \mathcal{V}(\mW) + 
  (g_1^2 + g_2^2)\, \mathcal{V}(\mZ)
 \;, \la{defV}
\ee
a straightforward computation yields 
\ba
 \mathcal{V}(m) & = & 
 -\frac{1}{4\pi^2} \biggl\{ 
 \int_0^\infty \! {\rm d}q \, \nF{}(q) \,
 \biggl[
   q + \frac{m^2}{8 k} 
   \ln \biggl| 
             \frac{ m^2 - 4 k q }{m^2 + 4 k q } 
       \biggr|  \,   
 \biggr]
 \nn 
 & + & 
 \int_{m}^\infty \! {\rm d}\epsilon \, \nB{}(\epsilon) \, 
 \biggl[
   \sqrt{\epsilon^2 - m^2} + 
  \frac{m^2}{8 k } 
   \ln \biggl| 
  \frac{ m^2 - 4 k^2 - 4 k \sqrt{\epsilon^2 - m^2}  }
       { m^2 - 4 k^2 + 4 k \sqrt{\epsilon^2 - m^2}  }
       \biggr|\,      
 \biggr]
 \, \biggr\}
 \;, 
 \la{resV}
\ea
where $\nF{}$ and $\nB{}$ are the Fermi and Bose distributions, respectively. 
This is a limit of the results in ref.~\cite{ReSigma}.
At high temperatures the potential can be approximated as 
\be
 \mathcal{V}(m) \; 
  \stackrel{m \ll \pi T}{\approx} \;
 - \frac{1}{4\pi^2} \int_0^\infty \! {\rm d}q \, q \, 
  \bigl[\nF{}(q) + \nB{}(q) \bigr] 
 = -\frac{T^2}{16}
 \;, \la{VhighT}
\ee
which directly leads to \eq\nr{Vell}~\cite{weldon}. 
At low temperatures we get 
\be
 \mathcal{V}(m) \; 
  \stackrel{m \gg \pi T}{\approx} \;
  \frac{4 k^2}{3\pi^2 m^4} \int_0^\infty \! {\rm d}q \, q^3 \, 
  \nF{}(q) 
 = \frac{7\pi^2 T^4 k^2}{90 m^4}
 \;, \la{VlowT}
\ee
which corresponds to the result in ref.~\cite{ReSigmaold}
(cf.\ also refs.~\cite{ReS2,ReS3}).

For a numerical evaluation, the $q$-integral in \eq\nr{resV} can be 
divided into two ranges as 
$
 \int_0^\infty \! {\rm d}q = 
 \int_0^{q_*} \! {\rm d}q + 
  \int_{q_*}^\infty \! {\rm d}q
$
and the $\epsilon$-integral as
$
 \int_{m}^\infty \! {\rm d}\epsilon = 
 \int_{m}^{\epsilon_*} \! {\rm d}\epsilon + 
  \int_{\epsilon_*}^\infty \! {\rm d}\epsilon
$,
where
\be
 q_* \;\equiv\; \frac{m^2}{4k} 
 \;, \quad
 \epsilon_* \;\equiv\; 
 \sqrt{m^2 + \biggl(
 \frac{m^2 - 4 k^2}{4k}
 \biggr)^2}
 \;. 
\ee
Then numerical evaluation poses no problems; the result is illustrated
in \fig\ref{fig:reSigma}. 

\begin{figure}[t]

\hspace*{-0.1cm}
\centerline{%
  \epsfxsize=7.5cm\epsfbox{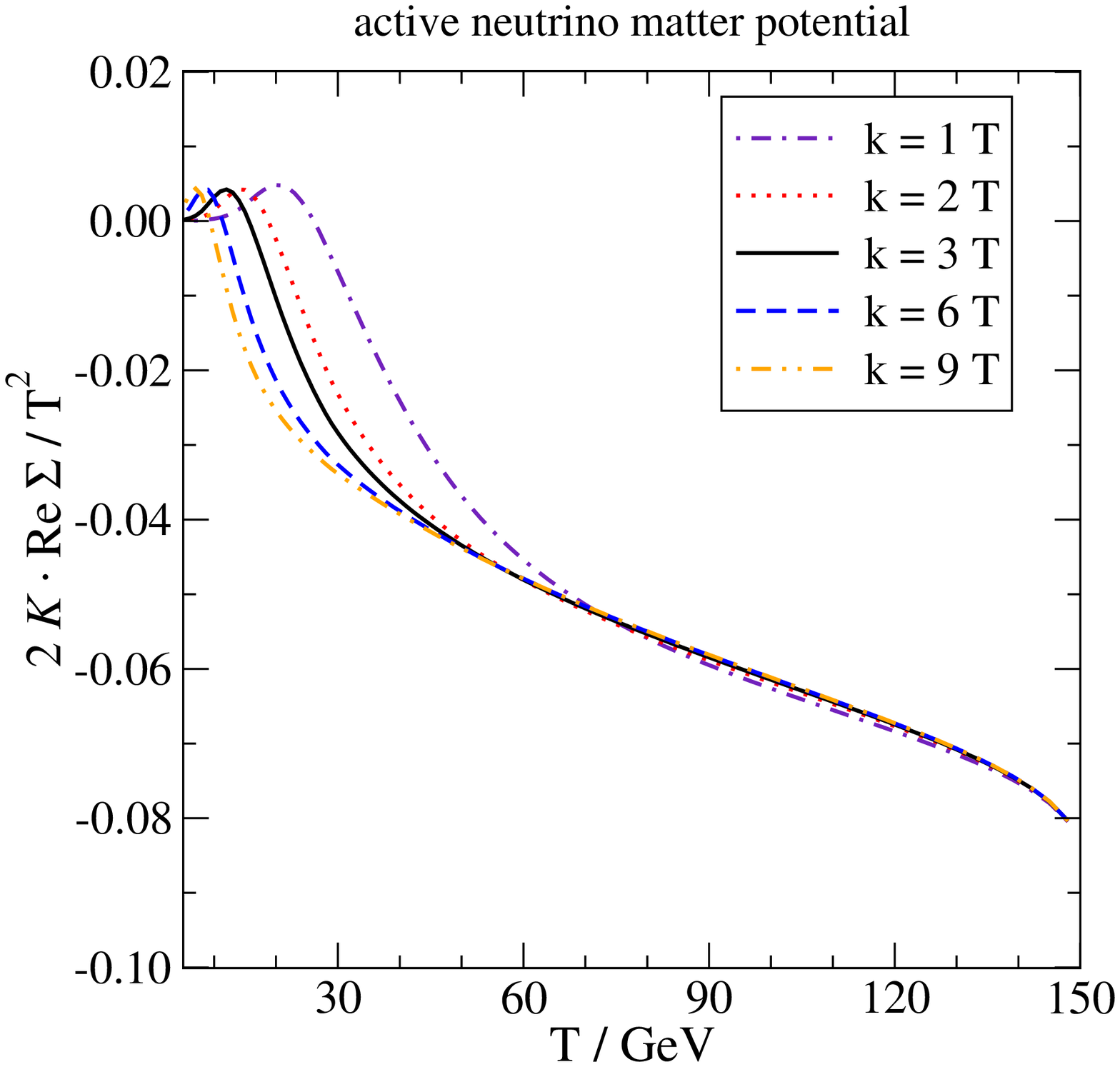}%
  \hspace{0.1cm}
  \epsfxsize=7.5cm\epsfbox{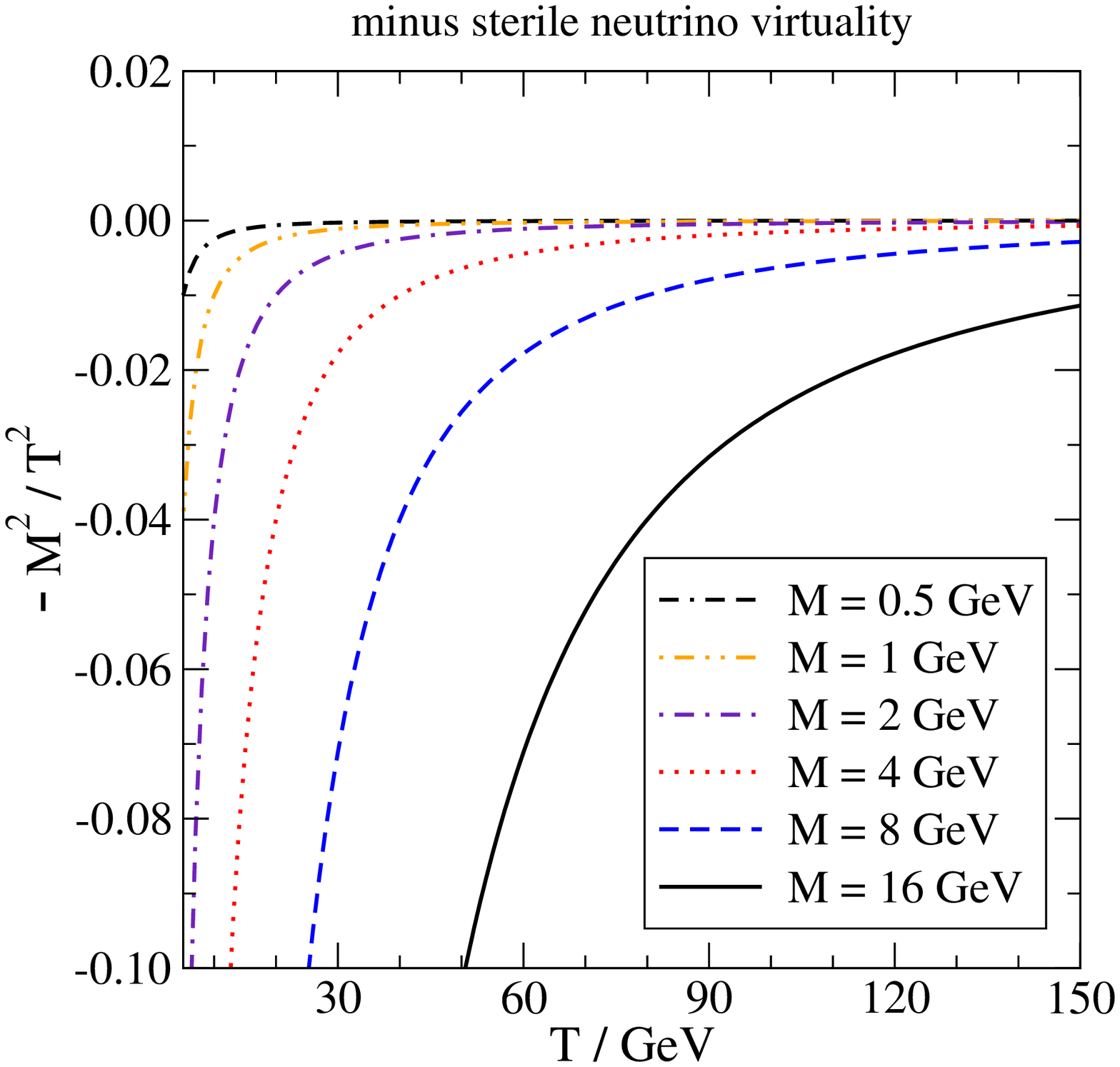}
}

\caption[a]{\small
 Quantities characterizing the active neutrino propagator, 
 \eq\nr{master0}. 
 Left: the real part of the active neutrino self-energy, 
 or ``finite-temperature matter potential'', 
 $2 \mathcal{K}\cdot \re \Sigma/T^2$, from 
 \eq\nr{defV}. The high-temperature limit is 
 given by \eq\nr{Vell}, and the low-temperature limit 
 corresponds to the Fermi model, cf.\ \eq\nr{VlowT}.
 Right: minus the sterile neutrino mass squared in the 
 same units. The crossing of the two curves implies  
 a ``resonant'' conversion from active to 
 sterile neutrinos, however the resonance is parametrically
 fairly broad because of a large width $\ko \Gamma$,
 cf.~\eq\nr{indirect_est} and \fig\ref{fig:imSigma}.  
}

\la{fig:reSigma}
\end{figure}

%
\subsection{Interaction rate from $1+n\leftrightarrow 2+n$ scatterings}
\la{ss:lpm_broken_indir}

We now move on to $\Gamma$ as defined by \eq\nr{Gamma_def}, 
frequently called the active neutrino 
width or damping or interaction rate. 
The $1+n\leftrightarrow 2+n$ contributions to $\Gamma$ are 
illustrated in \fig\ref{fig:lpm}(b), and correspond physically 
to the decays $W^\pm,Z^0,\gamma \to \ell N$. 
{}From the kinematics point of view these processes are
similar to those appearing in the direct contribution
of \se\ref{ss:lpm_broken}, if we simply
replace the Goldstone modes by gauge fields; the difference is 
that the cubic coupling is now $g$ rather than $h$. Thus, in the Feynman
$R_\xi$ gauge in which gauge field propagators are similar to 
scalar propagators and no additional structures appear,\footnote{%
 In other gauges the longitudinal 
 combination $Q_\mu Q_\nu / \mW^2$
 appears in the gauge field propagator. 
 This leads to large $\rmO(\pi^2/g^2)$ effects for 
 $q \sim \pi T, \mW \sim gT$ and violates 
 our power counting setup.
 }  
the parametric magnitude of these processes is 
$\delta \ko \Gamma \sim g^2 m^2 \sim g^4 T^2$.
This turns out to be of 
NNLO compared with the contribution from $2\to 2$ scatterings, 
and is therefore negligible at high temperatures.  

At low temperatures, when $\mW \gsim \ko \sim \pi T$, 
there is no need for resummation, cf.\ \se\ref{ss:lowT12}.\footnote{%
 Resummation becomes important when the 
 ultrarelativistic $1\leftrightarrow 2$ and the full
 $1+n \leftrightarrow 2+n$ LPM lines
 depart from each other in \fig\ref{fig:direct12}, i.e.\ $T\gsim 60$~GeV.} 
Then the relevant $1\leftrightarrow 2$ processes  are the decays 
of the $W^\pm$ and $Z^0$ gauge bosons. We can write a Born
rate like in \eq\nr{tree}, 
\be
 \ko\, \Gamma_{ }^{\rmi{Born}} \; = \; 
 (g_1^2 + g_2^2)\, \mathcal{F}(\mZ) + 
 2 g_2^2 \, \mathcal{F}(\mW)
 \;, \la{tree2}
\ee
where $\mathcal{F}$ is from \eq\nr{calF}. 
It is appropriate to remark 
that $\Gamma$ is gauge independent only on  
the mass-shell of {active neutrinos}, 
i.e.\ $M\to 0$, 
in accordance with \eq\nr{Gamma_def}. Thereby we obtain
\be
 \mathcal{F}(m) \; \stackrel{M\to 0}{\longrightarrow} \;
 \frac{m^2 T}{32\pi k}
 \ln \Biggl\{ 
   \frac{1 + e^{-\frac{m^2}{4 k T}}}
        {1 - e^{-\frac{1}{T}( k+\frac{m^2}{4 k } )}}
 \Biggr\}
 \;. \la{tree3}
\ee
The contribution of \eq\nr{tree2} in this limit
is illustrated in \fig\ref{fig:imSigma} (``Born $1\leftrightarrow 2$''), 
and it represents the dominant process for $T \lsim 30$~GeV.

\begin{figure}[t]

\hspace*{-0.1cm}
\centerline{%
  \epsfxsize=7.5cm\epsfbox{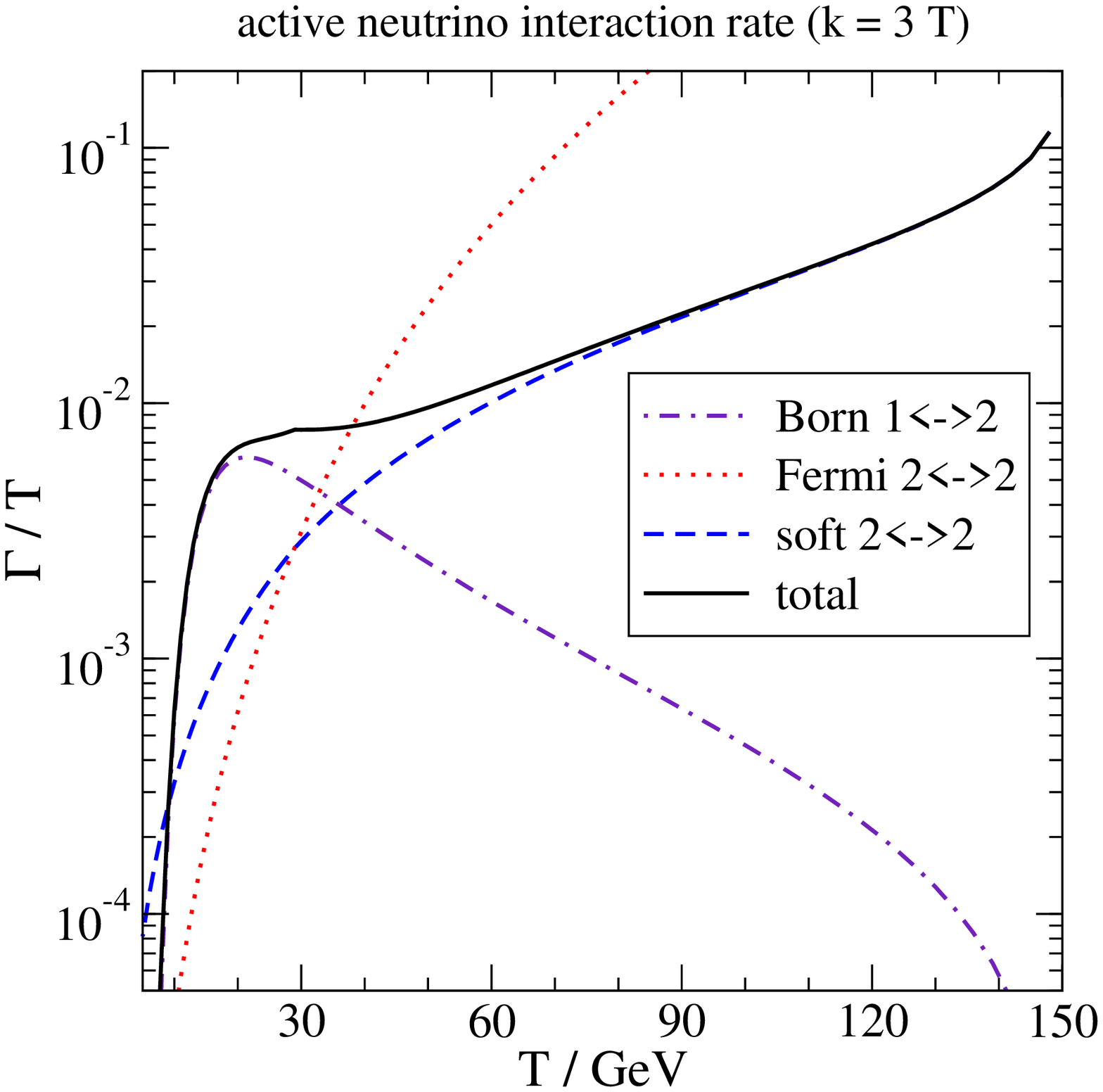}%
  \hspace{0.1cm}
  \epsfxsize=7.5cm\epsfbox{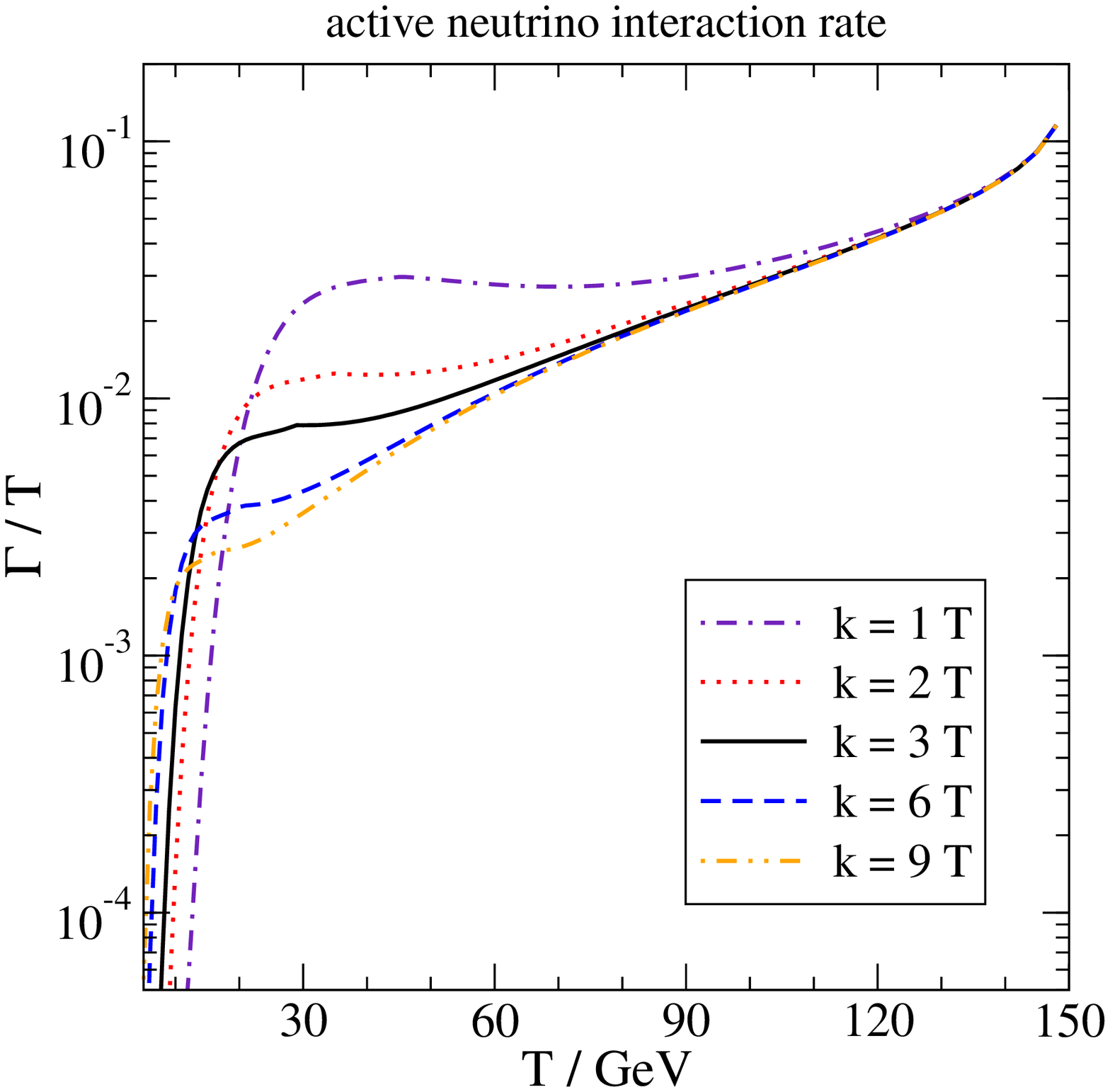}
}

\caption[a]{\small
 The active neutrino interaction rate, defined in \eq\nr{Gamma_def}. 
 Shown are the Born rate from \eq\nr{tree2} (``Born $1\leftrightarrow 2$''), 
 the Fermi model result for $2\leftrightarrow 2$ scatterings  
 from \eq\nr{indirect_Fermi} (``Fermi $2\leftrightarrow 2$''), and
 the soft $2\leftrightarrow 2$ scattering  contribution from 
 \eq\nr{indirect_HTL_gauge_soft} 
 (``soft $2\leftrightarrow 2$''). The total result has been obtained 
 by taking the smaller between the Fermi and the soft $2\leftrightarrow 2$
 scattering results, which limits both to their ranges of applicability, 
 and adding to it the Born $1\leftrightarrow 2$ rate.
 On the right, the total rate is shown for a number of momenta. 
}

\la{fig:imSigma}
\end{figure}

%
\subsection{Interaction rate from $2\leftrightarrow 2$ scatterings with
            hard momentum transfer}
\la{ss:hard_t}

We now turn to the $2\to 2$ contribution to $\Gamma$. 
Proceeding first with Feynman diagrams, the result can be written
in a form analogous to the direct contribution in \eq\nr{boltzmann}: 
\ba
 2 \nF{}(\ko)
 \left. \ko\Gamma \right|^\rmi{hard}_\rmi{$2\to 2$} 
 & = & 
 \int \! {\rm d}\Omega^{ }_{2\to 2} \; 
 \Bigl\{\; 
  \nB{}(p_1) \, \nB{}(p_2) \, \bigl[1 - \nF{}(k_1)\bigr]
  \, \fr12 \sum |\mathcal{M}^{ }_\rmi{d}|^2  
 \nn & & \hspace*{1.6cm} + \, 
   \nB{}(p_1) \, \nF{}(p_2) \, \bigl[1 + \nB{}(k_1)\bigr]
  \, \sum |\mathcal{M}^{ }_\rmi{e}|^2 
 \nn & & \hspace*{1.6cm} + \, 
   \nF{}(p_1) \, \nF{}(p_2) \, \bigl[1 - \nF{}(k_1)\bigr]
  \, \sum |\mathcal{M}^{ }_\rmi{f}|^2 
 \; \Bigr\} 
 \;. \la{boltzmann2}
\ea
The corresponding diagrams are shown in \fig\ref{fig:processes}.
In the massless limit (this will be rectified below), we obtain
\ba
 \sum |\mathcal{M}^{ }_\rmi{d}|^2 & \equiv & 
 - \biggl( 6 g_2^4 + 
   \frac{g_1^4 + 3 g_2^4}{2} \nS^{ }\biggr)
 \, \biggl(  \frac{u^2 + t^2}{s^2}  \biggr)
 +  \frac{\nS^{ }}{2} \bigl( g_1^4 + 3 g_2^4 \bigr)
 \nn & & \; + \, 
 \biggl( \frac{g_1^2 + 3 g_2^2}{2} \biggr)^2
 \biggl( \frac{u}{t} + \frac{t}{u} \biggr)
 \;, \la{M*M2} \\ 
 \sum |\mathcal{M}^{ }_\rmi{e}|^2 & \equiv & 
  \biggl( 6 g_2^4 + 
   \frac{g_1^4 + 3 g_2^4}{2} \nS^{ }\biggr)
 \, \biggl(  \frac{u^2 + s^2}{t^2}  \biggr)
 -  \frac{\nS^{ }}{2} \bigl( g_1^4 + 3 g_2^4 \bigr)
 \nn & & \; - \, 
  \biggl( \frac{g_1^2 + 3 g_2^2}{2} \biggr)^2
 \biggl( \frac{u}{s} + \frac{s}{u} \biggr)
 \;, \\ 
 \sum |\mathcal{M}^{ }_\rmi{f}|^2 & \equiv & 
  \biggl( 3 g_2^4 + \fr{5}3 g_1^4 \biggr)
  \nG^{} \biggl(  \frac{u^2 + s^2}{t^2} + \frac{u^2 + t^2 }{s^2}
 + \frac{t^2 + s^2}{u^2}  \biggr)
 \nn & & \; + \, 
  \fr34 \bigl( g_1^4 + 6 g_1^2 g_2^2 - 3 g_2^4 \bigr)
 \;. \la{M*M4}
\ea 
In order to simplify 
the last equation we have 
symmetrized the integrand in $p_1 \leftrightarrow p_2$
and made use of the identity
$
 u^2/(st)+t^2/(su)+s^2/(ut) = 3
$.
If the phase space integrals were finite (which they are not), 
\eqs\nr{M*M2}--\nr{M*M4}
would suggest that $\ko\Gamma \sim g^4 T^2$.

%
\begin{figure}[t]

\begin{eqnarray*}
  \sum |\mathcal{M}^{ }_\rmi{d}|^2 
 & = & 
 \hspace*{0.45cm}
 \biggl| \nprocK \hspace*{-2mm} \biggr|^2
 \;\; + \;\; 
 \biggl| 
 \nprocL
  + 
 \nprocM
 + 
 \nprocMp
 \hspace*{-2mm} \biggr|^2 
 \\[3mm] 
  \sum |\mathcal{M}^{ }_\rmi{e}|^2 
 & = & 
 \hspace*{0.45cm}
 \biggl| \nprocJ \hspace*{-2mm} \biggr|^2
 \;\; + \;\; 
 \biggl| 
 \nprocG
  + 
 \nprocH
  + 
 \nprocIp \hspace*{-2mm} \biggr|^2
 \\[3mm] 
  \sum |\mathcal{M}^{ }_\rmi{f}|^2 
 & = & 
 \hspace*{0.45cm}
 \biggl| 
 \nprocN
  + 
 \nprocO \hspace*{-2mm} \biggr|^2
 \;\; + \;\; 
 \biggl| 
 \nprocP
  + 
 \nprocPp \hspace*{-2mm} \biggr|^2
\end{eqnarray*}

\caption[a]{\small 
 $2\to 2$ scattering processes contributing to \eqs\nr{M*M2}--\nr{M*M4}. 
 The notation is like in \fig\ref{fig:lpm}.
} 
\la{fig:processes}
\end{figure}
%

In analogy with \eq\nr{direct_hard}, the phase space can be reduced
into a 2-dimensional one:
\ba
 \left. \ko \Gamma \right|^\rmi{hard}_\rmi{$2\to 2$} 
 & = & 
 \frac{1}{(4\pi)^3\ko}
 \int_{\ko}^{\infty} \! {\rm d} \qp \int_0^{\ko} \! {\rm d} \qm 
 \Bigl\{ 
  \bigl[\nB{}(q_0) + \nF{}(q_0 - \ko) \bigr]  \,  \Xi_\rmi{$s$1} 
 \nn & & \quad
 + \, 
  \bigl[\nF{}(q_0) + \nB{}(q_0 - \ko) \bigr]  \,  \Xi_\rmi{$s$2} 
 \Bigr\} 
 \nn  
 & + & 
 \frac{1}{(4\pi)^3\ko}
 \int_{0}^{\ko} \! {\rm d} \qp \int_{-\infty}^{0} \! {\rm d} \qm  
 \Bigl\{ 
  \bigl[1 + \nB{}(q_0) - \nF{}(\ko - q_0) \bigr]  \,  
  \Xi_\rmi{$t$1} 
 \nn & & \quad
 + \, 
  \bigl[1 - \nF{}(q_0) + \nB{}(\ko - q_0) \bigr]  \,  
  \Xi_\rmi{$t$2} 
 \Bigr\} 
 \;. \la{k0Gamma}
\ea
The most important case $\Xi_\rmi{$t$1}$ corresponds to 
bosonic $t$-channel exchange. 
The functions appearing in \eq\nr{k0Gamma} are given 
in appendix~C. 

The $s$-channel functions $  \Xi_\rmi{$s$1} $ and   $\Xi_\rmi{$s$2} $
remain finite in the whole integration range. In contrast,  
the $t$-channel functions
$  \Xi_\rmi{$t$1}$ and
$  \Xi_\rmi{$t$2}$ 
have non-integrable divergences at $q,q_0 \ll \ko$: 
\ba
 \Xi_\rmi{$t$1} 
 & = &
 \Bigl( 6g_2^4 + \frac{g_1^4 + 3 g_2^4}{2} \nS \Bigr)
 \biggl\{ 
   \frac{\pi^2 T^2 q_0}{3 q^4} \Bigl[3(q_0 - 2\ko)^2 - q^2 \Bigr]
 \nn & & \quad
   + \, \frac{\ko(q_0 - \ko)T}{q^4}\Bigl[ 6 q_0 q + 
    (q^2 - 3 q_0^2)\ln\frac{q+q_0}{q-q_0} \Bigr] 
   + \frac{\ko^2 q_0 (q^2 - q_0^2)}{2 q^4}
    \biggr\}
  \nn 
  & + & 
 \Bigl( 6g_2^4 + \fr{10}3 g_1^4 \Bigr) \nG 
 \biggl\{  
   \frac{\pi^2 T^2 q_0}{6 q^4} \Bigl[3(q_0 - 2\ko)^2 - q^2 \Bigr]
   - \frac{\ko^2 q_0 (q^2 - q_0^2 )}{2 q^4}
  \biggr\} 
  + \rmO \bigl( 1 \bigr)
  \;, \la{asyt1} \hspace*{8mm} \\  
 \Xi_\rmi{$t$2} 
 & = &  \Bigl( \frac{g_1^2 + 3g_2^2}{2} \Bigr)^2 \, 
  \frac{\ko \pi^2 T^2}{q^2}
  + \rmO\Bigl( \frac{1}{q} \Bigr) 
 \;. \la{indirect_div_lepton}
\ea
Here all terms that need to be subtracted in order for the integrals
to be finite have been shown; in \eq\nr{asyt1} one more order is needed, 
because the multiplier of $\Xi_\rmi{$t$1}$ in \eq\nr{k0Gamma}
contains the divergent factor $\nB{}(q_0)$.

It may be noted that \eq\nr{indirect_div_lepton} has precisely
the same type of logarithmic divergence as \eq\nr{direct_div}.
In contrast, \eq\nr{asyt1} leads to
a power-divergent integral. This indicates that our
naive estimate concerning the magnitude of $\ko \Gamma$ is incorrect; 
in fact soft gauge scatterings boost the width, so that its correct
magnitude is $\ko\Gamma \sim g^4 T^4 / (g^2 T^2) \sim g^2 T^2$. 
We now turn to the determination of this IR contribution.

%
\subsection{Interaction rate from $2\leftrightarrow 2$ scatterings with
             soft momentum transfer}
\la{ss:soft_t}

The particle mediating soft
$t$-channel exchange in \fig\ref{fig:processes}
can be either a gauge boson or a lepton. 
However only one particle can be soft at a time. 
We start by briefly discussing the simpler case that the exchanged particle  
is a lepton, because the analysis is then analogous to that 
in \eqs\nr{direct_HTL_pre}-\nr{direct_HTL_soft}, however in the 
end this contribution will turn out to be parametrically subdominant
(it amounts to an NNLO contribution). 

When the lepton is soft, the gauge boson is hard. 
A hard gauge boson can be treated like a free massless
particle in the symmetric phase. Then the HTL contribution looks 
much like in \eqs\nr{direct_HTL_pre} and \nr{direct_HTL}, 
\ba
 \left. \ko \Gamma \right|^{\rmii{HTL}\rmi{-lepton}}
 _\rmi{$2\to 2$}
 & = & 
 \frac{g_1^2 + 3 g_2^2}{4}
 \int_{-\infty}^{\infty} \! {\rm d}q_0 
 \int_\vec{q} \frac{-2 \mathcal{K}\cdot\rho_{\ell}(q_0,\vec{q})}
 {|\vec{k-q}|}
 \nn[2mm] 
 & & \; \times \, 
  \bigl[ 1 - \nF{}(q_0) + \nB{}(\ko - q_0)  \bigr]
 \, \delta(\ko - q_0 - |\vec{k-q}|)
  \la{indirect_HTL_lepton_pre} \\[2mm]
& \approx & 
 \frac{g_1^2 + 3 g_2^2}{8\pi^2}
 \int_{-\infty}^{\ko} \! {\rm d}q_0 
 \int_{|q_0|}^{2\ko - q_0} \! {\rm d}q\, q\, q_0 \, 
 (\hat{\rho}_s - \hat{\rho}_0) (q_0,q) 
 \, \Bigl[ \nB{}(\ko) + \fr12 \Bigr]
 \;. \hspace*{8mm} \la{indirect_HTL_lepton}
\ea
We again treat this in two different ways. A subtraction term is obtained
by expanding like in \eq\nr{lepton_expand}, and yields
\be
 \left. \ko \Gamma \right|
  ^{\rmii{HTL}\rmi{-lepton},\,\rmi{expanded}}
  _\rmi{$2\to 2$} = 
 \frac{g_1^2 + 3 g_2^2}{32\pi}
 \int_{-\infty}^{\ko} \! {\rm d}q_0 
 \int_{|q_0|}^{2\ko - q_0} \! {\rm d}q\,  \frac{m_\ell^2}{q^2}
 \, \Bigl[ \nB{}(\ko) + \fr12\, \Bigr]
 \;. \la{indirect_HTL_lepton_exp}
\ee
This agrees with the divergence 
originating from \eq\nr{indirect_div_lepton}. 
The philosophy is to subtract \eq\nr{indirect_HTL_lepton_exp}
from \eq\nr{k0Gamma}, and add the corresponding
``soft'' contribution from \eq\nr{indirect_HTL_lepton} 
in its proper form. Following \eq\nr{direct_HTL_soft}, 
we readily obtain 
\ba
 \left. \ko \Gamma \right|^{\rmii{HTL}\rmi{-lepton},\,\rmi{soft}}
 _\rmi{$2\to 2$}
 & \approx & 
 \frac{(g_1^2 + 3 g_2^2) m_\ell^2}{16\pi} \ln\Bigl(\frac{2\ko}{m_\ell} \Bigr)
 \, \Bigl[ \nB{}(\ko) + \fr12\, \Bigr]
 \;, \la{indirect_HTL_lepton_soft}
\ea
where terms of $\rmO(m_\ell/\ko)$ were omitted. 
With \eq\nr{indirect_HTL_lepton_exp} subtracted and 
\eq\nr{indirect_HTL_lepton_soft} added, the indirect 
contribution from the function $ \Xi_\rmi{$t$2} $ to
$
 \left. \ko \Gamma \right|^{ }
 _\rmi{$2\to 2$}
$ 
is finite and of $\rmO(g^4 T^2)$. 
 
If, in contrast, the lepton is hard, it can be treated like a free massless
particle in the symmetric phase. The exchanged gauge boson needs now to 
be HTL resummed. We get\footnote{%
 Vertex corrections are of higher order and can be omitted.  
 } 
\ba
 & & 
 \hspace*{-1.0cm}
 \left. \ko \Gamma \right|^{\rmii{HTL}\rmi{-gauge}}
 _\rmi{$2\to 2$} 
 \; = \;
  \int_{-\infty}^{\infty} \! {\rm d}q_0 
 \int_\vec{q} 
 \frac{
 \, \bigl[ 1 + \nB{}(q_0) - \nF{}(\ko - q_0) \bigr]
 \, \delta(\ko - q_0 - |\vec{k-q}|) 
 }
 {2 |\vec{k-q}|}
 \nn[2mm] 
 & & \; \times \, 
 \biggl\{
 2 g_2^2 \biggl[  
 \frac{\ko^2 q_\perp^2 
 (\rho^{ }_\rmii{T2} - \rho^{ }_\rmii{E2})(q_0,{q}) }{q^2} 
  +\frac{(q^2 - q_0^2) \rho^{ }_\rmii{T2} (q_0,{q}) }{2 } 
 \biggr]
 \, + \, 
 (g_1^2 + g_2^2) \bigl[  
      2 \to Z 
 \bigr]
  \biggr\}
 \nn[2mm] 
 & = & 
 \frac{1}{8\pi^2\ko}
 \int_{-\infty}^{\ko} \! {\rm d}q_0 
 \int_{|q_0|}^{2\ko - q_0} \! {\rm d}q\, q\,  
 \bigl[ 1 + \nB{}(q_0) - \nF{}(\ko - q_0) \bigr]
 \nn[2mm] 
 & & \; \times \, 
 \biggl\{
 2 g_2^2 \biggl[  
 \frac{\ko^2 q_\perp^2 
 (\rho^{ }_\rmii{T2} - \rho^{ }_\rmii{E2})(q_0,{q}) }{q^2} 
  +\frac{(q^2 - q_0^2) \rho^{ }_\rmii{T2} (q_0,{q}) }{2 } 
 \biggr]
 \, + \, 
 (g_1^2 + g_2^2) \bigl[  
      2 \to Z 
 \bigr]
  \biggr\} 
 \;, \nn \la{indirect_HTL_gauge}
\ea
where (for $k =\ko$)
\be
 q_\perp^2 
 \; \equiv \;
    q^2 - q_\parallel^2
 \; \equiv \; q^2 - \frac{(\vec{k}\cdot\vec{q})^2}{k^2} 
 \; = \; \frac{(q^2 - q_0^2)[(q_0 - 2\ko )^2 - q^2] }{4\ko^2}
 \;. \la{qperp2}
\ee
The HTL spectral functions 
$\rho^{ }_\rmii{T$2$}$, $\rho^{ }_\rmii{E$2$}$, 
$\rho^{ }_\rmii{T$Z$}$, $\rho^{ }_\rmii{E$Z$}$
are given in appendix~B. 

We again treat the soft contribution
in \eq\nr{indirect_HTL_gauge} in two different ways. 
A subtraction term is obtained by going to the symmetric phase, 
like in the computation based on \eqs\nr{M*M2}--\nr{M*M4}, 
and expanding like in \eq\nr{lepton_expand}. 
For $q,q_0 \gg g T$ this yields
\be
 \frac{\ko^2 q_\perp^2 
 (\rho^{ }_\rmii{T$i$} - \rho^{ }_\rmii{E$i$})(q_0,{q}) }{q^2} 
  +\frac{(q^2 - q_0^2) \rho^{ }_\rmii{T$i$} (q_0,{q}) }{2 } 
 \; \approx \; \frac{\pi m_\rmii{E$i$}^2q_0}{16 q^5}
 \, \bigl[ 3(q_0 - 2\ko)^2 - q^2 \bigr] \, 
 \;, \quad 
 i \in \{ 1,2\} 
 \;.  
\ee
Thereby 
\ba
 && \hspace*{-2cm}
 \left. \ko \Gamma \right|
  ^{\rmii{HTL}\rmi{-gauge},\,\rmi{expanded}}
  _\rmi{$2\to 2$} \; = \;  
  \frac{ g_1^2 m_\rmii{E1}^2 + 3  g_2^2 m_\rmii{E2}^2 }{128\pi\ko}
 \nn & \times & 
 \int_{-\infty}^{\ko} \! {\rm d}q_0 
 \int_{|q_0|}^{2\ko - q_0} \! {\rm d}q\,  
 \bigl[1 + \nB{}(q_0) - \nF{}(\ko - q_0) \bigr]
 \, \frac{q_0}{q^4}
 \, \bigl[ 3(q_0 - 2\ko)^2 - q^2 \bigr] \, 
 \;. \la{indirect_HTL_gauge_exp}
\ea
Inserting the masses from \eq\nr{Debye}
this agrees exactly with the leading $1/q^4$ divergence 
as shown in \eq\nr{asyt1}. 

The philosophy is thus to subtract \eq\nr{indirect_HTL_gauge_exp}
from \eq\nr{k0Gamma}, and add the corresponding
``soft'' contribution from \eq\nr{indirect_HTL_gauge} 
in its proper form. The soft contribution originates
from $q,q_0 \sim m^{ }_\rmii{E$i$} \ll \ko$.
Changing variables from $q$ to 
$q_\perp$ (cf.\ \eq\nr{qperp2}) 
the leading contribution 
from \eq\nr{indirect_HTL_gauge} 
(in an expansion in $\rmO(m_\rmii{E$i$}^2/\ko^2)$)
becomes
\ba
 && \hspace*{-1cm}
 \left. \ko \Gamma \right|^{\rmii{HTL}\rmi{-gauge},\,\rmi{soft}}
 _\rmi{$2\to 2$} \nn 
 & \approx & 
 \frac{\ko T}{8\pi^2}
 \int_{-\infty}^{\ko} \! {\rm d}q_0 
 \int_{0}^{2\ko} \! {\rm d}q_\perp\, q_\perp\, 
 \frac{2g_2^2( {\rho}^{ }_\rmii{T2} - {\rho}^{ }_\rmii{E2} ) + 
  (g_1^2 + g_2^2)( {\rho}^{ }_\rmii{T$Z$} - {\rho}^{ }_\rmii{E$Z$} ) }{q_0} 
 \, \frac{q_\perp^2}{q_\perp^2 + q_0^2}
 \;. \hspace*{8mm} \la{HTL_g_soft} 
\ea
At this point we can change the order of integrations
like in \eq\nr{direct_HTL_soft}
and make use of a sum rule 
derived in refs.~\cite{sum1,sum2}, 
\ba
  \int_{-\infty}^{\infty} \! \frac{{\rm d} q_0}{\pi}
    \frac{  \rho_\rmii{T2} - \rho_\rmii{E2} }{q_0} 
 \Bigl(q_0,\sqrt{q_0^2 + q_\perp^2}\Bigr) 
  \frac{q_\perp^2}{ q_\perp^2 + q_0^2}
  \; = \; 
  \frac{1}
  {q_\perp^2 + \mW^2} - 
   \frac{1}
  {q_\perp^2 + \mWt^2}   
 \;. 
\ea
This structure corresponds to that in the Matsubara zero-mode 
sector, and equals the integrand in \eq\nr{VWy}. 
Similarly, the $Z$ channel case leads to the integrand in 
\eq\nr{VZy}. 
The integral over $q_\perp$ can now be carried out, yielding
\ba
 & &  \hspace*{-1.5cm} 
 \left. \ko \Gamma \right|^{\rmii{HTL}-\rmi{gauge},\,\rmi{soft}}
 _\rmi{$2\to 2$}
  =    
 \frac{ \ko T}{8\pi} 
 \biggl\{ 
 2 g_2^2 \ln
   \frac{ \mWt }{ \mW } 
  \nn 
  & + & 
    (g_1^2 + g_2^2)
 \biggl[
   \cos^2(\theta - \tilde\theta)
   \ln \frac{ \mZt }{ \mZ } 
 + 
   \sin^2(\theta - \tilde\theta)
   \ln \frac{ \mQt }{ \mZ } 
 \biggr]
 \biggr\}
 \; + \; 
 \rmO\Bigl( \frac{g^2 T m_\rmii{E$2$}^2}{\ko} \Bigr)
 \;. \hspace*{6mm}
 \la{indirect_HTL_gauge_soft} 
\ea
We note that \eq\nr{indirect_HTL_gauge_soft}
is of $\rmO(g^2 T^2)$ and is finite in the broken phase.
In the notation of \eq\nr{V4x4}, 
$
\left. \Gamma \right|^{\rmii{HTL}-\rmi{gauge},\,\rmi{soft}}_{2\to 2} = 
 2 \Gamma^{ }_\rmii{$W$}(0) + \Gamma^{ }_\rmii{$Z$}(0)
$. 
Eq.~\nr{indirect_HTL_gauge_soft} is among our main results. 

Given that \eq\nr{indirect_HTL_gauge_soft} represents an 
IR sensitive result, being dominated by momentum transfer
of $\rmO(gT)$, it could experience large 
radiative corrections. In fact, as is typical of observables
determined by the Debye scale, these are only suppressed
by $\rmO(g/\pi)$. We have not computed these NLO corrections, 
but a way to do this is outlined in appendix~D. Let us stress again
that, in contrast, the contribution from hard momentum transfer 
is of $\rmO(g^4 T^2)$, i.e.\ NNLO, once the IR sensitive parts
have been subtracted and treated properly. 

The result of \eq\nr{indirect_HTL_gauge_soft} is illustrated
numerically in \fig\ref{fig:imSigma} (``soft $2\leftrightarrow 2$''). 
It dominates the active neutrino interaction rate at $T \gsim 40$~GeV.

%
\subsection{Limit of low temperatures}
\la{ss:indirect_Fermi}

At low temperatures, $\mW^{ }\gsim gT$, the mass ratios appearing 
in \eq\nr{indirect_HTL_gauge_soft} behave as 
$
 { \mWt^2 } / { \mW^2 } \approx 1 +
 m_\rmii{E2}^2 / { \mW^2 } 
$, so that the interaction rate decreases as 
$
 \ko \Gamma \sim \ko\, g^4 T^3 / { \mW^2 }
$.
This is not the correct low-temperature asymptotics, however; 
the approximations made break down when 
$ \mW \gg g T$, and the correct form is
$
 \ko \Gamma \sim \ko^2\, g^4 T^4 / { \mW^4 }
$. 
This asymptotics can be 
computed within the Fermi model. The corresponding results
have been tabulated, up to $T\sim 10$~GeV, on the web page
related to ref.~\cite{numsm}. For completeness we specify here
the result for 5~GeV $\lsim T \lsim 30$~GeV in a form which 
is easily amenable to a numerical evaluation. 

Adding the contribution of the bottom quark to the processes listed
in ref.~\cite{numsm}; going to the limit $M\to 0$ in which the result
is gauge independent; and making use of the same variables as in 
\eqs\nr{not1}--\nr{direct_hard}, we obtain 
\ba
 \left. \ko 
 \Gamma \right|^\rmi{Fermi}_\rmi{$2\to 2$} 
 & = & 
 \frac{1}{(4\pi)^3\ko}
 \int_{\ko}^{\infty} \! {\rm d} \qp \int_0^{\ko} \! {\rm d} \qm 
 \, 
  \bigl[\nB{}(q_0) + \nF{}(q_0 - \ko) \bigr]  \,  \Xi_\rmi{$s$1} 
 \nn  
 & + & 
 \frac{1}{(4\pi)^3\ko}
 \int_{0}^{\ko} \! {\rm d} \qp \int_{-\infty}^{0} \! {\rm d} \qm 
 \, 
  \bigl[1 + \nB{}(q_0) - \nF{}(\ko - q_0) \bigr]  \,  
  \Xi_\rmi{$t$1} 
 \;. \hspace*{5mm} \la{indirect_Fermi}
\ea
The functions appearing read
\ba
 \Xi_\rmi{$s$1} 
 & = & 
  16 \, A \, G_\rmii{F}^2 \, (q_0^2 - q^2)^2  \, 
  \bigl[ q + 2 T (\lnfplus - \lnfminus) \bigr]
 \;, \\ 
 \Xi_\rmi{$t$1} 
 & = & 
  16 \, A \, G_\rmii{F}^2 \, (q_0^2 - q^2)^2  \, 
  \bigl[ 2 T (\lnfminus - \lnfplus) \bigr]
 \;, 
\ea
where $G_\rmii{F}^{ }$ is the Fermi constant, and 
\be
 A \equiv \frac{15}{2} - 2 s^2 + 12 s^4 
 + \Nc \, \biggl[
  \frac{5}{2} - \frac{14 s^2}{3} + \frac{44 s^4}{9} 
  + 2 \Bigl(
    |V^{ }_\rmi{ud}|^2 + 
    |V^{ }_\rmi{us}|^2 + 
    |V^{ }_\rmi{cd}|^2 + 
    |V^{ }_\rmi{cs}|^2 
      \Bigr)
 \biggr]
 \;.
\ee
Here $s = \sin\theta$ is the weak mixing angle, 
defined like below \eq\nr{theta_w}, 
and $ |V^{ }_\rmi{ub}|^2 $ and $ |V^{ }_\rmi{cb}|^2 $ have been omitted
as vanishingly small. 
Numerically, the integral evaluates to 
$\ko \Gamma \approx 10 G_\rmii{F}^2 T^4 \ko^2$.
The result is illustrated in \fig\ref{fig:imSigma}
(``Fermi $2\leftrightarrow 2$'').

%
\section{Numerical results}
\la{se:num}

\begin{figure}[t]

\hspace*{-0.1cm}
\centerline{%
 \epsfxsize=7.5cm\epsfbox{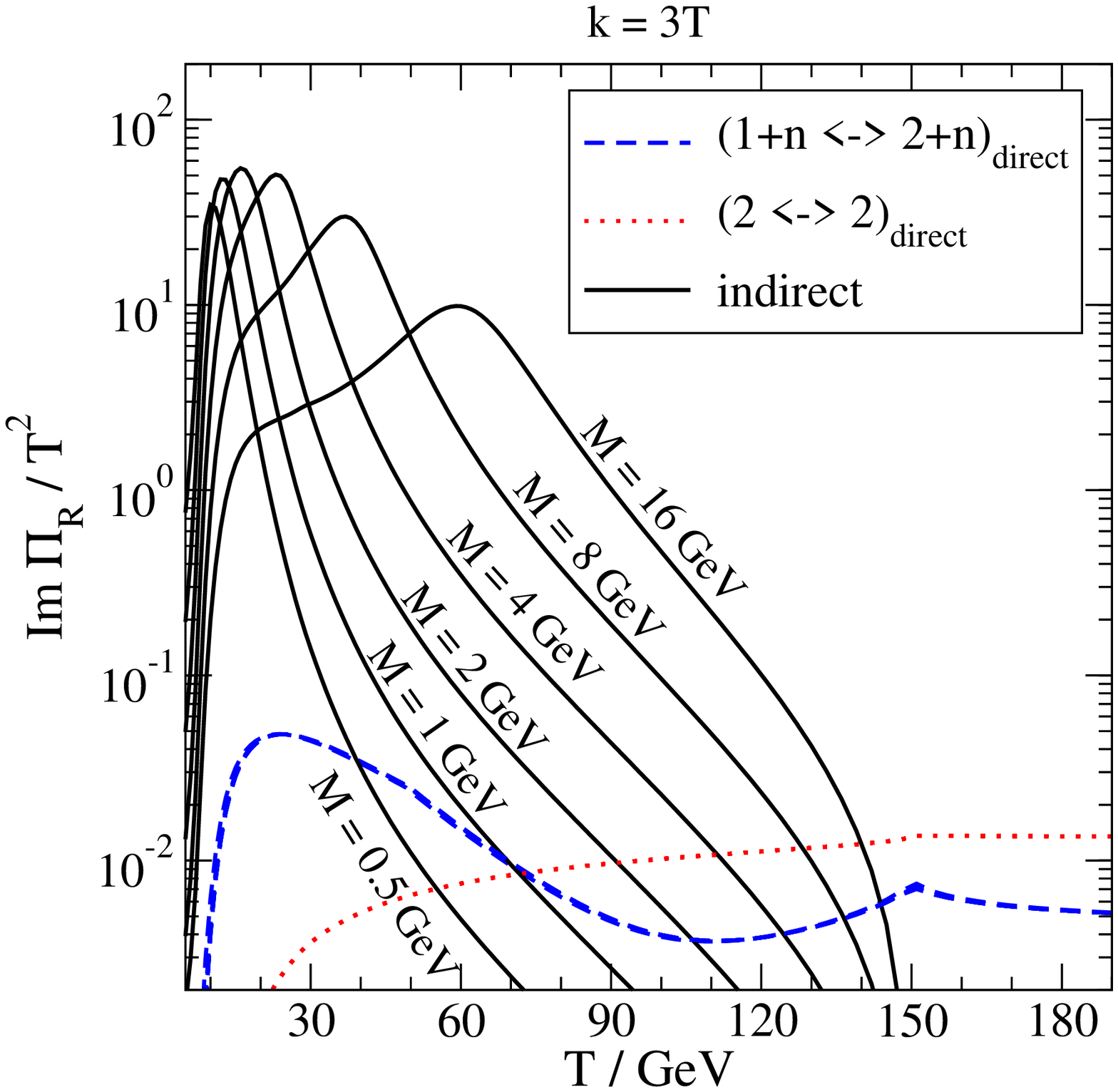}%
 \hspace{0.1cm}
 \epsfxsize=7.5cm\epsfbox{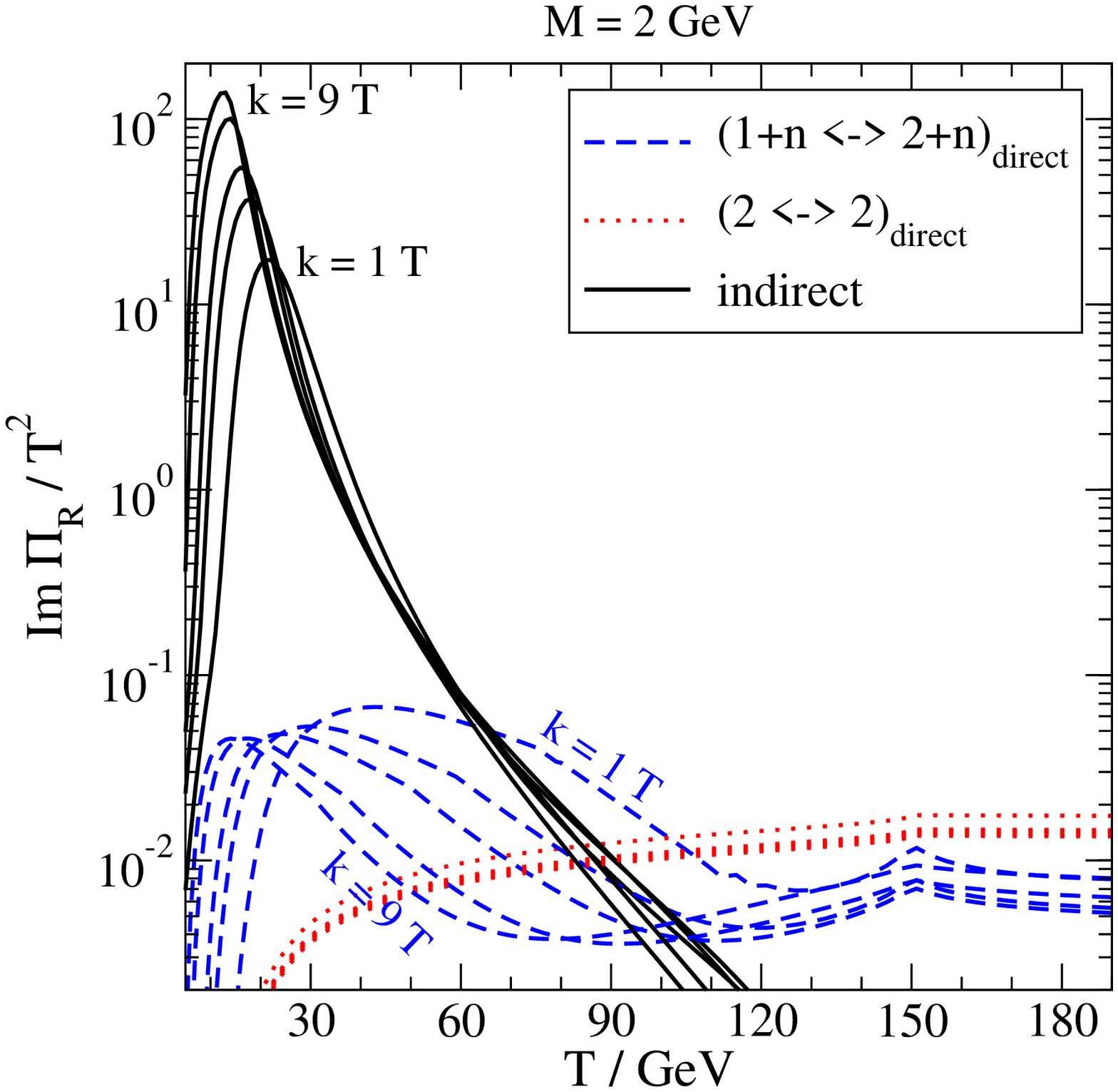}
}

\caption[a]{\small
 Various contributions to $\im \Pi^{ }_\rmii{R}/T^2$. 
 Shown are the direct $1+n\leftrightarrow 2+n$ scatterings 
 from \se\ref{se:dir12} (dashed lines); 
 the direct $2\leftrightarrow 2$ scatterings from 
 \se\ref{se:dir22} (dotted lines); 
 as well as the full indirect
 contribution from \se\ref{se:indirect} (solid lines). 
 Left: fixed $k = 3 T$ and 
 masses $M/\mbox{GeV} \in\{ 0.5,...,16\}$.
 Right: fixed $M = 2$~GeV and 
 momenta $k/T \in \{1,2,3,6,9\}$. 
}

\la{fig:contrs}
\end{figure}

\begin{figure}[t]

\hspace*{-0.1cm}
\centerline{%
 \epsfxsize=7.5cm\epsfbox{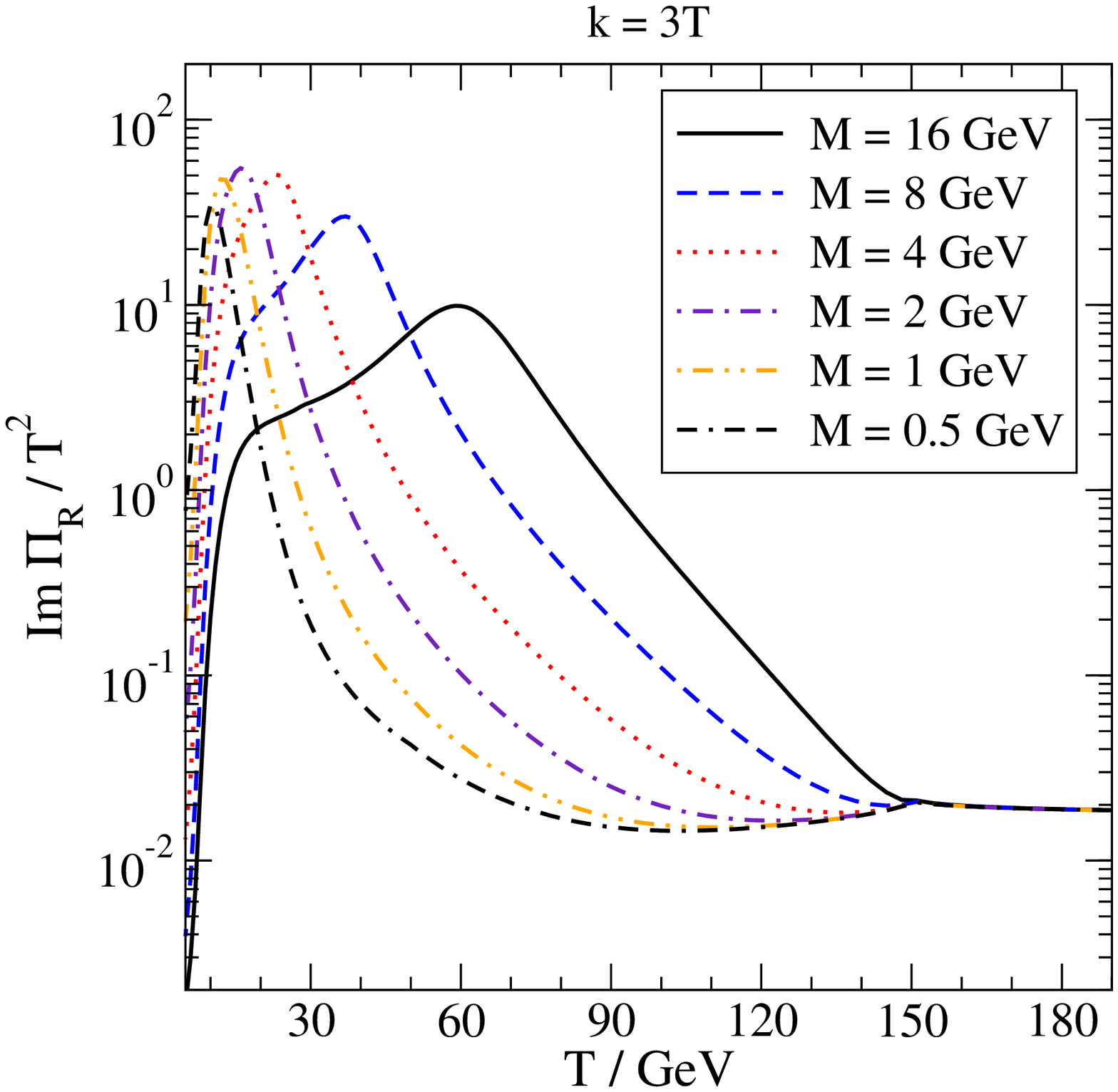}%
 \hspace{0.1cm}
 \epsfxsize=7.5cm\epsfbox{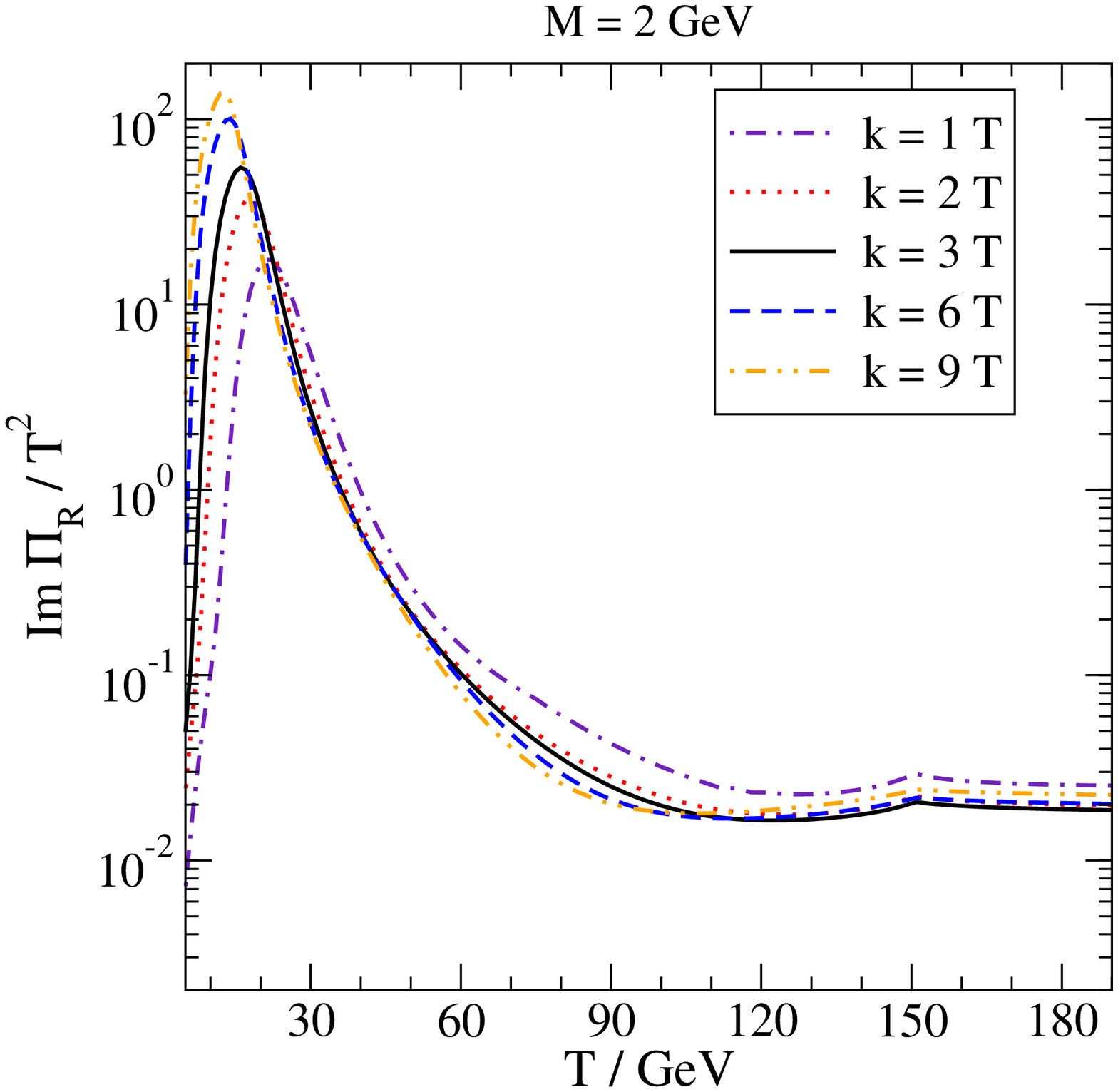}
}

\caption[a]{\small
 The full $\im \Pi^{ }_\rmii{R}/T^2$, obtained by 
 summing together the contributions from \fig\ref{fig:contrs}. 
 Left: fixed $k = 3 T$.
 Right: fixed $M = 2$~GeV.
 Corresponding results for $T< 10$~GeV and $T> 160$~GeV have been 
 tabulated in ref.~\cite{interpolation}.  
}

\la{fig:full}
\end{figure}

\begin{figure}[t]

\hspace*{-0.1cm}
\centerline{%
 \epsfxsize=7.5cm\epsfbox{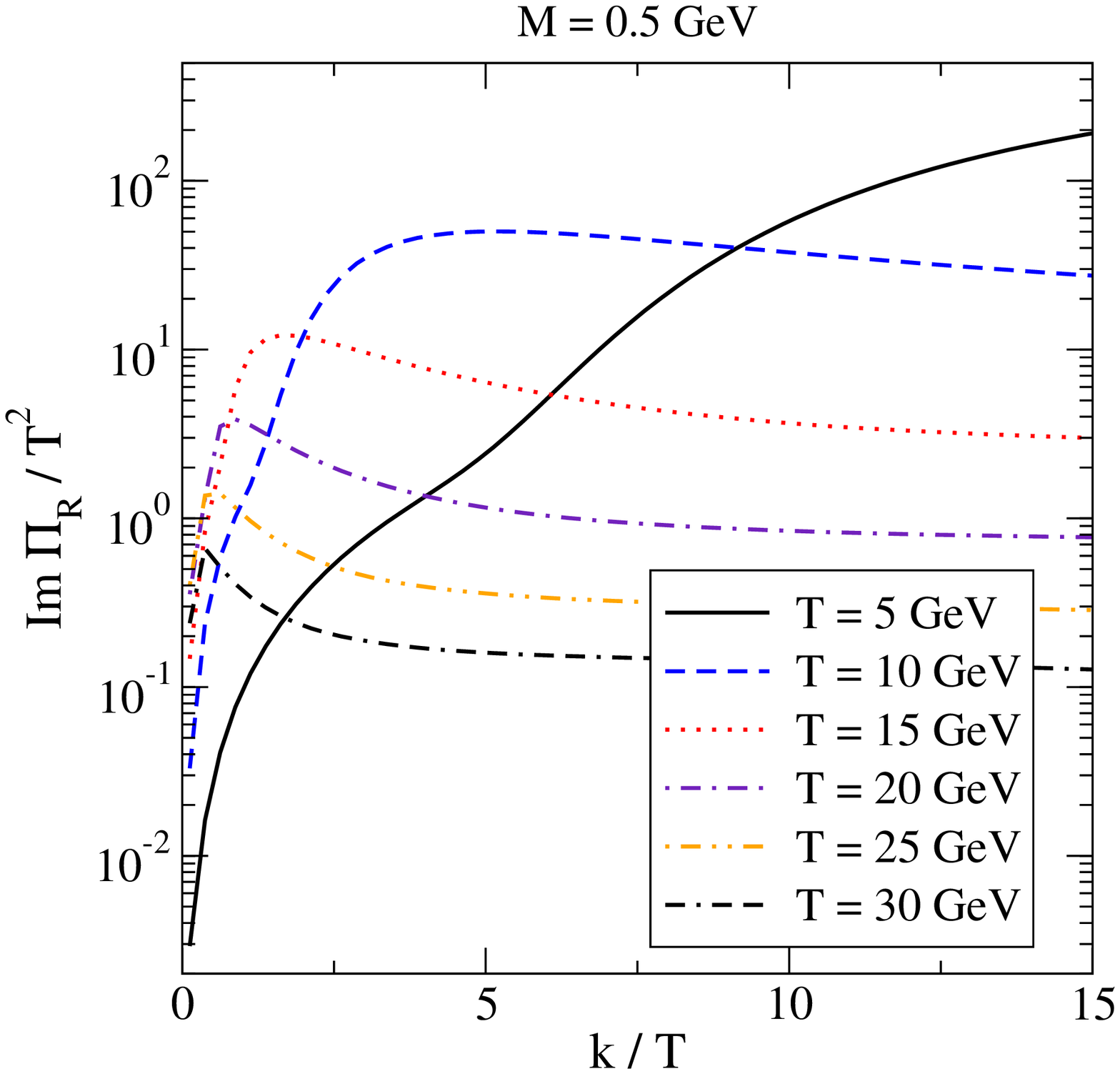}%
 \hspace{0.1cm}
 \epsfxsize=7.5cm\epsfbox{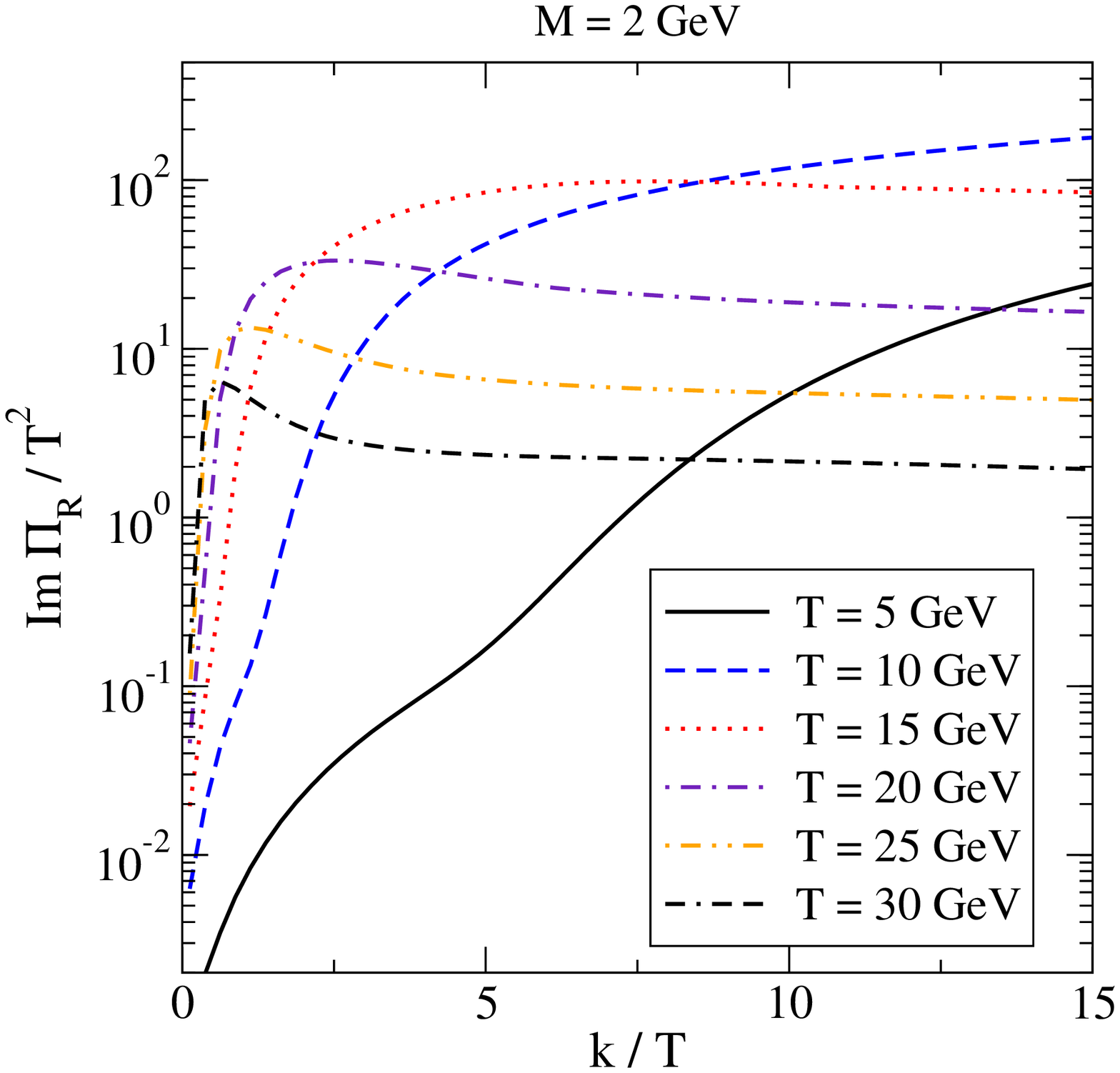}
}


\caption[a]{\small
 The dependence of $\im \Pi^{ }_\rmii{R}/T^2$ on $k$ 
 for $M = 0.5$~GeV (left) and $M = 2$~GeV (right). 
 The spectra at these and other temperatures
 can be downloaded as explained in footnote~\ref{fn:midT}.
}

\la{fig:spectra}
\end{figure}

The contributions of \ses\ref{se:dir12}
(direct $1+n\leftrightarrow 2+n$ scatterings), 
\ref{se:dir22} 
(direct $2\leftrightarrow 2$ scatterings), 
and~\ref{se:indirect} 
(indirect contributions), 
are collected together into \fig\ref{fig:contrs}. 
It is immediately clear that the indirect contribution 
dominates at low temperatures by several orders of 
magnitude. However, the smaller $M$ is, the sooner
does the direct contribution take over, 
because the indirect rate is proportional to $M^2$, 
cf.\ \eq\nr{indirect}.

The total rate, obtained by summing together the direct and indirect
contributions from \fig\ref{fig:contrs}, 
is shown in \fig\ref{fig:full}. The $k$-dependence is
illustrated in more detail for $M = 0.5$~GeV and
$M=2.0$~GeV in \fig\ref{fig:spectra}.\footnote{%
 Tabulated results for full spectra can be downloaded from \la{fn:midT}
 {www.laine.itp.unibe.ch/production-midT/}.
 This web site also lists the active neutrino interaction rate
 shown in \fig\ref{fig:imSigma}(right) and the lepton number
 washout rate shown in \fig\ref{fig:equil}(right). 
 } 
Various physical 
quantities can be obtained by weighting these rates appropriately
and integrating over the spectrum
(cf.\ \eqs\nr{gammaI}, \nr{nI_prod} and \nr{gamma_ab}). Because
of the appearance of the Fermi distribution 
in these weights, 
the phase space is dominated by $k\sim 3 T$.
Two examples of physically relevant quantities are 
discussed in the next section. 

%
\section{Conclusions}
\la{se:concl}

\begin{figure}[t]

\hspace*{-0.1cm}
\centerline{%
 \epsfxsize=7.5cm\epsfbox{equil_k3_sol.eps}%
 \hspace{0.1cm}
 \epsfxsize=7.5cm\epsfbox{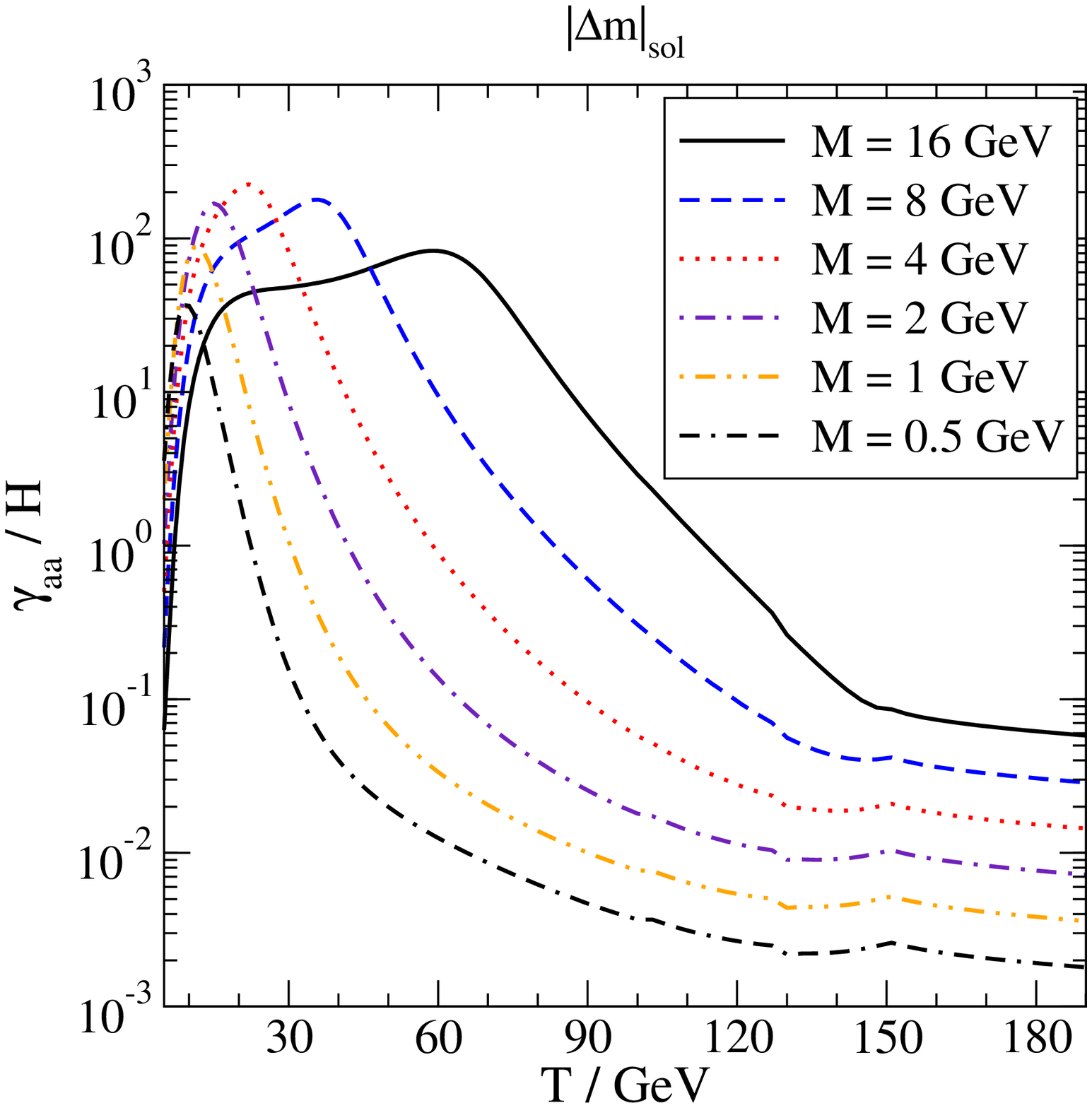}
}

\caption[a]{\small
 Left: The right-handed neutrino equilibration rate 
 compared with the Hubble rate, cf.\ \eq\nr{pheno}, 
 for $k = 3 T$. Active neutrino
 masses correspond to 
 $|\Delta m|^{ }_\rmi{sol} \approx 8.7 \times 10^{-3}$~eV;  
 for the atmospheric neutrino 
 value $|\Delta m|^{ }_\rmi{atm} \approx 4.9\times 10^{-2}$~eV
 the rate is 
 faster by a factor $\sim 5.6$.
 Right: The (diagonal) lepton number washout rate compared with 
 the Hubble rate, cf.\ \eq\nr{na_prod}. 
}

\la{fig:equil}
\end{figure}

For $T < 160$~GeV so that the Higgs mechanism is operative, 
the equilibration rate of a right-handed neutrino of mass $M$
(cf.\ \eq\nr{gammaI})
can be split into ``direct'' and ``indirect'' contributions
(cf.\ \eq\nr{splitup}). These correspond to 
different types of scatterings as illustrated in \figs\ref{fig:lpm} and 
\ref{fig:types}. In the ultrarelativistic regime, where all
particle masses are $\ll \pi T$, 
the indirect contribution can in turn be expressed in terms
of the left-handed neutrino ``asymptotic thermal mass'', $m_\ell$, and
``interaction rate'', $\Gamma$, as indicated by \eq\nr{indirect}. 
At lower temperatures the general structure remains intact even though 
$m_\ell$ gets replaced by a more complicated (momentum-dependent) 
function, as has been reviewed in \se\ref{ss:potential}. 

We have shown that in the regime $T \gsim 40$~GeV, the active neutrino
interaction rate $\Gamma$ is dominated by $t$-channel
scatterings mediated by soft gauge boson exchange
(referred to as the ``soft $2\leftrightarrow 2$'' contribution
in \fig\ref{fig:imSigma}). 
In this situation 
$\Gamma $ is ``large'', 
$\Gamma  \sim g^2 T/\pi$. 
The explicit expression is fairly simple, 
cf.\ \eq\nr{indirect_HTL_gauge_soft}. This large contribution
originates from contributions sensitive
to momenta $\sim gT$ which would be quadratically infrared 
divergent without the appropriate HTL resummation. There is 
also a subleading (linear) infrared divergence 
in \eq\nr{asyt1} whose origin can also be understood 
(cf.\ appendix~D). 

For the masses $M \sim 0.5...2.0$~GeV, relevant for the SHiP
experiment~\cite{ship}, the right-handed neutrino equilibration
rate peaks at low temperatures, $T \sim 5...30$~GeV
(cf.\ \fig\ref{fig:full}). 
In this regime the active neutrino
interaction rate
$\Gamma$ is dominated by $1\rightarrow 2$ decays
(cf.\ \fig\ref{fig:imSigma}) and $\im \Pi^{ }_\rmii{R}$
is dominated by the indirect contribution 
(cf.\ \fig\ref{fig:contrs}). It is again possible to express
the dominant contribution to $\Gamma$ in a simple analytic form, 
cf.\ \eqs\nr{tree2} and \nr{tree3}. 

In order to illustrate the physics significance
of these results, let us first 
compare the right-handed neutrino equilibration
rate $\gamma^{ }_{I\vec{k}}$ from \eq\nr{gammaI} with the Hubble rate 
$H = \sqrt{8\pi e/(3 m_\rmi{Pl}^{2})} $, where $e$ is the energy density
of the universe and $m_\rmi{Pl}$ is the Planck mass.  For simplicity 
we consider a seesaw scenario with hierarchical neutrinos, and 
assume that only one neutrino Yukawa coupling contributes to 
a given mass difference. Then active neutrino mass differences
are of the form 
$
 |\Delta m| = |h^{ }_{Ia}|^2 v^2 / (2 M)
$, 
whereby we can eliminate $|h^{ }_{Ia}|^2$ from $\gamma^{ }_{I\vec{k}}$ to get
\be
 \frac{\gamma^{ }_{I\vec{k}}}{H} \; = \; 
 1.39 \times 10^5 \times
 \left| \frac{\Delta m}{\mbox{eV}}\right| \times 
 \frac{M}{\ko} \times
 \frac{\im \Pi^{ }_\rmii{R}}{\sqrt{e(T)}}
 \;. \la{pheno} 
\ee
Inserting $e$ as tabulated in ref.~\cite{crossover}  
(cf.\ also ref.~\cite{dono}), the result is illustrated 
in \fig\ref{fig:equil}(left). We conclude that in the mass range 
$M \sim 0.5...16$~GeV
right-handed neutrinos do equilibrate at temperatures
above $T=5$~GeV. Increasing the mass above 4~GeV 
decreases the peak equilibration rate but broadens the temperature
range in which the rate is substantial. 

Turning to our main observable, the lepton number washout
rate from \eq\nr{na_prod},  
the flavour-diagonal part of the result is shown  
in \fig\ref{fig:equil}(right). The flavour non-diagonal
rate is an order of magnitude slower because
of the smaller inverse susceptibility, 
cf.\ \fig\ref{fig:susc}(right). The flavour-diagonal rate exceeds 
the Hubble rate for all masses considered. However
we note that this equilibration dynamics rapidly switches off
in the range $T \lsim 4$~GeV in which
dark matter computations have been 
carried out~\cite{numsm,dmpheno,hirata}. 

The results of \fig\ref{fig:equil}(right)
indicate that leptogenesis based on right-handed
neutrinos with few GeV masses remains an interesting possibility, 
because these degrees of freedom do not equilibrate 
at $T \gsim 130$~GeV when sphaleron processes are 
active~\cite{sphaleron}. 
In contrast it is difficult
to generate a large lepton asymmetry for low temperatures, 
which could boost dark matter production
in the scenario of ref.~\cite{sf}, because at $T \lsim 30$~GeV
lepton number violating reactions are in equilibrium and therefore 
an efficient washout process takes place. 
It should be acknowledged, 
however, that we have not performed
a detailed phenomenological scan of the whole parameter space, so the
existence of fine-tuned regions where the window 
may remain open cannot be excluded. 
The numerical results tabulated as explained in footnote~\ref{fn:midT} should 
hopefully permit for further work to be carried out in this direction. 

%
\section*{Acknowledgements}

M.L.\ thanks D.~B\"odeker and M.~Shaposhnikov for valuable discussions. 
The work was supported by the Swiss National Science Foundation
(SNF) under grant 200020-155935, and by the V\"ais\"al\"a Foundation.

%
\appendix
\renewcommand{\thesection}{Appendix~\Alph{section}}
\renewcommand{\thesubsection}{\Alph{section}.\arabic{subsection}}
\renewcommand{\theequation}{\Alph{section}.\arabic{equation}}

%
\section{Lepton number susceptibility matrix}
\la{app:A}

\begin{figure}[t]

\hspace*{-0.1cm}
\centerline{%
 \epsfxsize=7.5cm\epsfbox{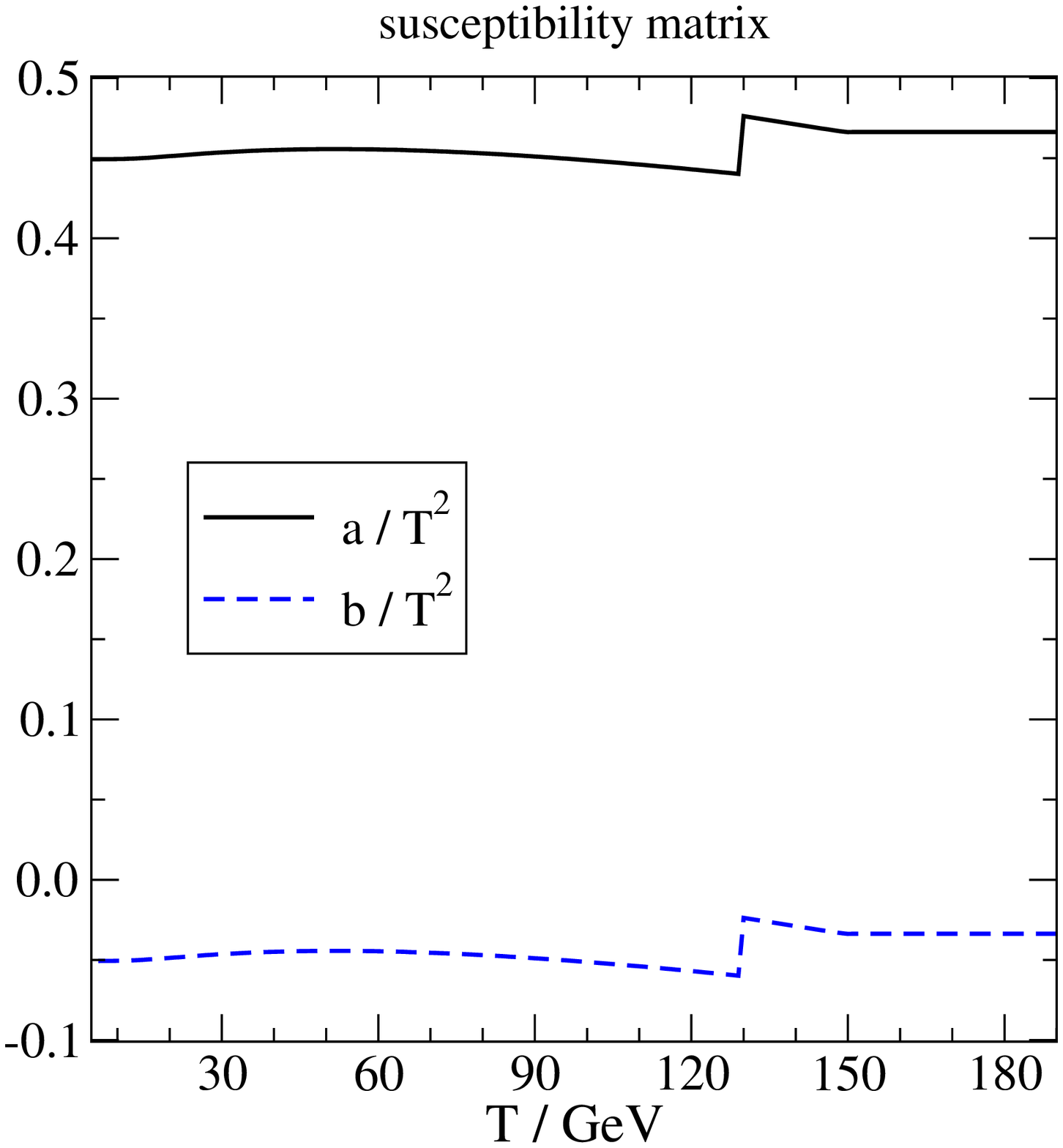}%
 \hspace{0.1cm}
 \epsfxsize=7.5cm\epsfbox{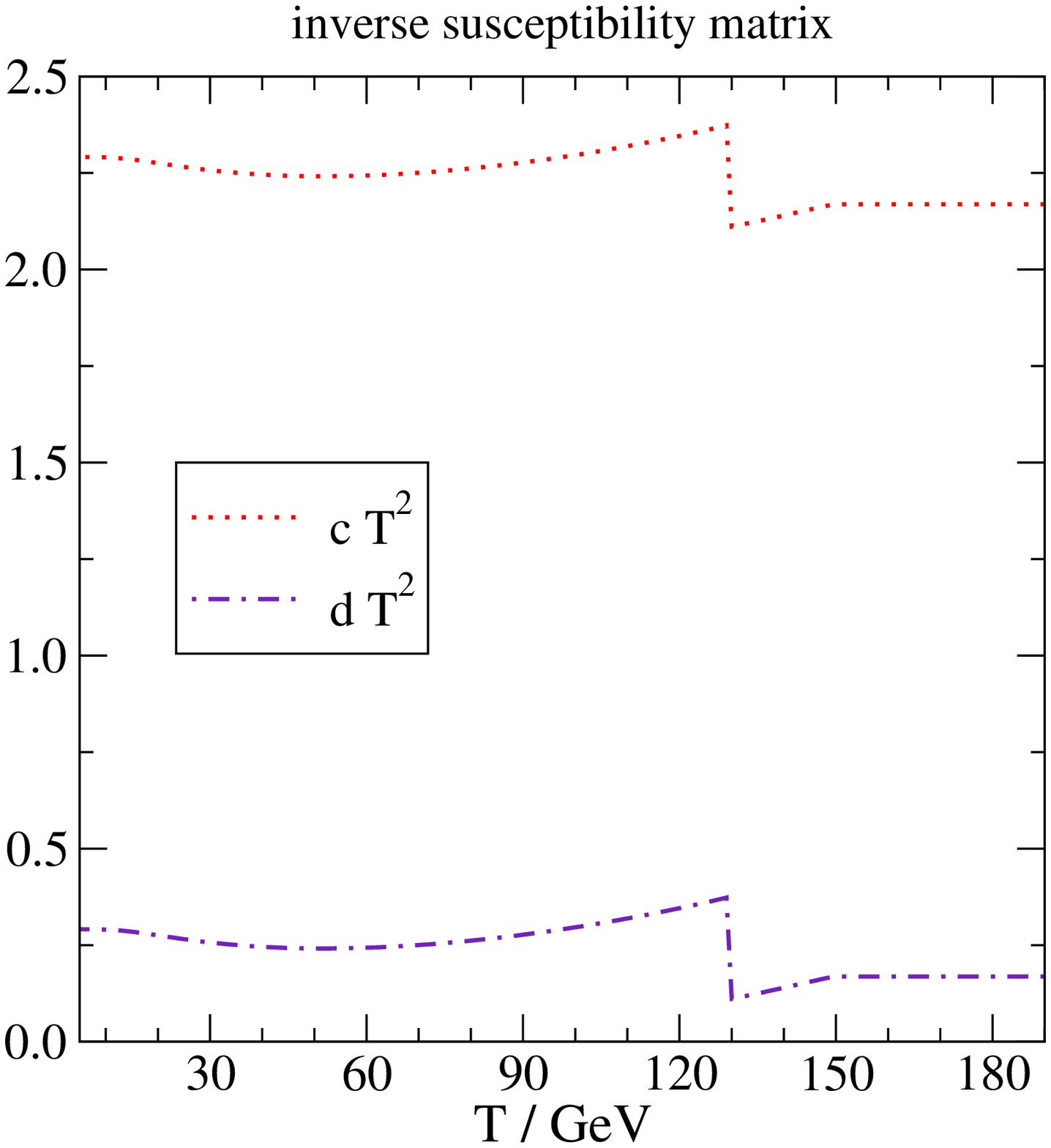}
}


\caption[a]{\small
 The coefficients $a$ and $b$ determining the lepton number
 susceptibility matrix (cf.\ \eq\nr{susc}) and 
 $c$ and $d$ determining its inverse (cf.\ \eq\nr{inv_susc}). 
 The apparent discontinuity at $T =  130$~GeV originates from the fact  
 that the $B+L$ violating rate 
 was assumed to be in equilibrium at $T > 130$~GeV and 
 out of equilibrium at $T < 130$~GeV. Treating this regime precisely
 requires solving a non-equilibrium problem with a finite 
 $B+L$ violating rate~\cite{sphaleron}, however in practice it  
 may be sufficient to solve separate non-equilibrium problems 
 on both sides and connect the solutions continuously. 
}

\la{fig:susc}
\end{figure}

Here we compute the susceptibility matrix defined below
\eq\nr{gamma_ab} to leading order in Standard Model couplings. 
Two regimes are considered: 
$5\,\mbox{GeV} \lsim T \lsim 130$~GeV
so that $B+L$ violation is out of thermal equilibrium~\cite{sphaleron}; 
and $T \gsim 130$~GeV
so that $B+L$ violation is in equilibrium. 
The methods of the computation have been discussed
in refs.~\cite{washout,dmpheno}, whereas the general approach
dates back to ref.~\cite{khlebnikov}.

Given that at $T \gsim 5$~GeV all lepton flavours are degenerate, 
the susceptibility matrix 
$\Xi^{ }_{ab} = \partial n^{ }_a / \partial {\mu^{ }_b} 
$
takes the form 
\be
 \Xi = \left( 
 \begin{array}{ccc}
  a & b & b \\ 
  b & a & b \\ 
  b & b & a 
 \end{array}
 \right) 
 \;. \la{susc}
\ee
Its inverse, playing a role in \eq\nr{gamma_ab}, reads
\be
 \Xi^{-1} = \left( 
 \begin{array}{ccc}
  c & d & d \\ 
  d & c & d \\ 
  d & d & c 
 \end{array}
 \right) 
 \;, \quad
 c = \frac{a+b}{(a-b)(a+2b)} 
 \;, \quad
 d = \frac{-b}{(a-b)(a+2b)}
 \;. \la{inv_susc}
\ee
The functions $a, ..., d$ 
are plotted in \fig\ref{fig:susc}. We now give
details concerning the computation. 

The computation proceeds by assigning chemical potentials $\mu_a$
to the different lepton densities; the chemical potential 
$\mu^{ }_{q}$ to the quark number density; and by denoting 
the zero components of the gauge potentials by 
\be
 \muY^{ } \equiv i g^{ }_1 {B}^{ }_0 \;, \quad
 \muA^{ } \equiv i g^{ }_2 {A}^{3}_0
 \;. 
\ee
In the regime $T > 130$~GeV, the quark chemical potential is 
eliminated through the sphaleron constraint 
$\mu^{ }_{q} = -\fr19 \sum_a \mu_a $, whereas
for $T < 130$~GeV it is expressed in terms of a baryon 
chemical potential as $\mu_q = \mu^{ }_\rmii{$B$}/3$; 
subsequently we make a Legendre transform to an ensemble
with a fixed baryon density $n^{ }_\rmii{$B$}$ and then set
$n^{ }_\rmii{$B$} \to 0$ in comparison with lepton densities. 
The gauge potentials $\mu^{ }_\rmii{$Y$}$ and $\mu^{ }_\rmii{$A$}$
are eliminated by requiring charge neutrality, 
$\partial p / \partial \mu^{ }_\rmii{$Y$} = 
\partial p / \partial \mu^{ }_\rmii{$A$} = 0$, where
$p = - \Omega/V$ is the pressure,  
$\Omega$ is the grand canonical potential, 
and $V$ is the volume. Subsequently,
$\Xi^{ }_{ab} = \frac{\partial^2 p }{\partial \mu_a\partial \mu_b}$. 

The contributions of various particle species  to $p$
are parametrized by the susceptibilities
\ba
 \chi^{ }_\rmii{F}(m) & \equiv & 
 \int_\vec{p} \bigl[ - 2 \nF{}'(E) \bigr] 
 \; = \; 
 \frac{m^2}{\pi^2} \sum_{n=1}^{\infty} (-1)^{n+1}
  K^{ }_2 \Bigl( \frac{n m}{T}\Bigr)
 \;\; \stackrel{m\to 0}{\to} \;\; \frac{T^2}{6}
 \;, \la{chif} \\ 
 \chi^{ }_\rmii{B}(m) & \equiv & 
 \int_\vec{p} \bigl[ - 2 \nB{}'(E) \bigr] 
 \; = \; 
 \frac{m^2}{\pi^2} \sum_{n=1}^{\infty} 
  K^{ }_2 \Bigl( \frac{n m}{T}\Bigr)
 \;\; \stackrel{m\to 0}{\to} \;\; \frac{T^2}{3}
 \;, \la{chib} 
\ea
where $K^{} _2$ is a modified Bessel function. 
At leading order\footnote{%
 We stress that the gauge boson contribution is treated 
 consistently only in the regimes $\mW \ll \pi T$ and $\mW \gg \pi T$. 
 For $\mW \sim gT$ the mass dependence amounts to a correction of 
 $\rmO(g)$ which is not correctly represented by this expression. 
 For $\mW \sim \pi T$ the susceptibilities are parametrically
 of the same order as the unknown $\rmO(g^2)$ corrections and do 
 not constitute any theoretically well-defined subset thereof. 
 } 
we obtain
\ba
 p(T,\mu) - p(T,0) & = & 
 \chi^{ }_\rmii{F}(0) 
 \, \biggl[ 
   12 \mu_q^2 + 4 \muY^{ } \mu_q + \sum_a \frac{3\mu_a^2}{2}
  - 2 \muY^{ }\sum_a \mu_a + \frac{9 \muA^2}{4} + \frac{49\muY^2}{12}
 \biggr]
 \nn 
 & + & 
 \chi^{ }_\rmii{F}(m_t) 
 \, \biggl[ 
   3 \mu_q^2
  - \frac{3 \muA^{ }\mu_q}{2} 
  + \frac{5 \muY^{ } \mu_q}{2}
  + \frac{3 \muA^2}{8} - \frac{\muA^{ }\muY^{ }}{4} + \frac{17\muY^2}{24}
 \biggr]
 \nn 
 & + & 
 \chi^{ }_\rmii{F}(m_b) 
 \, \biggl[ 
   3 \mu_q^2
  + \frac{3 \muA^{ }\mu_q}{2} 
  - \frac{\muY^{ } \mu_q}{2}
  + \frac{3 \muA^2}{8} + \frac{\muA^{ }\muY^{ }}{4} + \frac{5\muY^2}{24}
 \biggr]
 \nn 
 & + & 
  \Bigl[ \chi^{ }_\rmii{B}(m_\phi) + \chi^{ }_\rmii{B}(\mZ) \Bigr]
   \, \frac{(\muA + \muY)^2}{16}
  \; + \; 
  \chi^{ }_\rmii{B}(\mW)
 \, \biggl[
   \frac{9\muA^2}{8} - \frac{\muA^{ }\muY^{ }}{4} + \frac{\muY^2}{8} 
 \biggr]
 \nn 
 & + & 
  \frac{ v^2 (\muA + \muY)^2}{8}
 \; + \; 
 \rmO\Bigl( \frac{g v \mu^2 T}{4\pi} ,
            \frac{g^2 v^2 \mu^2}{16\pi^2} , \mu^4\Bigr) 
 \;, \la{delta_pmu}
\ea
where $ v^2  \simeq - m_\phi^2 / \lambda$ 
is the thermal Higgs expectation value
and $\mu$ denotes generically all chemical potentials. 
It can be checked that for $\mu_q, \mu_a \to 0$ and taking 
the temperature to be larger than all masses, this expression
reproduces the Debye masses in \eq\nr{Debye}. 

As far as the coefficients in \eqs\nr{susc} and \nr{inv_susc} are
concerned, 
for the regime $T > 130$~GeV 
a straightforward minimization leads to 
\be
 c - d = \frac{1}{a - b} = \frac{1}{3 \chi^{ }_\rmii{F}(0)}
 \;.  \la{aminusb}
\ee
The other linear combinations have more complicated expressions, for 
instance 
\be
 \frac{a}{2} =  
 \ch{11}^{ } + \frac{
 \ch{1Y}^2 \ch{AA}^{ } 
 - \ch{1A}^{ } \ch{1Y}^{ } \ch{AY}^{ }
 + \ch{1A}^2 \ch{YY}^{ }}
 {\ch{AY}^2 - 4 \ch{AA}^{ } \ch{YY}^{ } }
 \;, \la{ao2}
\ee
where the $\chi$'s are combinations of susceptibilities coupling 
to different chemical potentials: 
\ba
 \ch{11}^{ }  & = & \frac{89 \chF^{ }(0)}{54}
  + \frac{\chF^{ }(m_t) + \chF^{ }(m_b)}{27}
 \;, \la{ch11} \\ 
 \ch{1A}^{ }  & = & 
 \frac{\chF^{ }(m_t) - \chF^{ }(m_b)}{6}
 \;, \\ 
 \ch{1Y}^{ }  & = & - \frac{22 \chF^{ }(0)}{9}
  + \frac{\chF^{ }(m_b) - 5 \chF^{ }(m_t) }{18}
 \;, \\ 
 \ch{AA}^{ }  & = & 
 \frac{ v^2 }{8} 
  + \frac{9 \chF^{ }(0)}{4}
  + \frac{3[\chF^{ }(m_t) + \chF^{ }(m_b)] }{8}
  + \frac{\chB^{ }(m_\phi) + \chB^{ }(\mZ)
   + 18 \chB^{ }(\mW) }{16}
 \;, \hspace*{8mm} \\ 
 \ch{AY}^{ }  & = & 
  \frac{ v^2 }{4} 
  + \frac{\chF^{ }(m_b) - \chF^{ }(m_t) }{4}
  + \frac{\chB^{ }(m_\phi) + \chB^{ }(\mZ)
   - 2 \chB^{ }(\mW) }{8}
 \;, \\ 
 \ch{YY}^{ }  & = & 
 \frac{ v^2 }{8} 
  + \frac{49 \chF^{ }(0)}{12}
  + \frac{17 \chF^{ }(m_t) + 5 \chF^{ }(m_b) }{24}
  + \frac{\chB^{ }(m_\phi) + \chB^{ }(\mZ)
   + 2 \chB^{ }(\mW) }{16}
 \;. \hspace*{8mm} \la{chYY}
\ea
Going to the symmetric phase, it can be checked that 
the resulting expressions for $c$ and $d$ agree with the 
leading-order results given in ref.~\cite{washout}. 

For the regime $T < 130$~GeV, we may first eliminate $\mu_q$ from the 
requirement of vanishing baryon density, whereby \eqs\nr{ch11}--\nr{chYY}
get replaced with 
\ba
 \ch{11}^{ }  & = & \frac{3 \chF^{ }(0)}{2}
 \;, \quad 
 \ch{1A}^{ }  \; = \; 
  0 
 \;, \quad
 \ch{1Y}^{ }  \; = \; - 2 \chF^{ }(0)
 \;, \\ 
 \ch{AA}^{ }  & = & 
 \frac{ v^2 }{8} 
  + \frac{15 \chF^{ }(0)}{4}
  + \frac{\chB^{ }(m_\phi) + \chB^{ }(\mZ)
   + 18 \chB^{ }(\mW) }{16}
 \nn 
 & & \; 
   +\, \frac{3[\chF^{2}(m_t) + 6 \chF^{ }(m_t) \chF^{ }(m_b)
   +  \chF^{2}(m_b) - 32 \chF^{2}(0)] }
  {16[4 \chF^{ }(0) + \chF^{ }(m_t) + \chF^{ }(m_b) ]}
 \;, \\ 
 \ch{AY}^{ }  & = & 
 \frac{ v^2 }{4} 
  + \frac{\chB^{ }(m_\phi) + \chB^{ }(\mZ)
   -2 \chB^{ }(\mW) }{8}
   + \frac{3[\chF^{ }(m_t) - \chF^{ }(m_b)]^2 }
  {8[4 \chF^{ }(0) + \chF^{ }(m_t) + \chF^{ }(m_b) ]}
 \;, \hspace*{8mm} \\ 
 \ch{YY}^{ }  & = & 
 \frac{ v^2 }{8} 
  + \frac{21 \chF^{ }(0)}{4}
  + \frac{\chB^{ }(m_\phi) + \chB^{ }(\mZ)
   + 2 \chB^{ }(\mW) }{16}
 \nn 
 & & \; 
   +\, \frac{3[\chF^{2}(m_t) + 6 \chF^{ }(m_t) \chF^{ }(m_b)
   +  \chF^{2}(m_b) - 32 \chF^{2}(0)] }
  {16[4 \chF^{ }(0) + \chF^{ }(m_t) + \chF^{ }(m_b) ]}
 \;.
\ea
Eq.~\nr{aminusb} and the form of \eq\nr{ao2} remain unchanged. 
At low temperatures the expressions agree with those given 
in ref.~\cite{dmpheno}. 
Numerical results are shown in \fig\ref{fig:susc}.

The numerical uncertainties of the leading-order susceptibilities
have been discussed in refs.~\cite{washout,susc}. Because of effects
of the QCD gauge coupling on quark number susceptibilities, and because
of infrared sensitive bosonic effects only suppressed by $\rmO(g)$, 
uncertainties are expected to be on the $\sim 20$\% level.

%
\section{Hard Thermal Loop resummed leptons and gauge bosons}
\la{app:B}

For completeness we list in this appendix the Hard Thermal Loop 
resummed~\cite{htl1,htl2} 
spectral functions corresponding to the lepton and gauge
field propagators in the regime where the masses of these particles,
including the contribution from the Higgs mechanism, 
are parametrically at most of $\rmO(gT)$. 

The lepton spectral function, defined as a four-vector originating from
the imaginary part of the retarded propagator, has the form
\ba
 \rho^{ }_\ell (q_0,\vec{q}) & \equiv & 
 \Bigl( q_0\, \hat{\rho}_0 (q_0,q)
 ,\vec{q}\, \hat{\rho}_s (q_0,q)
 \, \Bigr)
 \;, \la{rhoell} \\
 \hat{\rho}^{ }_0(q_0,q) & = & 
 \im \Biggl\{ \frac{1 - \frac{m_\ell^2 L}{2q_0}}
 {\bigl[q_0 - \frac{ m_\ell^2 L}{2}\bigr]^2 -
  \bigl[q + \frac{m_\ell^2 (1-q_0 L)}{2 q}\bigr]^2} \Biggr\}
 \;, \la{rho0}
 \\
 \hat{\rho}^{ }_s(q_0,q) & = & 
 \im \Biggl\{ \frac{ 1 + \frac{m_\ell^2 (1-q_0 L)}{2 q^2}  }
 {\bigl[q_0 - \frac{ m_\ell^2 L}{2}\bigr]^2 -
  \bigl[q + \frac{m_\ell^2 (1-q_0 L)}{2 q}\bigr]^2} \Biggr\}
 \;, \la{rhos}
\ea
where $L \equiv \frac{1}{2q} \ln \frac{q_0 + q}{q_0 - q}$ 
and $q_0$ has a small positive imaginary part, so that 
$\im L = -\pi/(2q)$ for $q > q_0$. The ``asymptotic'' 
thermal mass appearing in these equations is 
given in \eq\nr{mell}. 

The gauge field spectral function also 
contains two independent structures, associated with the 
(imaginary-time) projectors
\ba
 \mathbbm{P}^\rmii{T}_{\mu\nu}(Q) & \equiv &  \delta_{\mu i} \delta_{\nu j}
 \biggl( \delta_{ij} - \frac{q_i q_j}{q^2} \biggr)
 \;, \la{PT} \\ 
 \mathbbm{P}^\rmii{E}_{\mu\nu}(Q) & \equiv & 
 \delta_{\mu\nu} - \frac{Q_\mu Q_\nu }{Q^2}
 - \mathbbm{P}^\rmii{T}_{\mu\nu}(Q)
 \;. \la{PE}
\ea
For $W^\pm$ the spectral functions 
corresponding to these projections read
\ba
 \rho^{ }_\rmii{T$2$}(q_0,q) & = & 
 \im \biggl\{ 
 \frac{1}{
  q^2 - q_0^2 + \mW^2 +  \Pi^{ }_\rmii{T2}
 }
 \biggr\}
 \;, \quad
 \Pi^{ }_\rmii{T$2$} \; \equiv \;
  \frac{m_\rmii{E$2$}^2}{2 q^2}\Bigl[
  q_0^2 + q_0(q^2-q_0^2) L \Bigr] 
 \;, \la{SigT} \\
 {\rho}^{ }_\rmii{E$2$}(q_0,q) & = & 
 \im \biggl\{ 
 \frac{1}{
  q^2 - q_0^2 + \mW^2 +  \Pi^{ }_\rmii{E2}
 }
 \biggr\}
 \;, \quad
 \Pi^{ }_\rmii{E$2$} \; \equiv \; 
  \frac{ m_\rmii{E$2$}^2 (q^2 - q_0^2)}{q^2}
  \Bigl[ 1 - q_0\, L \Bigr] 
 \;, \hspace*{5mm} \la{SigE} 
\ea
where again $q_0$ has a small positive imaginary part, 
and $m_\rmii{E$2$}^2$ is from \eq\nr{Debye}. 
The $Z$-channel spectral function, appearing
in \eqs\nr{indirect_HTL_gauge}
and \nr{HTL_g_soft}, is more complicated 
because the self-energies lead to a different 
mixing angle than in vacuum. 
It can be expressed as 
\be
 \rho^{ }_\rmii{T$Z$} \; \equiv \; 
 \im \biggl\{  
 \frac{q^2 - q_0^2 + c^2 \Pi^{ }_\rmii{T1}
           + s^2 \Pi^{ }_\rmii{T2} }
 {(q^2 - q_0^2)^2
 + (q^2 - q_0^2) (\mZ^2 + \Pi^{ }_\rmii{T1} + \Pi^{ }_\rmii{T2})
 + \mZ^2 (
             c^2 \Pi^{ }_\rmii{T1}
           + s^2 \Pi^{ }_\rmii{T2} ) 
  + \Pi^{ }_\rmii{T1}\, \Pi^{ }_\rmii{T2} }
 \biggr\}^{ }_{ }
 \;, \la{rhoZ}
\ee
and correspondingly for $\rho^{ }_\rmii{E$Z$}$, where
the weak mixing angles $c,s$ have been 
defined around \eq\nr{theta_w}. 
The self-energy $ \Pi^{ }_\rmii{T1} $ is like
$\Pi^{ }_\rmii{T2}$ in \eq\nr{SigT} but with the 
replacement $m^{ }_\rmii{E2} \to m^{ }_\rmii{E1}$.
In the symmetric phase, 
i.e.\ $\mZ \to 0 $, the $Z$ channel 
spectral function simplifies into
\be
  \rho^{ }_\rmii{T$Z$}  \; = \; 
  s^2  \rho^{ }_\rmii{T1} + 
  c^2  \rho^{ }_\rmii{T2}
 \;, 
\ee
so that 
$
 (g_1^2 + g_2^2) \rho^{ }_\rmii{T$Z$} = 
 \sum_{i=1}^2 g_i^2 \rho^{ }_\rmii{T$i$}
$.
Finally, we note that the photon and 
the mixed photon-$Z$ propagators can 
be expressed in a form similar to \eq\nr{rhoZ}, 
\ba
 \rho^{ }_\rmii{T$Q$} & \equiv & 
 \im \biggl\{  
 \frac{q^2 - q_0^2 + \mZ^2 + s^2 \Pi^{ }_\rmii{T1}
           + c^2 \Pi^{ }_\rmii{T2} }
 { \Delta  }
 \biggr\}^{ }_{ }
 \;, \\
 \rho^{ }_\rmii{T$ZQ$} & \equiv & 
 \im \biggl\{  
 \frac{c s (\Pi^{ }_\rmii{T1} - 
           \Pi^{ }_\rmii{T2}) }
 { \Delta   }
 \biggr\}^{ }_{ }
 \;, 
\ea
where $\Delta$ is the denominator of \eq\nr{rhoZ}.

%
\section{Integrated matrix elements for indirect
 $2\leftrightarrow 2$ processes}
\la{app:C}

We list in this appendix the functions defined
in \eq\nr{k0Gamma}, obtained after carrying out all but two of the
phase space integrals in \eq\nr{boltzmann2}. Making use of the notation
of \eqs\nr{not1}--\nr{qpm} we get
\ba
 \Xi_\rmi{$s$1} 
 & = & 
  \frac{\nS}{4} \bigl( g_1^4 + 3 g_2^4 \bigr)  \, 
  \bigl[ q + 2 T (\lnbplus - \lnbminus) \bigr]
 \nn 
  & + & \fr34 \bigl( g_1^4 + 6 g_1^2 g_2^2 - 3 g_2^4 \bigr)  \, 
  \bigl[ q + 2 T (\lnfplus - \lnfminus) \bigr]
 \nn 
  & - & 
  \Bigl( 6g_2^4 + \frac{g_1^4 + 3 g_2^4}{2} \nS \Bigr)  
 \biggl\{ 
  \frac{q}{3} + \frac{T}{2q^2} 
  \bigl[ 
   q^2 + (q_0 - 2\ko)^2 
  \bigr]    (\lnbplus - \lnbminus) 
 \nn 
 & & \; 
   + \, \frac{T^2}{q^3} 
   \bigl[ 
   q^2 -3 (q_0 - 2\ko)^2 
  \bigr]
  \Bigl[    \libplus{2} + \libminus{2} 
   + \, \frac{2 T}{q} 
    \bigl(  \libplus{3} - \libminus{3}  \bigr) 
  \Bigr]
 \biggr\} 
 \nn 
  & + & 
  \Bigl( 6g_2^4 + \fr{10}3 g_1^4 \Bigr) \nG  
 \biggl\{ 
  \frac{q}{3} + \frac{T}{2q^2} 
  \bigl[ 
   q^2 + (q_0 - 2\ko)^2 
  \bigr]    (\lnfplus - \lnfminus) 
 \nn 
 & & \; 
   + \, \frac{T^2}{q^3} 
   \bigl[ 
   q^2 -3 (q_0 - 2\ko)^2 
  \bigr]
  \Bigl[    \lifplus{2} + \lifminus{2} 
   + \, \frac{2 T}{q} 
    \bigl(  \lifplus{3} - \lifminus{3}  \bigr)
  \Bigr]
 \biggr\} 
 \;, \\ 
 \Xi_\rmi{$s$2} 
 & = & 
  - \frac{\nS}{2} \bigl( g_1^4 + 3 g_2^4 \bigr)  \, 
  \bigl[ q + T (\lnbplus + \lnfplus - \lnbminus - \lnfminus) \bigr]
 \nn 
 & + & 
 \Bigl( \frac{ g_1^2 + 3g_2^2}{2} \Bigr)^2 \, 
 \biggl\{
  \frac{q}{2} + \frac{T}{q} 
  \bigl[ 
   (\ko - \qm)(\lnfplus - \lnbminus) + (\ko - \qp)(\lnfminus - \lnbplus)
  \bigr] 
 \nn 
 & & \; 
   + \, \frac{T^2}{q^2} (2 \ko - q_0) 
    \bigl(  \libplus{2} + \lifminus{2} - \lifplus{2} - \libminus{2}  \bigr)
 \biggr\}
 \;, \\ 
 \Xi_\rmi{$t$1} 
 & = & 
  \Bigl( 6g_2^4 + \frac{g_1^4 + 3 g_2^4}{2} \nS \Bigr)  
 \biggl\{ 
  \frac{T}{2q^2} 
  \bigl[ 
   q^2 + (q_0 - 2\ko)^2 
  \bigr]    (\lnbplus - \lnbminus) 
 \nn 
 & & \; 
   + \, \frac{T^2}{q^3} 
   \bigl[ 
   q^2 -3 (q_0 - 2\ko)^2 
  \bigr]
  \Bigl[    \libplus{2} - \libminus{2} 
   + \, \frac{2 T}{q} 
    \bigl(  \libplus{3} - \libminus{3}  \bigr)
  \Bigr]
 \biggr\} 
 \nn 
  & - & 
  \Bigl( 6g_2^4 + \fr{10}3 g_1^4 \Bigr) \nG  
 \biggl\{ 
  \frac{T}{2q^2} 
  \bigl[ 
   q^2 + (q_0 - 2\ko)^2 
  \bigr]    (\lnfplus - \lnfminus) 
 \nn 
 & & \; 
   + \, \frac{T^2}{q^3} 
   \bigl[ 
   q^2 -3 (q_0 - 2\ko)^2 
  \bigr]
  \Bigl[    \lifplus{2} - \lifminus{2} 
   + \, \frac{2 T}{q} 
    \bigl(  \lifplus{3} - \lifminus{3}  \bigr)
  \Bigr]
 \biggr\} 
 \;, \\ 
 \Xi_\rmi{$t$2} 
 & = & 
 \Bigl( \frac{ g_1^2 + 3g_2^2}{2} \Bigr)^2 \, 
 \biggl\{
  \frac{T}{q} 
  \bigl[ 
   (\ko - \qm)(\lnfplus - \lnbminus) + (\ko - \qp)(\lnfminus - \lnbplus)
  \bigr] 
 \nn 
 & & \; 
   + \, \frac{T^2}{q^2} (2 \ko - q_0) 
    \bigl(  \libplus{2} + \libminus{2} - \lifplus{2} - \lifminus{2}  \bigr)
 \biggr\}
 \;.
\ea

%
\section{Towards soft momentum transfer at next-to-leading order}
\la{app:D}

In \se\ref{ss:soft_t} we accounted for the leading divergence 
at $q,q_0 \ll \ko$ given
in \eq\nr{asyt1}. Let us now show that the origin of the next-to-leading
divergence can be understood as well. 

If we change variables from $q$ to $q_\perp$, defined in \eq\nr{qperp2}, 
and integrate over $q_0$, then $\Gamma$ can be expressed as an integral
over ${q}_\perp$. It turns out that the integrand is equivalent to 
the ``transverse collision kernel'', $C(q_\perp)$, determined up to 
NLO in QCD in the domain $q_\perp \sim m^{ }_\rmii{E}$
in ref.~\cite{sum2}, or the elastic scattering cross section,   
${\rm d}\Gamma^{ }_\rmi{el} / {\rm d}^2\vec{q}_\perp$, 
determined up to $\rmO(g^4)$ for  
$q_\perp \gg m^{ }_\rmii{E}$ in ref.~\cite{ax}.
Concretely, \eq(20) of ref.~\cite{sum2} can for 
$q_\perp \gg m^{ }_\rmii{E}$ be expanded as 
\be
 C(q_\perp) \; 
 \stackrel{q_\perp \gg m^{ }_\rmii{E}}{=} 
 \; g^2 T \CF \, 
 \biggl[
  \frac{m_\rmii{E}^2}{q_\perp^4} - \frac{g^2 T \CA}{16 q_\perp^3} + 
  \rmO\bigl( g^4 T^2 \bigr) 
 \biggr]
 \;,
 \la{sch_asym}
\ee
where $\CF \equiv (\Nc^2 -1)/(2\Nc)$ and 
$\CA \equiv \Nc$ are Casimir factors related
to the fundamental and adjoint representation, respectively. 
We now show that the NLO divergence in \eq\nr{asyt1} 
amounts precisely to the NLO term in \eq\nr{sch_asym}.

Let us introduce a scale $\Lambda$ in the range $gT \ll \Lambda \ll \ko$
and consider the contribution to \eq\nr{k0Gamma} from momenta 
$q \le \Lambda$. In this domain $q_\perp$ can be approximated
as in \eq\nr{qperp}, and we can change variables according to 
\ba
 \int_0^\Lambda \! {\rm d}q \int_{-q}^{q} \! {\rm d}q_0 \, f(q_0,q) 
 & = & 
 \int_0^\Lambda \! {\rm d}q_\perp \, q_\perp 
 \int_{-\sqrt{\Lambda^2 - q_\perp^2}}^{\sqrt{\Lambda^2 - q_\perp^2}}
  {\rm d}q_0 \, 
 \frac{ f\bigl(  q_0, \sqrt{q_\perp^2 + q_0^2} \bigr) }
      {\sqrt{q_\perp^2 + q_0^2}} 
 \nn
 & = & 
 \int_0^\Lambda \! {\rm d}q_\perp \, q_\perp 
 \int_1^{\Lambda/q_\perp}  {\rm d}x \, \sum_{\sigma = \pm}
 \frac{f( \sigma\, q_\perp \sqrt{x^2 - 1}, q_\perp x)
 }{\sqrt{x^2 - 1}} 
 \;.  \hspace*{8mm}
\ea
The leading and NLO divergences originate from a domain where we can 
approximate 
\be
 1 + \nB{}(q_0) - \nF{}(\ko - q_0) \approx \frac{T}{q_0}
 \;.
\ee
Thereby we are left with the integrals 
\ba
 \int_1^{\Lambda/q_\perp} \! {\rm d}x \, \frac{1}{x^4 \sqrt{x^2 - 1}}
 & = & 
 \fr23 
 \; + \; \rmO\Bigl(\frac{q_\perp}{\Lambda}\Bigr)\;, \\ 
 \int_1^{\Lambda/q_\perp} \! {\rm d}x \,
 \biggl[ 
   \frac{6}{x^3 \sqrt{x^2-1}} + \frac{3-2x^2}{x^4(x^2-1)}
 \ln \frac{x+\sqrt{x^2-1}}{x-\sqrt{x^2-1}}
 \biggr]
 & = &
 \frac{\pi^2}{2} 
 \; + \; \rmO\Bigl(\frac{q_\perp}{\Lambda}\Bigr)
 \;. 
\ea
Inserting these, the leading and NLO infrared divergences to $\Gamma$
become
\be
 \Gamma^\rmi{hard,expanded}_\rmi{$2\to 2$} = \int_0^\Lambda \! 
 \frac{{\rm d}^2\vec{q}_\perp}{(2\pi)^2}
 \, \biggl\{ 
   \frac{g_1^2T }{4} 
     \biggl[ 
       \frac{m_\rmii{E1}^2}{q_\perp^4}
      - \frac{g_1^2 T (\fr{\nS}2)}{16 q_\perp^3}
     \biggr] 
   + 
   \frac{3g_2^2T }{4} 
     \biggl[ 
       \frac{m_\rmii{E2}^2}{q_\perp^4}
      - \frac{g_2^2 T (2 + \fr{\nS}2)}{16 q_\perp^3}
     \biggr] 
 \biggr\} 
 \;. 
\ee
The non-Abelian part agrees exactly with \eq\nr{sch_asym} after 
inserting $\CF = 3/4$ and $\CA = 2$; the contribution from the Higgs 
field has a similar structure but a different group theory factor. 

We conclude that accounting
properly for the subleading divergence in \eq\nr{asyt1} would 
require an NLO computation similar to that performed in ref.~\cite{sum2}
but generalized to the broken phase and including the contribution
of the Higgs field. 

%
\section{Fixing the parameters}
\la{app:E}

Given that the computation presented in the main 
body of this paper is a leading-order one, without NLO logarithms 
to cancel the renormalization scale dependence, the choices that 
we make for the running parameters have some numerical significance. 
Our general approach follows that in ref.~\cite{interpolation}.
The running $\msbar$ parameters are fixed in terms of 
physical quantities (pole masses, Fermi constant) at a scale 
$\bmu^{ }_0 \equiv \mZ$, making use of the NLO relations specified 
in ref.~\cite{generic}. The Higgs mass is set to $\mH = 125$~GeV. 
The renormalization scale appearing in 
the thermal computations is chosen as $\bmu \equiv \pi T$. 
If $\bmu > \bmu^{ }_0$, the couplings are evolved through 
1-loop renormalization group equations. If $\bmu < \bmu^{ }_0$, 
whereby we find ourselves in a massive regime, the couplings 
are not evolved at all. We note that the strong gauge coupling, 
which should logically be evolved with 5 quark flavours in this regime, 
does not appear in our expressions at the current order. The Higgs
expectation value is determined from the (resummed) 
leading-order expression $v^2 = - m_\phi^2 / \lambda$, where 
$m_\phi^2$ is the effective thermal Higgs mass parameter
from \eq\nr{mell}. With this procedure the electroweak 
``transition'' takes in our results place at $T \approx 150$~GeV, 
whereas a non-perturbative investigation of the crossover region
postpones it to $T \approx 160$~GeV~\cite{dono}.   

%

\end{document}